\documentclass[useAMS,usenatbib]{mn2e}
\usepackage{graphicx}
\title[A new determination of the primordial He abundance]
{A new determination of the primordial He abundance using   
the He~{\sc i} $\lambda$10830\AA\ emission line: cosmological implications}
\author[Y. I. Izotov et al.]{Y.~I.~Izotov$^{1}$,
T.~X.~Thuan$^{2}$ and N.~G.~Guseva$^{1}$
\\
$^{1}$Main Astronomical Observatory,
                     Ukrainian National Academy of Sciences,
                     Zabolotnoho 27, Kyiv 03680,  Ukraine\\
$^{2}$Astronomy Department, University of Virginia, P.O. 
                     Box 400325, Charlottesville, VA 22904, USA\\
}
\begin{document}

\date{Accepted 1988 December 15. Received 1988 December 14; in original form 1988 October 11}

\pagerange{\pageref{firstpage}--\pageref{lastpage}} \pubyear{2012}

\maketitle

\label{firstpage}

\begin{abstract}
We present near-infrared spectroscopic observations of the high-intensity 
He~{\sc i} $\lambda$10830\AA\ emission line in 45 low-metallicity H~{\sc ii} 
regions. We combined these NIR data with spectroscopic data in the optical 
range to derive the primordial He abundance. The use of the
He~{\sc i} $\lambda$10830\AA\ line, the intensity of which is very sensitive 
to the density of the H~{\sc ii} region, greatly improves the determination of
the physical conditions in the He$^+$ zone. This results in a considerably 
tighter $Y$ -- O/H linear regression compared to all previous studies.
We extracted a final sample of 28 H~{\sc ii} regions with  H$\beta$ equivalent 
width EW(H$\beta$) $\geq$ 150\AA, excitation parameter
O$^{2+}$/O $\geq$ 0.8, and with helium mass 
fraction $Y$ derived with an accuracy better than 3\%. With this final sample 
we derived a primordial $^4$He mass fraction $Y_{\rm p}$ = 0.2551 $\pm$ 0.0022. 
The derived value of $Y_{\rm p}$ is higher than the one 
predicted by the standard big bang nucleosynthesis (SBBN) model.
Using our derived $Y_{\rm p}$ together with
 D/H = (2.53$\pm$0.04)$\times$10$^{-5}$, and 
the $\chi^2$ technique, we found that
the best agreement between these light element abundances is achieved
in a cosmological model with a baryon mass density $\Omega_{\rm b} h^2$ = 
0.0240$\pm$0.0017 (68\% CL), $\pm$0.0028 (95.4\% CL), $\pm$0.0034 (99\% CL) 
and an effective number of neutrino species $N_{\rm eff}$ = 
3.58$\pm$0.25 (68\% CL), $\pm$0.40 (95.4\% CL), $\pm$0.50 (99\% CL).
A non-standard value of $N_{\rm eff}$ is preferred at the 99\% CL, implying the 
possible existence of additional types of neutrino species. 
\end{abstract}

\begin{keywords}
galaxies: abundances --- galaxies: irregular --- 
galaxies: ISM --- cosmology: cosmological parameters.
\end{keywords}

\section[]{Introduction}\label{sec:intro}

In the standard theory of big bang nucleosynthesis (SBBN), given the 
number of light neutrino species, the abundances of light elements D, $^3$He, 
$^4$He  (hereafter He) and $^7$Li depend only on one cosmological 
parameter, the baryon-to-photon number ratio $\eta$, which is related 
to the baryon density parameter $\Omega_{\rm b}$, the present ratio of the 
baryon mass density to the critical density of the Universe, by
the expression 10$^{10}$$\eta$ = 273.9 $\Omega_{\rm b}h^2$,
where $h$~=~$H_0$/100~km~s$^{-1}$~Mpc$^{-1}$ and 
$H_0$ is the present value of the Hubble parameter \citep{St05,St12}. 

Because of the strong dependence of the D/H abundance ratio on $\Omega_{\rm b}h^2$ while 
the He mass fraction depends only logarithmically on the baryon density, 
deuterium is the light element of choice for determining the baryon mass
fraction. Its abundance can accurately be  measured in high-redshift 
low-metallicity QSO 
Ly$\alpha$ absorption systems. Although the data are still scarce -- there 
are only ten absorption systems for which such a D/H measurement has 
been carried out \citep{PC12} -- the measurements appear to converge to a mean 
primordial value D/H~$\sim$~(2.5 - 2.9)~$\times$~10$^{-5}$,
which corresponds to $\Omega_{\rm b}h^2$~$\sim$~0.0222 - 0.0223 
\citep{Io09,N12,PC12,C14}.
This estimate of $\Omega_{\rm b}h^2$ is in excellent agreeement with the value 
of 0.0221 - 0.0222 obtained from studies of the fluctuations of the cosmic 
microwave background (CMB) with {\sl WMAP} and {\sl Planck} \citep{K11,A13}.

However, although deuterium is sufficient to derive the baryonic mass density from BBN, 
accurate measurements of the primordial  He abundance $Y_{\rm p}$ are also needed as it plays a crucial role in determining cosmological parameters. First, it is needed to check the consistency of SBBN since this requires knowledge of the primordial abundances of at least two different relic elements. Second, while He is not a sensitive baryometer, its primordial abundance is much 
more sensitive to a non-standard, early Universe expansion rate and thus provides stronger constraints on non-standard physics, as compared to other primordial light elements. Thus the primordial helium abundance can help constrain the effective number of neutrino species and hence place restrictions on the presence of any "dark radiation" component \citep[e.g. "sterile" neutrinos, ][]{D14}. It also put constraints on any possible lepton asymmetry 
(an excess of neutrinos over antineutrinos, or vice versa) that may exist
\citep{St12}.

\begin{figure*}
\hbox{
\includegraphics[width=6.0cm,angle=-90.]{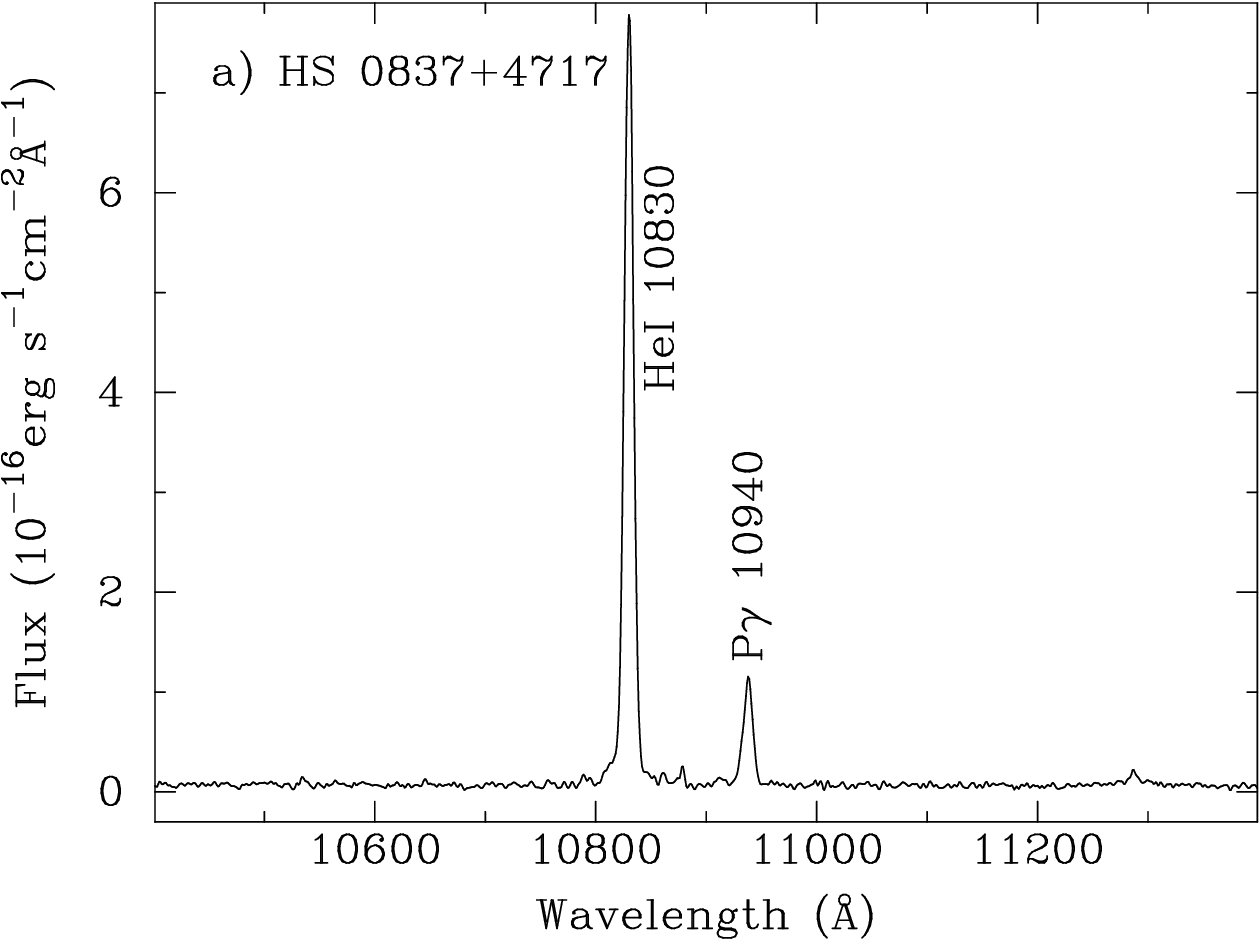}
\hspace{0.3cm}\includegraphics[width=6.0cm,angle=-90.]{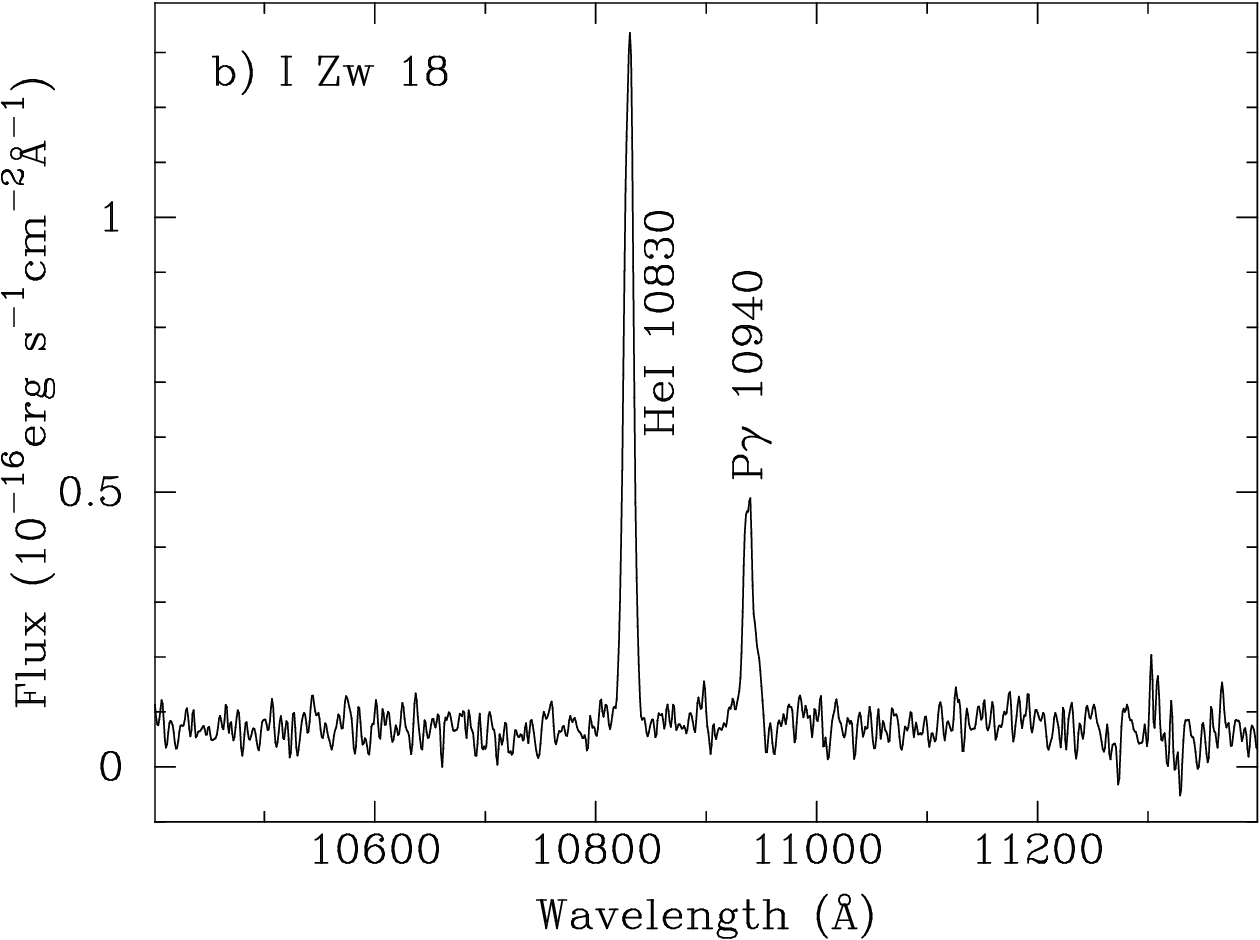}
}
\caption{Representative APO spectra of H~{\sc ii} regions showing the  
He~{\sc i} $\lambda$10830\AA\ and P$\gamma$10940\AA\ emission lines. The spectrum in 
(a) is that of a high-density H~{\sc ii} region while the one in (b) is that of 
low-density H~{\sc ii} region.}
\label{fig1}
\end{figure*}

To detect small deviations from SBBN and constrain cosmological parameters,
the primordial He abundance $Y_{\rm p}$ has to be determined to a level of 
accuracy of better than one percent. The primordial abundance of He 
can in principle be derived accurately from observations of 
the helium and hydrogen emission lines from low-metallicity blue compact dwarf 
(BCD) galaxies, which have undergone little chemical evolution. 

To attain the above accuracy, many conditions have to be met.
In particular, the observational data should be of good quality and the observational sample 
should be large to reduce statistical uncertainties. The primary concern of our group over the last two decades has been to gather such good data for large galaxy samples.
\citet{ITL94,ITL97} and 
\citet{IT98,IT04} have obtained high signal-to-noise
spectra of a total of 86 low-metallicity extragalactic H~{\sc ii} regions in 77 galaxies \citep{IT04}, 
constituting the HeBCD sample.
Later, \citet{I09,I11b} and \citet{G11}
have used archival data to collect a sample of 75 Very Large Telescope (VLT) spectra of low-metallicity H~{\sc ii} regions. Finally,
1442 high-quality spectra of low-metallicity H~{\sc ii} regions have been  
extracted from the Sloan Digital Sky Survey (SDSS) Data Release 7 (DR7) \citep{Ab09}. 
These were chosen to have the [O~{\sc iii}] $\lambda$4363 emission line measured with an accuracy better than 25\%,  
and to have all strongest He~{\sc i} emission lines in the optical range,
$\lambda$3889, $\lambda$4471, $\lambda$5876, $\lambda$6678, and $\lambda$7065, measured with good accuracy.
Because of the large sizes of these samples, it is now generally agreed 
that the accuracy of the determination of the primordial He
abundance is limited at present more by systematic uncertainties and biases
than by statistical errors.   

There are many known effects that need to be corrected for  
to transform the observed He~{\sc i} line intensities into a He abundance. 
Neglecting or misestimating them may lead 
to systematic errors in the $Y_{\rm p}$ determination that can be larger than the 
statistical errors.
Different empirical methods have been used to derive the He primordial abundance 
\citep[e.g. ][]{ITL94,ITL97,IT10,Pe07,A10,A11,A12,Av13}.  All of them make use of analytical fits for  
various physical processes 
\citep[e.g. ][]{I06}, including fits of
He~{\sc i} and H~{\sc i} emissivities and of effects that make the 
observed line intensities deviate from their recombination values. 
These effects include for instance collisional and fluorescent enhancements 
of He~{\sc i} recombination lines and underlying He~{\sc i} stellar absorption 
lines. A detailed discussion of the role of each of these various effects is 
given in \citet{I07}. 
Most of the known systematic effects have generally been 
taken into account in the most recent
work on the determination of the helium abundance\citep[e.g. ][]{I07,IT10,Pe07,A10,A12,Av13}. 

Based on the HeBCD sample, \citet{IT10} 
found $Y_{\rm p}$ = 0.2565~$\pm$~0.003 (1$\sigma$). Using a restricted subsample of 22 objects selected 
from the same HeBCD sample of \citet{I07}, \citet{A12} derived a value 
$Y_{\rm p}$ = 0.2534~$\pm$~0.0083. Subsequent to this work, a new set of He~{\sc i} line emissivities was put forward by \citet{P13}. With this new emissitivity set, \citet{I13} derived $Y_{\rm p}$ = 
0.254~$\pm$~0.003 (1$\sigma$) based on a large sample, while  \citet{Av13} obtained
 $Y_{\rm p}$ = 0.2465~$\pm$~0.0097 with the same restricted sample. We note
that, \citet{IT10} and \citet{I13}, by using much larger samples of 
H~{\sc ii} regions as compared to \citet{A12} and \citet{Av13}, 
determined $Y_{\rm p}$ with considerably smaller statistical errors.
Taking into account the statistical and systematic errors in the $Y_{\rm p}$ 
determination, one can conclude that these latest determinations of 
$Y_{\rm p}$ are broadly \citep[at the 1$\sigma$ level for the value by ][]{Av13} consistent with the prediction 
of SBBN based on the cosmic background (CMB) measurements of Planck, $Y_{\rm p}$ = 0.2477~$\pm$~0.0001 \citep{A13}.

\citet{I13} have also   
checked the validity of the above overall procedure for determining the He 
abundance by using photoionisation CLOUDY models \citep{F98,F13}.
Since a photoionisation code such as CLOUDY takes into account all the
processes affecting the He~{\sc i} line intensities, it should produce in principle
model H~{\sc ii} regions that are very similar in properties to real  H~{\sc ii} 
regions. \citet{I13} showed that the empirical method used by, e.g. \citet{IT10}, 
does reproduce very well the input CLOUDY helium mass fraction $Y$. However,
the physical conditions, as characterized by the electron temperature $T_{\rm e}$(He$^+$) and
the electron number density $N_{\rm e}$(He$^+$) are rather poorly reproduced, if
only the 5 strongest $\lambda$3889,  $\lambda$4471, $\lambda$5876,
$\lambda$6678, and $\lambda$7065 He~{\sc i} emission lines in the optical range 
are used for the $\chi^2$ minimisation
in the determination of best $Y$ values. \citet{I13} suggested that adding
the strong near-infrared He~{\sc i} $\lambda$10830 emission line may greatly
improve the determination of physical conditions and $Y$ because of the
very strong dependence of the intensity of this line on electron number density, and hence 
diminish the systematic uncertainties.

In this paper, we present for the first time near-infrared (NIR)
observations of a large sample of high-excitation low-metallicity galaxies,
aiming to improve the primordial He abundance determination by including the
He~{\sc i} $\lambda$10830 emission line, as suggested by \citet{I13}.
In sect. \ref{sec:obs}, we describe the observations and data reduction of
our sample. In Sect. \ref{real}, we discuss the method for $Y$ determination. 
In Sect. \ref{primo}, we present the linear regressions $Y$ -- O/H and derive
the primordial He mass fraction $Y_{\rm p}$. Cosmological implications of 
the derived $Y_{\rm p}$ are discussed in Sect. \ref{cosmo}. In particular, we
obtain the effective number of neutrino species $N_{\rm eff}$ and discuss 
constraints on the possible presence of ``dark radiation''. Section
\ref{summary} summarises our results.

\begin{figure}
\hbox{
\includegraphics[width=6.0cm,angle=-90.]{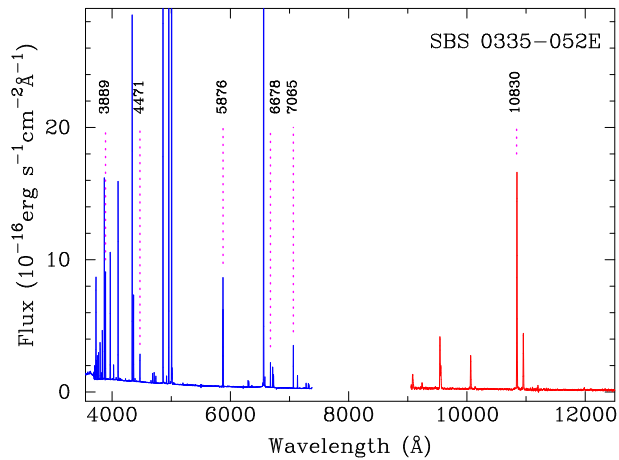}
}
\caption{Redshift-corrected spectra of SBS 0335$-$052E in the optical range
(blue) and in the near-infrared range (red) obtained with the 8.2m Very Large
Telescope (VLT) and the 8.4m Large Binocular Telescope (LBT), respectively.
The He~{\sc i} emission lines are marked by vertical dotted lines and labelled.}
\label{fig2}
\end{figure}

\begin{table*}
\centering
\begin{minipage}{170mm}
\caption{General characteristics.}
\label{tab1}
\begin{tabular}{@{}lcccccl@{}}\hline
Name             &R.A. (J2000)&Dec. (J2000)&Redshift&12+logO/H&References$^{\rm a}$&Other names \\
\hline
CGCG 007$-$025   &09:44:01.9  &$-$00:38:32 & 0.0048 &   7.79  &  1,2     &SHOC 270 \\
Haro 3           &10:45:22.4  &$+$55:57:37 & 0.0031 &   8.37  &  1,2     &NGC 3353, Mrk 35, SBS 1042$+$562 \\
HS 0837$+$4717   &08:40:29.9  &$+$47:07:10 & 0.0420 &   7.62  &  1,2     &SHOC 220 \\
HS 1734$+$5704   &17:35:01.2  &$+$57:03:09 & 0.0472 &   8.10  &  2       &SHOC 579 \\
I Zw 18 SE       &09:34:02.0  &$+$55:14:28 & 0.0026 &   7.23  &  2,3,4   &UGCA 116, Mrk 116, SBS 0930$+$554 \\
J0024$+$1404     &00:24:26.0  &$+$14:04:10 & 0.0142 &   8.40  &  2       &HS0021$+$1347 \\
J0115$-$0051     &01:15:33.8  &$-$00:51:31 & 0.0056 &   8.38  &  2       &H~{\sc ii} region in NGC 450 \\
J0301$-$0052     &03:01:49.0  &$-$00:52:57 & 0.0073 &   7.59  &  2       & \\
J0519$+$0007     &05:19:02.6  &$+$00:07:30 & 0.0444 &   7.43  &  1,5     & \\
J1038$+$5330     &10:38:44.9  &$+$53:30:05 & 0.0032 &   8.37  &  2       &H~{\sc ii} region in NGC 3310 \\
J1203$-$0342     &12:03:51.1  &$-$03:42:35 & 0.0130 &   8.49  &  2       & \\
J1253$-$0313     &12:53:06.0  &$-$03:12:59 & 0.0228 &   8.04  &  2       &SHOC 391 \\
J1624$-$0022     &16:24:10.1  &$-$00:22:03 & 0.0313 &   8.20  &  2       &SHOC 536 \\
KUG 0952$+$418   &09:55:45.6  &$+$41:34:30 & 0.0157 &   8.40  &  2       &HS 0952$+$4148 \\
Mrk 36           &11:04:58.3  &$+$29:08:23 & 0.0022 &   7.81  &  2,6     &UGCA 225, Haro 4, KUG 1102+294  \\
Mrk 59           &12:59:00.3  &$+$34:50:42 & 0.0026 &   8.03  &  2,3     &NGC 4861, Arp 209, I Zw 49  \\
Mrk 71           &07:28:42.8  &$+$69:11:21 & 0.0004 &   7.89  &  3       &NGC 2363 \\
Mrk 162          &11:05:08.1  &$+$44:44:47 & 0.0215 &   8.12  &  2,6     &KUG 1102+450, CG 1366 \\
Mrk 209          &12:26:15.9  &$+$48:29:37 & 0.0009 &   7.81  &  2,3     &UGCA 281, I Zw 36, Haro 29 \\
Mrk 259          &13:28:44.0  &$+$43:55:51 & 0.0279 &   8.08  &  2       &HS 1326+4411 \\
Mrk 450          &13:14:48.3  &$+$34:52:51 & 0.0029 &   8.25  &  1,2     &UGC 8323, VV 616, HS 1312+3508 \\
Mrk 490          &15:46:30.7  &$+$45:59:54 & 0.0089 &   8.21  &  2       & \\
Mrk 689          &15:36:19.4  &$+$30:40:56 & 0.0058 &   8.15  &  2       &CG 1307 \\
Mrk 930          &23:31:58.3  &$+$28:56:50 & 0.0183 &   8.09  &  6       & \\
Mrk 1315         &12:15:18.6  &$+$20:38:27 & 0.0028 &   8.28  &  1,2     &KUG 1212+209B  \\
Mrk 1329         &12:37:03.0  &$+$06:55:36 & 0.0054 &   8.29  &  1,2     &IC 3591, Tol 1234+072 \\
Mrk 1448         &11:34:45.7  &$+$50:06:03 & 0.0260 &   8.27  &  2       &SBS 1132+503 \\
Mrk 1486         &13:59:50.9  &$+$57:26:23 & 0.0338 &   7.95  &  2       &SBS 1358+576 \\
NGC 1741         &05:01:38.3  &$-$04:15:25 & 0.0135 &   8.11  &  6       &Mrk 1089, Arp 259, VV 524, VV 565 \\
SBS 0335$-$052E  &03:37:44.0  &$-$05:02:40 & 0.0135 & 7.14-7.30  &  4,5,6,7 & \\
SBS 0940$+$544   &09:44:16.6  &$+$54:11:34 & 0.0055 &   7.46  &  8,9 & \\
SBS 1030$+$583   &10:34:10.1  &$+$58:03:49 & 0.0076 &   7.81  &  2,3 &Mrk 1434, KUG 1030+583 \\
SBS 1135$+$581   &11:38:35.7  &$+$57:52:27 & 0.0032 &   8.07  &  8   &Mrk 1450, VII Zw 415 \\
SBS 1152$+$579   &11:55:28.3  &$+$57:39:52 & 0.0179 &   7.90  &  2,8 &Mrk 193, VII Zw 415 \\
SBS 1222$+$614   &12:25:05.4  &$+$61:09:11 & 0.0024 &   7.98  &  2,3 & \\
SBS 1415$+$437   &14:17:01.4  &$+$43:30:05 & 0.0020 &   7.60  &  8,10& \\
SBS 1428$+$457   &14:30:12.2  &$+$45:32:32 & 0.0078 &   8.41  &  2   &CG 453 \\
SBS 1437$+$370   &14:39:05.4  &$+$36:48:22 & 0.0019 &   7.94  &  3   &Mrk 475 \\
Tol 1214$-$277   &12:17:17.1  &$-$28:02:33 & 0.0260 &   7.55  &5,11,12&Tol 21 \\
Tol 65           &12:25:46.9  &$-$36:14:01 & 0.0094 &   7.55  &5,11,12&Tol 1223-359 \\
Tol 2138$-$405   &21:41:21.8  &$-$40:19:06 & 0.0581 &   8.03  &  5   & \\
Tol 2146$-$391   &21:49:48.2  &$-$38:54:09 & 0.0294 &   7.83  &  5   & \\
UM 311           &01:15:34.4  &$-$00:51:46 & 0.0056 &   7.83  &  2,6 &SHOC 56 \\
\hline
\end{tabular}

$^{\rm a}$References of optical spectra.

{\bf References}: (1) \citet{IT04}; (2) SDSS data base; (3) \citet{ITL97}; (4) \citet{I99}; (5) \citet{G11};
(6) \citet{IT98}; (7) \citet{I09}; (8) \citet{ITL94}; (9) \citet{I07}; (10) \citet{G03}; (11) \citet{ICG01};
(12) \citet{I04}.
\end{minipage}
\end{table*}

\begin{table*}
\centering
\begin{minipage}{170mm}
\caption{He~{\sc i} $\lambda$10830\AA\ and P$\gamma$ $\lambda$10940\AA\ 
emission-line parameters.}
\label{tab2}
\begin{tabular}{@{}lrrcrrcl@{}}\hline
Name    &\multicolumn{2}{c}{He~{\sc i} $\lambda$10830\AA }&& \multicolumn{2}{c}{P$\gamma$ $\lambda$10940\AA }&$F$(He~{\sc i})/$F$(P$\gamma$) &Telescope/Instrument \\ \cline{2-3} \cline{5-6}
\\
 &\multicolumn{1}{c}{$F$$^{\rm a}$}&\multicolumn{1}{c}{EW$^{\rm b}$}& &\multicolumn{1}{c}{$F$$^{\rm a}$}&\multicolumn{1}{c}{EW$^{\rm b}$}& & \\ \hline
CGCG 007$-$025       &  150.0$\pm$0.8& 383$\pm$2&&  40.7$\pm$0.5& 160$\pm$2& 3.69$\pm$0.05& 3.5m APO/Triplespec \\
Haro 3               &  786.0$\pm$1.7& 221$\pm$1&& 320.8$\pm$1.3&  90$\pm$1& 2.45$\pm$0.02& 3.5m APO/Triplespec \\
HS 0837$+$4717       &  162.1$\pm$0.6&1190$\pm$5&&  21.8$\pm$0.4& 146$\pm$4& 7.43$\pm$0.13& LBT/Lucifer \\
HS 1734$+$5704       &  170.9$\pm$2.6& 708$\pm$2&&  47.8$\pm$0.8& 146$\pm$2& 3.58$\pm$0.08& 3.5m APO/Triplespec \\
I Zw 18 SE           &   27.4$\pm$0.4& 209$\pm$4&&   9.5$\pm$0.4&  57$\pm$3& 2.90$\pm$0.12& LBT/Lucifer \\
J0024$+$1404         &   54.1$\pm$0.9&  40$\pm$1&&  20.6$\pm$0.8&  17$\pm$1& 2.63$\pm$0.11& 3.5m APO/Triplespec \\
J0115$-$0051         &  295.1$\pm$1.3& 600$\pm$2&& 116.2$\pm$0.9& 243$\pm$1& 2.54$\pm$0.02& 3.5m APO/Triplespec \\
J0301$-$0052         &   10.2$\pm$0.2& 150$\pm$3&&   5.1$\pm$0.2&  54$\pm$2& 2.01$\pm$0.08& LBT/Lucifer \\
J0519$+$0007         &   94.3$\pm$0.4& 927$\pm$5&&  13.9$\pm$0.3& 151$\pm$4& 6.81$\pm$0.13& LBT/Lucifer \\
J1038$+$5330         &  389.6$\pm$4.0&  27$\pm$1&& 151.8$\pm$3.7&  12$\pm$1& 2.57$\pm$0.07& 3.5m APO/Triplespec \\
J1203$-$0342         &   62.9$\pm$1.0&  42$\pm$1&&  25.4$\pm$0.8&  17$\pm$1& 2.48$\pm$0.09& 3.5m APO/Triplespec \\
J1253$-$0313         & 1045.0$\pm$1.8&1050$\pm$2&& 173.1$\pm$1.7& 149$\pm$1& 6.04$\pm$0.07& 3.5m APO/Triplespec \\
J1624$-$0022         &   47.6$\pm$0.8&  98$\pm$2&&  18.1$\pm$0.6&  50$\pm$2& 2.64$\pm$0.10& 3.5m APO/Triplespec \\
KUG 0952$+$418       &   46.8$\pm$0.8&  56$\pm$1&&  15.6$\pm$0.6&  18$\pm$1& 3.00$\pm$0.13& 3.5m APO/Triplespec \\
Mrk 36               &   89.1$\pm$0.9&  95$\pm$1&&  37.5$\pm$0.6&  42$\pm$1& 2.38$\pm$0.05& 3.5m APO/Triplespec \\
Mrk 59               &  601.7$\pm$1.5& 284$\pm$1&& 243.8$\pm$1.0& 125$\pm$1& 2.47$\pm$0.02& 3.5m APO/Triplespec \\
Mrk 71               & 2055.0$\pm$2.2& 978$\pm$1&& 573.2$\pm$1.4& 313$\pm$1& 3.59$\pm$0.01& 3.5m APO/Triplespec \\
Mrk 162              &  174.2$\pm$1.1& 105$\pm$1&&  66.3$\pm$1.1&  39$\pm$1& 2.63$\pm$0.05& 3.5m APO/Triplespec \\
Mrk 209              &  168.6$\pm$0.9& 376$\pm$2&&  55.3$\pm$0.6& 127$\pm$2& 3.05$\pm$0.04& 3.5m APO/Triplespec \\
Mrk 259              &  106.6$\pm$1.5& 101$\pm$2&&  35.5$\pm$1.5&  80$\pm$2& 3.01$\pm$0.06& 3.5m APO/Triplespec \\
Mrk 450              &  205.4$\pm$0.9& 711$\pm$4&&  69.3$\pm$0.7& 205$\pm$3& 2.96$\pm$0.03& 3.5m APO/Triplespec \\
Mrk 490              &   60.2$\pm$1.2& 467$\pm$3&&  20.3$\pm$1.0& 120$\pm$2& 2.97$\pm$0.15& 3.5m APO/Triplespec \\
Mrk 689              &   15.6$\pm$0.5&  52$\pm$1&&   6.6$\pm$0.5&  21$\pm$1& 2.35$\pm$0.18& 3.5m APO/Triplespec \\
Mrk 930              &  121.4$\pm$0.9&  77$\pm$1&&  47.5$\pm$0.9&  30$\pm$1& 2.56$\pm$0.05& 3.5m APO/Triplespec \\
Mrk 1315             &  169.6$\pm$0.9& 441$\pm$2&&  74.5$\pm$0.7& 179$\pm$4& 2.28$\pm$0.03& 3.5m APO/Triplespec \\
Mrk 1329             &  172.0$\pm$0.9& 313$\pm$1&&  60.4$\pm$0.7& 106$\pm$1& 2.85$\pm$0.04& 3.5m APO/Triplespec \\
Mrk 1448             &   80.6$\pm$1.0& 106$\pm$2&&  27.5$\pm$1.6&  36$\pm$2& 2.94$\pm$0.18& 3.5m APO/Triplespec \\
Mrk 1486             &   85.5$\pm$1.3&  71$\pm$1&&  30.1$\pm$0.6&  26$\pm$1& 2.84$\pm$0.07& 3.5m APO/Triplespec \\
NGC 1741 \#1         &  234.4$\pm$1.4& 138$\pm$1&&  87.5$\pm$1.1&  55$\pm$1& 2.68$\pm$0.04& 3.5m APO/Triplespec \\
NGC 1741 \#2         &  302.0$\pm$1.6& 175$\pm$1&& 108.0$\pm$1.2&  62$\pm$1& 2.80$\pm$0.04& 3.5m APO/Triplespec \\
SBS 0335$-$052E \#1+2&  382.7$\pm$0.6& 618$\pm$1&&  95.5$\pm$0.4& 163$\pm$1& 4.01$\pm$0.02& LBT/Lucifer         \\
SBS 0335$-$052E \#7  &   25.8$\pm$0.4& 432$\pm$8&&   6.7$\pm$0.3& 169$\pm$9& 3.86$\pm$0.18& LBT/Lucifer         \\
SBS 0940$+$544       &   37.7$\pm$0.4& 383$\pm$3&&  12.2$\pm$0.3& 139$\pm$4& 3.09$\pm$0.09& LBT/Lucifer         \\
SBS 1030$+$583       &   53.2$\pm$0.4& 101$\pm$1&&  20.8$\pm$0.4&  41$\pm$1& 2.55$\pm$0.05& LBT/Lucifer         \\
SBS 1135$+$581       &  180.3$\pm$1.0& 106$\pm$1&&  82.7$\pm$0.7&  60$\pm$1& 2.18$\pm$0.02& 3.5m APO/Triplespec \\
SBS 1152$+$579       &  169.9$\pm$0.6& 742$\pm$3&&  30.6$\pm$0.5& 137$\pm$3& 5.55$\pm$0.09& 3.5m APO/Triplespec \\
SBS 1222$+$614       &  134.2$\pm$0.9&  95$\pm$1&&  53.1$\pm$0.7&  43$\pm$1& 2.53$\pm$0.04& 3.5m APO/Triplespec \\
SBS 1415$+$437       &   94.9$\pm$0.6& 297$\pm$2&&  30.9$\pm$0.5&  95$\pm$2& 3.07$\pm$0.05& 3.5m APO/Triplespec \\
SBS 1428$+$457       &  154.0$\pm$0.9&  95$\pm$1&&  60.6$\pm$0.7&  37$\pm$1& 2.54$\pm$0.04& 3.5m APO/Triplespec \\
SBS 1437$+$370       &   84.6$\pm$1.0& 137$\pm$3&&  29.4$\pm$0.8&  40$\pm$2& 2.88$\pm$0.08& 3.5m APO/Triplespec \\
Tol 1214$-$277       &   14.4$\pm$0.1& 414$\pm$3&&   4.6$\pm$0.1& 145$\pm$3& 3.12$\pm$0.08& VLT/ISAAC \\
Tol 65               &   58.2$\pm$0.4& 778$\pm$3&&  12.4$\pm$0.2& 188$\pm$2& 4.70$\pm$0.07& VLT/ISAAC \\
Tol 2138$-$405       &  346.7$\pm$5.1&1042$\pm$9&&  59.0$\pm$2.8& 157$\pm$9& 5.88$\pm$0.29& NTT/SOFI \\
Tol 2146$-$391       &   60.5$\pm$0.5& 449$\pm$3&&  17.1$\pm$0.6& 134$\pm$2& 3.54$\pm$0.13& NTT/SOFI \\
UM 311               &   44.0$\pm$0.6& 223$\pm$4&&  20.2$\pm$0.5&  94$\pm$3& 2.18$\pm$0.05& 3.5m APO/Triplespec \\
\hline
\end{tabular}

$^{\rm a}$ Observed flux in units 10$^{-16}$ erg s$^{-1}$ cm$^{-2}$.

$^{\rm b}$ Equivalent width in \AA.







\end{minipage}
\end{table*}


%

\begin{table*}
\centering
\begin{minipage}{170mm}
\caption{Derived parameters for the He abundance determination.$^{\rm a}$}
\label{tab3}
\begin{tabular}{@{}lrcccrrcc@{}}\hline
Name             &O/H$\times$10$^5$&$Y$&$t_{\rm e}$(O {\sc iii})$^{\rm b}$&$t_{\rm e}$(He$^+$)$^{\rm b}$
&$N_{\rm e}$(S {\sc ii})$^{\rm c}$&$N_{\rm e}$(He$^+$)$^{\rm c}$&$\tau$($\lambda$3889)&$ICF$(He)$^{\rm d}$ \\
\hline
CGCG 007$-$025$^{\rm e}$       &  5.8$\pm$\,\,\,0.2&0.2568$\pm$0.0031& 1.64$\pm$0.02& 1.69$\pm$0.01& 124$\pm$\,\,\,38& 125$\pm$\,\,\,2& 1.29$\pm$0.03&0.9940\\
CGCG 007$-$025$^{\rm e}$       &  5.7$\pm$\,\,\,0.1&0.2587$\pm$0.0040& 1.70$\pm$0.02& 1.60$\pm$0.04& 153$\pm$\,\,\,57& 130$\pm$\,\,\,6& 1.92$\pm$0.01&0.9938\\
Haro 3               & 19.0$\pm$\,\,\,0.7&0.2637$\pm$0.0030& 1.02$\pm$0.01& 1.08$\pm$0.02& 175$\pm$\,\,\,32&  35$\pm$\,\,\,3& 2.14$\pm$0.10&1.0080\\
HS 0837$+$4717$^{\rm e}$       &  4.8$\pm$\,\,\,0.1&0.2455$\pm$0.0035& 1.93$\pm$0.02& 1.81$\pm$0.05& 343$\pm$\,\,\,93& 541$\pm$24     & 4.66$\pm$0.01&0.9941\\
HS 0837$+$4717$^{\rm e}$       &  5.1$\pm$\,\,\,0.1&0.2577$\pm$0.0046& 1.86$\pm$0.02& 1.75$\pm$0.01& 513$\pm$140     & 520$\pm$\,\,\,1& 5.00$\pm$0.01&0.9956\\
HS 0837$+$4717$^{\rm e}$       &  4.6$\pm$\,\,\,0.1&0.2602$\pm$0.0047& 1.91$\pm$0.03& 1.81$\pm$0.05& 287$\pm$106     & 484$\pm$22     & 5.00$\pm$0.03&0.9949\\
HS 1734$+$5704       & 14.8$\pm$\,\,\,0.3&0.2609$\pm$0.0035& 1.25$\pm$0.01& 1.19$\pm$0.01& 262$\pm$\,\,\,50& 208$\pm$\,\,\,6& 5.00$\pm$0.04&0.9972\\
I Zw 18 SE$^{\rm e}$           &  1.6$\pm$\,\,\,0.1&0.2515$\pm$0.0068& 1.86$\pm$0.06& 1.91$\pm$0.04&  10$\pm$\,\,\,10&  47$\pm$\,\,\,2& 0.03$\pm$0.19&0.9984\\
I Zw 18 SE$^{\rm e}$           &  1.7$\pm$\,\,\,0.2&0.2596$\pm$0.0084& 1.75$\pm$0.08& 1.81$\pm$0.03&  11$\pm$\,\,\,68&  43$\pm$\,\,\,1& 0.82$\pm$0.04&1.0004\\
I Zw 18 SE$^{\rm e}$           &  1.8$\pm$\,\,\,0.1&0.2446$\pm$0.0050& 1.70$\pm$0.03& 1.76$\pm$0.02&  55$\pm$\,\,\,70&  60$\pm$\,\,\,4& 0.14$\pm$0.14&0.9966\\
J0024$+$1404         & 19.7$\pm$\,\,\,3.6&0.2705$\pm$0.0127& 0.98$\pm$0.05& 1.05$\pm$0.02&  40$\pm$\,\,\,58&  56$\pm$10     & 0.03$\pm$0.10&1.0123\\
J0115$-$0051         & 19.2$\pm$\,\,\,1.5&0.2605$\pm$0.0060& 1.00$\pm$0.02& 1.07$\pm$0.01& 141$\pm$\,\,\,52&  58$\pm$\,\,\,3& 1.48$\pm$0.05&1.0035\\
J0301$-$0052         &  5.2$\pm$\,\,\,0.7&0.2542$\pm$0.0256& 1.60$\pm$0.10& 1.51$\pm$0.01&  69$\pm$283     &  11$\pm$\,\,\,1& 1.37$\pm$0.37&0.9962\\
J0519$+$0007$^{\rm e}$         &  3.2$\pm$\,\,\,0.1&0.2396$\pm$0.0047& 2.03$\pm$0.03& 1.90$\pm$0.09& 335$\pm$208     & 460$\pm$32     & 3.40$\pm$0.18&0.9943\\
J0519$+$0007$^{\rm e}$         &  3.3$\pm$\,\,\,0.1&0.2521$\pm$0.0027& 1.99$\pm$0.02& 1.88$\pm$0.01& 475$\pm$\,\,\,60& 467$\pm$\,\,\,1& 5.01$\pm$0.01&0.9955\\
J1038$+$5330         & 16.4$\pm$\,\,\,1.4&0.2473$\pm$0.0046& 1.00$\pm$0.03& 1.06$\pm$0.01& 200$\pm$\,\,\,40& 116$\pm$\,\,\,5& 0.45$\pm$0.01&1.0194\\
J1203$-$0342         & 26.1$\pm$14.8     &0.2718$\pm$0.0165& 0.89$\pm$0.16& 0.95$\pm$0.04&  76$\pm$\,\,\,70&  48$\pm$16     & 0.09$\pm$0.55&1.0177\\
J1253$-$0313         &  9.4$\pm$\,\,\,0.2&0.2512$\pm$0.0028& 1.38$\pm$0.01& 1.43$\pm$0.06& 495$\pm$\,\,\,58& 510$\pm$52     & 4.94$\pm$0.10&0.9965\\
J1624$-$0022         & 14.3$\pm$\,\,\,0.4&0.2577$\pm$0.0044& 1.17$\pm$0.01& 1.21$\pm$0.02&  58$\pm$\,\,\,38&  63$\pm$\,\,\,6& 1.54$\pm$0.02&1.0027\\
KUG 0952$+$418       & 28.0$\pm$\,\,\,5.2&0.2678$\pm$0.0161& 1.00$\pm$0.08& 0.97$\pm$0.04&  73$\pm$\,\,\,70& 172$\pm$24     & 1.78$\pm$0.03&1.0189\\
Mrk 36$^{\rm e}$               &  8.1$\pm$\,\,\,0.4&0.2501$\pm$0.0092& 1.51$\pm$0.04& 1.42$\pm$0.02&  28$\pm$\,\,\,49&  26$\pm$\,\,\,1& 3.04$\pm$0.01&1.0000\\
Mrk 36$^{\rm e}$               &  5.6$\pm$\,\,\,0.2&0.2493$\pm$0.0076& 1.54$\pm$0.03& 1.60$\pm$0.03&  46$\pm$\,\,\,56&  16$\pm$\,\,\,3& 0.25$\pm$0.09&0.9966\\
Mrk 59$^{\rm e}$               &  9.2$\pm$\,\,\,0.2&0.2555$\pm$0.0028& 1.35$\pm$0.01& 1.41$\pm$0.01&  84$\pm$\,\,\,32&  29$\pm$\,\,\,1& 1.05$\pm$0.07&0.9969\\
Mrk 59$^{\rm e}$               & 10.0$\pm$\,\,\,0.3&0.2465$\pm$0.0047& 1.31$\pm$0.01& 1.37$\pm$0.01&  83$\pm$\,\,\,62&  42$\pm$\,\,\,3& 0.52$\pm$0.06&0.9975\\
Mrk 71               &  6.8$\pm$\,\,\,0.2&0.2542$\pm$0.0026& 1.57$\pm$0.01& 1.62$\pm$0.02& 122$\pm$\,\,\,36& 126$\pm$\,\,\,5& 1.32$\pm$0.01&0.9956\\
Mrk 162              & 11.8$\pm$\,\,\,0.9&0.2586$\pm$0.0079& 1.17$\pm$0.03& 1.23$\pm$0.03&  80$\pm$\,\,\,54&  61$\pm$\,\,\,5& 0.07$\pm$0.09&1.0031\\
Mrk 209$^{\rm e}$              &  6.7$\pm$\,\,\,0.1&0.2475$\pm$0.0027& 1.63$\pm$0.02& 1.56$\pm$0.01&  75$\pm$\,\,\,34&  91$\pm$\,\,\,4& 0.01$\pm$0.08&0.9952\\
Mrk 209$^{\rm e}$              &  5.8$\pm$\,\,\,0.2&0.2643$\pm$0.0050& 1.64$\pm$0.02& 1.70$\pm$0.01&  68$\pm$\,\,\,70&  50$\pm$\,\,\,5& 0.03$\pm$0.05&0.9965\\
Mrk 259              & 13.9$\pm$\,\,\,0.9&0.2578$\pm$0.0089& 1.21$\pm$0.04& 1.16$\pm$0.05& 152$\pm$\,\,\,66& 132$\pm$12     & 0.10$\pm$0.29&1.0035\\
Mrk 450$^{\rm e}$              & 15.8$\pm$\,\,\,0.4&0.2523$\pm$0.0034& 1.17$\pm$0.01& 1.15$\pm$0.01& 128$\pm$\,\,\,34& 135$\pm$\,\,\,4& 2.97$\pm$0.03&0.9985\\
Mrk 450$^{\rm e}$              & 15.2$\pm$\,\,\,0.6&0.2643$\pm$0.0065& 1.16$\pm$0.02& 1.22$\pm$0.01& 171$\pm$\,\,\,86&  94$\pm$\,\,\,2& 2.34$\pm$0.01&1.0000\\
Mrk 490              & 18.7$\pm$\,\,\,5.0&0.2615$\pm$0.0206& 1.06$\pm$0.14& 1.02$\pm$0.05&  32$\pm$\,\,\,56& 159$\pm$22     & 0.48$\pm$1.01&1.0125\\
Mrk 689              & 12.3$\pm$\,\,\,1.0&0.2530$\pm$0.0106& 1.20$\pm$0.03& 1.26$\pm$0.02& 115$\pm$\,\,\,76&  32$\pm$\,\,\,1& 0.13$\pm$0.22&1.0033\\
Mrk 930              & 11.8$\pm$\,\,\,0.9&0.2590$\pm$0.0082& 1.23$\pm$0.04& 1.25$\pm$0.01&  59$\pm$\,\,\,46&  42$\pm$\,\,\,4& 0.01$\pm$0.07&1.0038\\
Mrk 1315$^{\rm e}$             & 16.4$\pm$\,\,\,0.4&0.2594$\pm$0.0029& 1.10$\pm$0.01& 1.15$\pm$0.01&  17$\pm$\,\,\,18&  10$\pm$\,\,\,3& 0.63$\pm$0.06&0.9977\\
Mrk 1315$^{\rm e}$             & 18.1$\pm$\,\,\,0.8&0.2615$\pm$0.0051& 1.06$\pm$0.01& 1.12$\pm$0.01&  86$\pm$\,\,\,60&  11$\pm$\,\,\,2& 0.55$\pm$0.05&0.9980\\
Mrk 1329$^{\rm e}$             & 18.0$\pm$\,\,\,0.5&0.2613$\pm$0.0029& 1.07$\pm$0.01& 1.10$\pm$0.01&  20$\pm$\,\,\,19& 106$\pm$\,\,\,4& 1.17$\pm$0.04&0.9994\\
Mrk 1329$^{\rm e}$             & 19.0$\pm$\,\,\,0.9&0.2560$\pm$0.0056& 1.04$\pm$0.02& 1.11$\pm$0.01&  87$\pm$\,\,\,59& 118$\pm$\,\,\,3& 0.72$\pm$0.10&0.9998\\
Mrk 1448             & 15.6$\pm$\,\,\,1.3&0.2599$\pm$0.0089& 1.08$\pm$0.03& 1.14$\pm$0.02&  61$\pm$\,\,\,59& 125$\pm$\,\,\,6& 0.07$\pm$0.16&1.0069\\
Mrk 1486             &  8.4$\pm$\,\,\,0.5&0.2631$\pm$0.0119& 1.40$\pm$0.04& 1.44$\pm$0.05&  18$\pm$\,\,\,50&  58$\pm$\,\,\,9& 0.76$\pm$0.05&0.9997\\
NGC 1741 \#1         & 11.0$\pm$\,\,\,1.7&0.2680$\pm$0.0088& 1.11$\pm$0.07& 1.16$\pm$0.01&  88$\pm$\,\,\,46&  72$\pm$\,\,\,4& 0.03$\pm$0.03&1.0093\\
SBS 0335$-$052E \#1+2$^{\rm e}$&  2.4$\pm$\,\,\,0.1&0.2562$\pm$0.0054& 2.00$\pm$0.04& 1.88$\pm$0.05& 313$\pm$180     & 103$\pm$\,\,\,5& 4.70$\pm$0.06&0.9948\\
SBS 0335$-$052E \#1+2$^{\rm e}$&  2.1$\pm$\,\,\,0.1&0.2611$\pm$0.0030& 2.00$\pm$0.02& 2.06$\pm$0.05& 321$\pm$\,\,\,69&  85$\pm$\,\,\,3& 4.74$\pm$0.02&0.9958\\
SBS 0335$-$052E \#1+2$^{\rm e}$&  2.3$\pm$\,\,\,0.1&0.2603$\pm$0.0031& 2.06$\pm$0.03& 1.96$\pm$0.02& 296$\pm$\,\,\,82&  92$\pm$\,\,\,1& 3.28$\pm$0.04&0.9956\\
SBS 0335$-$052E \#1+2$^{\rm e}$&  2.1$\pm$\,\,\,0.1&0.2543$\pm$0.0035& 2.08$\pm$0.03& 2.13$\pm$0.01& 237$\pm$104     &  90$\pm$\,\,\,1& 4.99$\pm$0.02&0.9967\\
SBS 0335$-$052E \#1+2$^{\rm e}$&  2.0$\pm$\,\,\,0.1&0.2559$\pm$0.0026& 2.03$\pm$0.02& 2.08$\pm$0.03&  10$\pm$\,\,\,10&  87$\pm$\,\,\,5& 5.00$\pm$0.01&0.9941\\
SBS 0335$-$052E \#1+2$^{\rm e}$&  2.1$\pm$\,\,\,0.1&0.2584$\pm$0.0026& 2.10$\pm$0.03& 1.98$\pm$0.04& 170$\pm$\,\,\,47& 119$\pm$\,\,\,4& 4.99$\pm$0.01&0.9940\\
SBS 0335$-$052E \#1+2$^{\rm e}$&  2.4$\pm$\,\,\,0.1&0.2604$\pm$0.0032& 2.03$\pm$0.03& 1.90$\pm$0.03& 421$\pm$\,\,\,88& 117$\pm$\,\,\,1& 4.98$\pm$0.01&0.9954\\
SBS 0335$-$052E \#1+2$^{\rm e}$&  2.4$\pm$\,\,\,0.1&0.2545$\pm$0.0033& 2.03$\pm$0.03& 1.90$\pm$0.03& 215$\pm$\,\,\,65& 140$\pm$\,\,\,3& 1.45$\pm$0.06&1.0003\\
SBS 0335$-$052E \#7$^{\rm e}$  &  1.7$\pm$\,\,\,0.1&0.2527$\pm$0.0048& 1.97$\pm$0.04& 1.85$\pm$0.04&  99$\pm$116     & 122$\pm$\,\,\,2& 0.02$\pm$0.03&0.9927\\
SBS 0335$-$052E \#7$^{\rm e}$  &  1.6$\pm$\,\,\,0.2&0.2457$\pm$0.0097& 2.00$\pm$0.10& 1.87$\pm$0.08& 352$\pm$442     & 135$\pm$10     & 0.03$\pm$0.09&0.9935\\
SBS 0940$+$544$^{\rm e}$       &  3.0$\pm$\,\,\,0.1&0.2601$\pm$0.0065& 1.99$\pm$0.04& 1.95$\pm$0.01& 167$\pm$118     &  45$\pm$\,\,\,1& 0.40$\pm$0.07&0.9938\\
SBS 0940$+$544$^{\rm e}$       &  3.3$\pm$\,\,\,0.1&0.2556$\pm$0.0037& 1.86$\pm$0.02& 1.87$\pm$0.04& 166$\pm$\,\,\,58&  56$\pm$\,\,\,3& 0.02$\pm$0.09&0.9947\\
SBS 1030$+$583$^{\rm e}$       &  7.5$\pm$\,\,\,0.2&0.2536$\pm$0.0052& 1.56$\pm$0.02& 1.47$\pm$0.02&  10$\pm$\,\,\,10&  43$\pm$\,\,\,5& 0.45$\pm$0.08&0.9997\\
SBS 1030$+$583$^{\rm e}$       &  6.4$\pm$\,\,\,0.4&0.2533$\pm$0.0119& 1.52$\pm$0.04& 1.57$\pm$0.04&  80$\pm$123     &  35$\pm$\,\,\,3& 0.03$\pm$0.33&0.9991\\
SBS 1135$+$581       & 12.8$\pm$\,\,\,0.3&0.2579$\pm$0.0034& 1.26$\pm$0.01& 1.22$\pm$0.01&  49$\pm$\,\,\,32&  10$\pm$\,\,\,2& 1.00$\pm$0.02&0.9996\\
SBS 1152$+$573$^{\rm e}$       &  9.1$\pm$\,\,\,0.2&0.2399$\pm$0.0029& 1.54$\pm$0.02& 1.45$\pm$0.01& 152$\pm$\,\,\,42& 481$\pm$\,\,\,6& 1.59$\pm$0.04&0.9963\\
SBS 1152$+$573$^{\rm e}$       &  8.3$\pm$\,\,\,0.4&0.2506$\pm$0.0076& 1.47$\pm$0.02& 1.53$\pm$0.04& 145$\pm$107     & 395$\pm$17     & 2.15$\pm$0.03&0.9968\\
SBS 1222$+$614       & 10.9$\pm$\,\,\,0.3&0.2495$\pm$0.0053& 1.40$\pm$0.02& 1.33$\pm$0.03&  74$\pm$\,\,\,50&  52$\pm$\,\,\,3& 0.51$\pm$0.01&0.9985\\
SBS 1415$+$437$^{\rm e}$       &  4.0$\pm$\,\,\,0.1&0.2522$\pm$0.0037& 1.64$\pm$0.02& 1.69$\pm$0.02&  56$\pm$\,\,\,39&  79$\pm$\,\,\,2& 0.83$\pm$0.05&0.9956\\
SBS 1415$+$437$^{\rm e}$       &  4.7$\pm$\,\,\,0.1&0.2510$\pm$0.0034& 1.70$\pm$0.02& 1.60$\pm$0.01& 108$\pm$\,\,\,42&  89$\pm$\,\,\,2& 1.09$\pm$0.03&0.9955\\
SBS 1415$+$437$^{\rm e}$       &  4.7$\pm$\,\,\,0.1&0.2511$\pm$0.0040& 1.68$\pm$0.02& 1.58$\pm$0.02&  83$\pm$\,\,\,42&  91$\pm$\,\,\,4& 0.97$\pm$0.01&0.9949\\
SBS 1428$+$457       & 21.9$\pm$\,\,\,2.3&0.2659$\pm$0.0095& 0.99$\pm$0.04& 1.04$\pm$0.03&  94$\pm$\,\,\,66&  59$\pm$\,\,\,7& 0.48$\pm$0.16&1.0130\\
\hline
\end{tabular}
\end{minipage}
\end{table*}

\setcounter{table}{2}

\begin{table*}
\centering
\begin{minipage}{170mm}
\caption{Continued.}
\begin{tabular}{@{}lrcccrrcc@{}}\hline
Name             &O/H$\times$10$^5$&$Y$&$t_{\rm e}$(O {\sc iii})$^{\rm b}$&$t_{\rm e}$(He$^+$)$^{\rm b}$
&$N_{\rm e}$(S {\sc ii})$^{\rm c}$&$N_{\rm e}$(He$^+$)$^{\rm c}$&$\tau$($\lambda$3889)&$ICF$(He)$^{\rm d}$ \\
\hline
SBS 1437$+$370       & 10.1$\pm$\,\,\,0.2&0.2574$\pm$0.0059& 1.41$\pm$0.02& 1.34$\pm$0.03&  76$\pm$\,\,\,36&  78$\pm$\,\,\,6& 0.72$\pm$0.07&0.9970\\
Tol 1214$-$277$^{\rm e}$       &  3.5$\pm$\,\,\,0.1&0.2563$\pm$0.0032& 1.96$\pm$0.02& 2.02$\pm$0.02& 258$\pm$\,\,\,74&  59$\pm$\,\,\,5& 0.02$\pm$0.07&0.9942\\
Tol 1214$-$277$^{\rm e}$       &  3.6$\pm$\,\,\,0.1&0.2518$\pm$0.0028& 1.95$\pm$0.02& 2.01$\pm$0.03& 188$\pm$\,\,\,63&  69$\pm$\,\,\,2& 0.77$\pm$0.04&0.9937\\
Tol 1214$-$277$^{\rm e}$       &  3.4$\pm$\,\,\,0.1&0.2527$\pm$0.0025& 1.98$\pm$0.02& 2.05$\pm$0.01&  10$\pm$\,\,\,10&  62$\pm$\,\,\,4& 0.50$\pm$0.07&0.9934\\
Tol 65$^{\rm e}$               &  3.4$\pm$\,\,\,0.1&0.2443$\pm$0.0033& 1.72$\pm$0.02& 1.78$\pm$0.03&  21$\pm$\,\,\,26& 242$\pm$\,\,\,6& 0.17$\pm$0.05&0.9949\\
Tol 65$^{\rm e}$               &  3.5$\pm$\,\,\,0.1&0.2440$\pm$0.0027& 1.70$\pm$0.02& 1.76$\pm$0.01& 171$\pm$\,\,\,41& 242$\pm$\,\,\,5& 1.21$\pm$0.06&0.9934\\
Tol 2138$-$405$^{\rm e}$       & 10.6$\pm$\,\,\,0.2&0.2548$\pm$0.0031& 1.39$\pm$0.01& 1.39$\pm$0.01& 339$\pm$\,\,\,47& 401$\pm$10     & 1.87$\pm$0.02&0.9960\\
Tol 2138$-$405$^{\rm e}$       & 11.9$\pm$\,\,\,0.4&0.2591$\pm$0.0046& 1.31$\pm$0.02& 1.37$\pm$0.05& 347$\pm$\,\,\,56& 399$\pm$40     & 2.53$\pm$0.04&0.9973\\
Tol 2146$-$391$^{\rm e}$       &  6.2$\pm$\,\,\,0.2&0.2553$\pm$0.0029& 1.57$\pm$0.02& 1.63$\pm$0.01& 165$\pm$\,\,\,40& 127$\pm$\,\,\,2& 1.20$\pm$0.11&0.9949\\
Tol 2146$-$391$^{\rm e}$       &  6.3$\pm$\,\,\,0.2&0.2481$\pm$0.0045& 1.59$\pm$0.02& 1.65$\pm$0.04& 183$\pm$\,\,\,54& 135$\pm$10     & 1.53$\pm$0.10&0.9946\\
UM 311$^{\rm e}$               & 23.3$\pm$\,\,\,2.4&0.2579$\pm$0.0065& 0.97$\pm$0.04& 0.98$\pm$0.01&  72$\pm$\,\,\,47&  10$\pm$\,\,\,6& 2.62$\pm$0.04&1.0020\\
UM 311$^{\rm e}$               & 27.8$\pm$\,\,\,2.7&0.2618$\pm$0.0060& 0.94$\pm$0.04& 0.91$\pm$0.01&  10$\pm$\,\,\,10&  13$\pm$\,\,\,1& 0.06$\pm$0.08&1.0181\\
\hline
\end{tabular}

$^{\rm a}$ Parameters are calculated for the case where all six He {\sc i} 
emission lines $\lambda$3889, $\lambda$4471, $\lambda$5876, $\lambda$6678, 
$\lambda$7065, and $\lambda$10830 have been used for the $\chi^2$ 
minimisation, and only the four He {\sc i}
emission lines $\lambda$4471, $\lambda$5876, $\lambda$6678, and $\lambda$10830 
have been used for the determination of $Y$.

$^{\rm b}$ $t_{\rm e}$ = 10$^{-4}$$T_{\rm e}$.

$^{\rm c}$ in cm$^{-3}$.

$^{\rm d}$ Uncertainties of $ICF$(He) are $\sim$0.0025 \citep{I13}.

$^{\rm e}$ Multiple entries for the same galaxy refer to independent sets of 
spectroscopic data for that galaxy.

\end{minipage}
\end{table*}



\section{The Sample}\label{sec:obs}

Our sample consists of 45 H~{\sc ii} regions in 43 low-metallicity 
emission-line galaxies. The general characteristics of the sample galaxies 
are shown in Table \ref{tab1}.

\subsection{Observations and data reduction}

\subsubsection{Spectroscopic data}

Our near-infrared spectroscopic data come from three sources. 

The majority of the near-infrared spectra of our sample galaxies were obtained with 
the 3.5 m Apache Point Observatory (APO) telescope, in conjunction
with the TripleSpec spectrograph, on a number of 
nights during the 2008 -- 2013 period. TripleSpec \citep{W04} is
a cross-dispersed NIR spectrograph that provides simultaneous
continuous wavelength coverage from 0.90 to 2.46 $\mu$m in five spectral 
orders during a single exposure. A 1\farcs1$\times$43\arcsec\ slit was
used, resulting in a resolving power of 3500.
During the course of each night, several A0V standard stars were observed for flux calibration and correction for telluric absorption. Spectra of Ar comparison arcs were also 
obtained for wavelength calibration. Since
all observed targets are smaller than the length of the slit, the
nod-on-slit technique was used to acquire the sky spectrum.
Objects were observed by nodding between two positions A
and B along the slit, following the ABBA sequence, and with an
integration time of 200 s or 300 s at each position.

The APO telescope is not large enough to obtain good NIR spectra for the 
faintest galaxies, with SDSS $g$ $\geq$ 17 mag. These also tend to be the 
lowest-metallicity galaxies which play an important role in the determination of 
$Y_{\rm p}$. We have thus observed eight low-metallicity H~{\sc ii} regions with 
the 8.4~m Large Binocular Telescope (LBT), in conjunction
with the Lucifer spectrograph, on several  
nights during the 2008 -- 2013 period. A 1\arcsec\ LS 600 slit 
was used, giving a resolving power of 8460 in the $J$-band . Various A0V standard stars were 
observed during each night for flux calibration and correction
for telluric absorption. Spectra of Ar comparison arcs were
also obtained for wavelength calibration.
All observed targets are smaller than the length of the slit, therefore the
nod-on-slit technique was used to acquire the sky spectrum.
Objects were observed by nodding between two positions A
and B along the slit, following the ABBA sequence, and with an
integration time of 240 s at each position.

Finally, we have retrieved from the European South Observatory (ESO) data archives VLT/ISAAC NIR spectra for two objects and NTT/SOFI NIR spectra for another two objects.

In addition, all galaxies listed in Table \ref{tab1} possess good optical 
spectra either obtained by our group in the past for the determination of the 
primordial helium abundance, or from the SDSS spectral data base. The optical 
data is described in \citet{I07} and in other references given in 
Table \ref{tab1}. The complementary optical data are necessary to derive 
physical conditions and abundances in the sample galaxies.  

\subsubsection{Reduction procedures}

We have carried out the reduction of the data according to the
following procedures. The two-dimensional spectra were first
cleaned for cosmic ray hits using the 
IRAF\footnote{IRAF is distributed by National Optical 
Astronomical Observatory, which is operated by the Association of Universities 
for Research in Astronomy, Inc., under cooperative agreement with the 
National Science Foundation.} routine CRMEDIAN. Then all A and B frames were 
separately co-added and the resulting B frame was subtracted from the 
resulting A frame. Finally, the (negative) spectrum at position B was adjusted 
to the (positive) spectrum at position A and subtracted from it. The
same reduction scheme was applied to the standard stars.
We then use the IRAF routines IDENTIFY, REIDENTIFY,
FITCOORD, and TRANSFORM to perform wavelength 
calibration and correction for distortion and tilt for each frame. For
all galaxies, a one-dimensional spectrum was extracted from the
two-dimensional frames using the APALL IRAF routine.

Flux calibration and correction for telluric absorption were
performed by first multiplying the one-dimensional spectrum of
each galaxy by the synthetic absolute spectral distribution of the
standard star, smoothed to the same spectral
resolution, and then by dividing the result by the observed one-dimensional
spectrum of the same star. Since there does not exist
published absolute spectral energy distributions of the standard
stars that were used, we have simply scaled the synthetic
absolute SED of the star Vega ($\alpha$ Lyrae), also of A0V spectral type, to the brightness of the standard star.

The emission-line fluxes were measured
using Gaussian fitting with the IRAF SPLOT routine. The  
line flux errors were estimated by Monte Carlo simulations in SPLOT, setting the number of trials to 200.  

Two representative spectra, one of a high-density H~{\sc ii} region (left) and 
the other of a low-density  one (right) are shown in Fig. \ref{fig1}. The fluxes 
and equivalent widths  of
the He~{\sc i} $\lambda$10830\AA\ emission line, needed for the $Y_{\rm p}$ 
determination, and of the P$\gamma$ $\lambda$10940\AA\ emission line, needed 
to adjust NIR and optical spectra, are presented in Table \ref{tab2}.

\subsection{Physical conditions and heavy element abundances}

We derived element abundances from the narrow emission-line fluxes, using the 
so-called direct method. This method is based on the determination of the 
electron temperature within the O$^{2+}$ zone from the 
[O~{\sc iii}] $\lambda$4363/($\lambda$4959 + $\lambda$5007)
line ratio. The fluxes in all optical spectra were
corrected for both extinction, using the reddening curve of \citet{C89},
and underlying hydrogen stellar absorption, derived simultaneously by an 
iterative procedure described by \citet{ITL94} and using the observed 
decrements of the narrow hydrogen Balmer lines. The extinction coefficient
$C$(H$\beta$) and equivalent width of hydrogen absorption lines EW(abs) are
derived in such a way to obtain the closest agreement between
the extinction-corrected and theoretical recombination hydrogen emission-line 
fluxes normalised to the H$\beta$ flux. It is assumed that EW(abs) is the same 
for all hydrogen lines. This assumption is justified by the evolutionary 
stellar population synthesis models of \citet{GD05}.

The physical conditions, and the ionic and total heavy element abundances in 
the H~{\sc ii} regions were derived following \citet{I06}. In particular, 
for O$^{2+}$, Ne$^{2+}$, and Ar$^{3+}$ abundances, we adopt the temperature
$T_{\rm e}$(O~{\sc iii}) directly derived from the 
[O~{\sc iii}] $\lambda$4363/($\lambda$4959 + $\lambda$5007) emission-line 
ratio. The electron temperatures $T_{\rm e}$(O~{\sc ii}) and
$T_{\rm e}$(S~{\sc iii}) were derived from the empirical relations by 
\citet{I06}. $T_{\rm e}$(O~{\sc ii}) was used for the calculation of 
O$^+$, N$^+$, S$^+$, and Fe$^{2+}$ abundances and $T_{\rm e}$(S~{\sc iii}) 
for the calculation of S$^{2+}$, Cl$^{2+}$, and Ar$^{2+}$ abundances.
The electron number densities $N_{\rm e}$(S~{\sc ii}) were obtained from the
[S~{\sc ii}] $\lambda$6717/$\lambda$6731 emission-line ratios. 
The low-density limit holds for the H~{\sc ii} regions that exhibit the 
emission lines considered here. The element abundances then do not
depend sensitively on $N_{\rm e}$. We use the ionisation correction
factors ($ICF$s) from \citet{I06} to correct for unseen
stages of ionisation and to derive the total O, N, Ne, S, Cl, Ar,
and Fe abundances.

The physical conditions and 
heavy-element abundances for the most of the objects 
in our sample were derived in several previous studies by our group. 
The references to  
these studies are shown in Table \ref{tab1}. We have listed 
the main physical parameters along with their uncertainties 
in Table \ref{tab3}. 
These uncertainties  were
derived from the uncertainties of the optical line intensities given in 
 our previous papers (references in Table \ref{tab1}),
 and from the uncertainties of the NIR line intensities given in Table 
\ref{tab2}. They were propagated to derive the oxygen and helium abundances.

Whenever there exists several observations
of the same object in the optical range, we treat those as
independent observations to increase the statistics. The optical data obtained by our group have been 
supplemented by spectra from the Sloan Digital Sky Survey (SDSS). 
In total, our sample consists of 75 optical
spectra which were combined with the 45
near-infrared spectra.

\subsection{Adjustment of NIR to optical spectra}

As the NIR and optical spectra were obtained with different telescopes and 
apertures at different times, we need to match them. 
The scale factor $f$ needed to 
adjust the intensity of the He~{\sc i} $\lambda$10830\AA\ NIR line to the 
H$\beta$ intensity is defined by the following equation: 
\begin{equation}
\frac{I({\rm He\ I}\ \lambda10830)}{I({\rm H}\beta)}= f\times
\frac{F({\rm He\ I}\ \lambda10830)}{F({\rm P}\gamma\ \lambda10940)} \label{f}
\end{equation} 
In this equation, $f$ is the theoretical value of the recombination line ratio 
P$\gamma$/H$\beta$ \citep{HS92} which depends mainly on the electron 
temperature, but little on the electron density. Thus in the $N_{\rm e}$ =  10$^2$-10$^4$ cm$^{-3}$ range, $f$=9.04 for 
$T_{\rm e}$ =  10$^4$K, and  $f$=8.11 for $T_{\rm e}$ =  2$\times$10$^4$K.
$I$(He~{\sc i}~$\lambda$10830)/$I$(H$\beta$) is the extinction-corrected
ratio, and  $F$(He~{\sc i}~$\lambda$10830)/$F$(P$\gamma$~$\lambda$10940) is
the observed ratio derived from Table \ref{tab2}. The He~{\sc i} $\lambda$10830 and the 
P$\gamma$ $\lambda$10940 emission lines are very close in wavelength, so 
that differential extinction between these two lines is negligible. Uncertainties of the 
$F$(He {\sc i} $\lambda$10830)/$F$(P$\gamma$ $\lambda$10940) ratios were propagated 
in the derivation of the He abundance.

Fig. \ref{fig2} shows an example of the matching of the optical and NIR 
spectra for one galaxy, the BCD SBS 0335--052E. 
The He~{\sc i} lines used for the determination of the primordial 
helium abundance are indicated. The Figure shows that the intensity 
of the NIR line is considerably higher than those of the optical lines.   

\begin{figure*}
\hbox{
\includegraphics[width=4.0cm,angle=-90.]{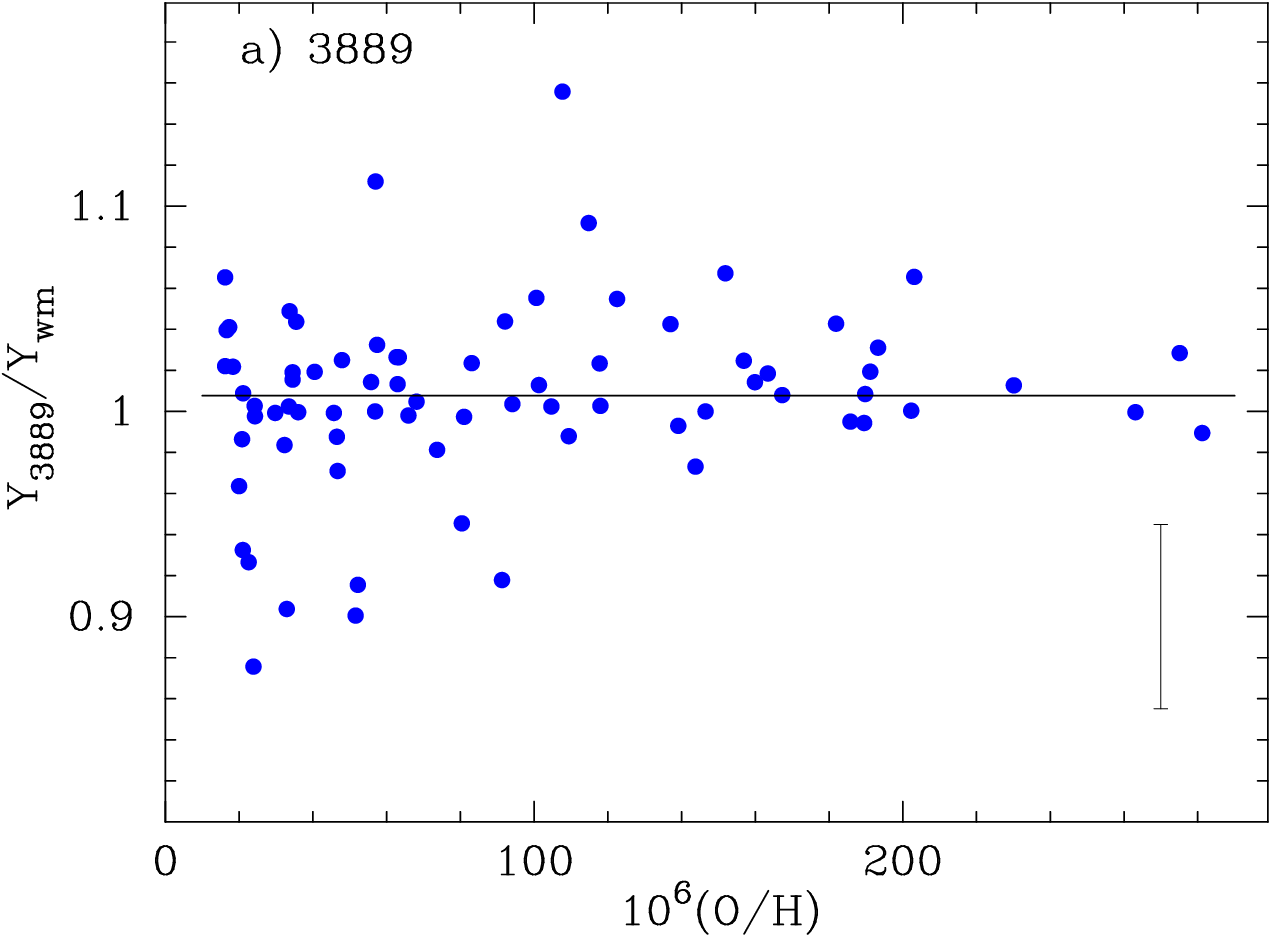}
\hspace{0.3cm}\includegraphics[width=4.0cm,angle=-90.]{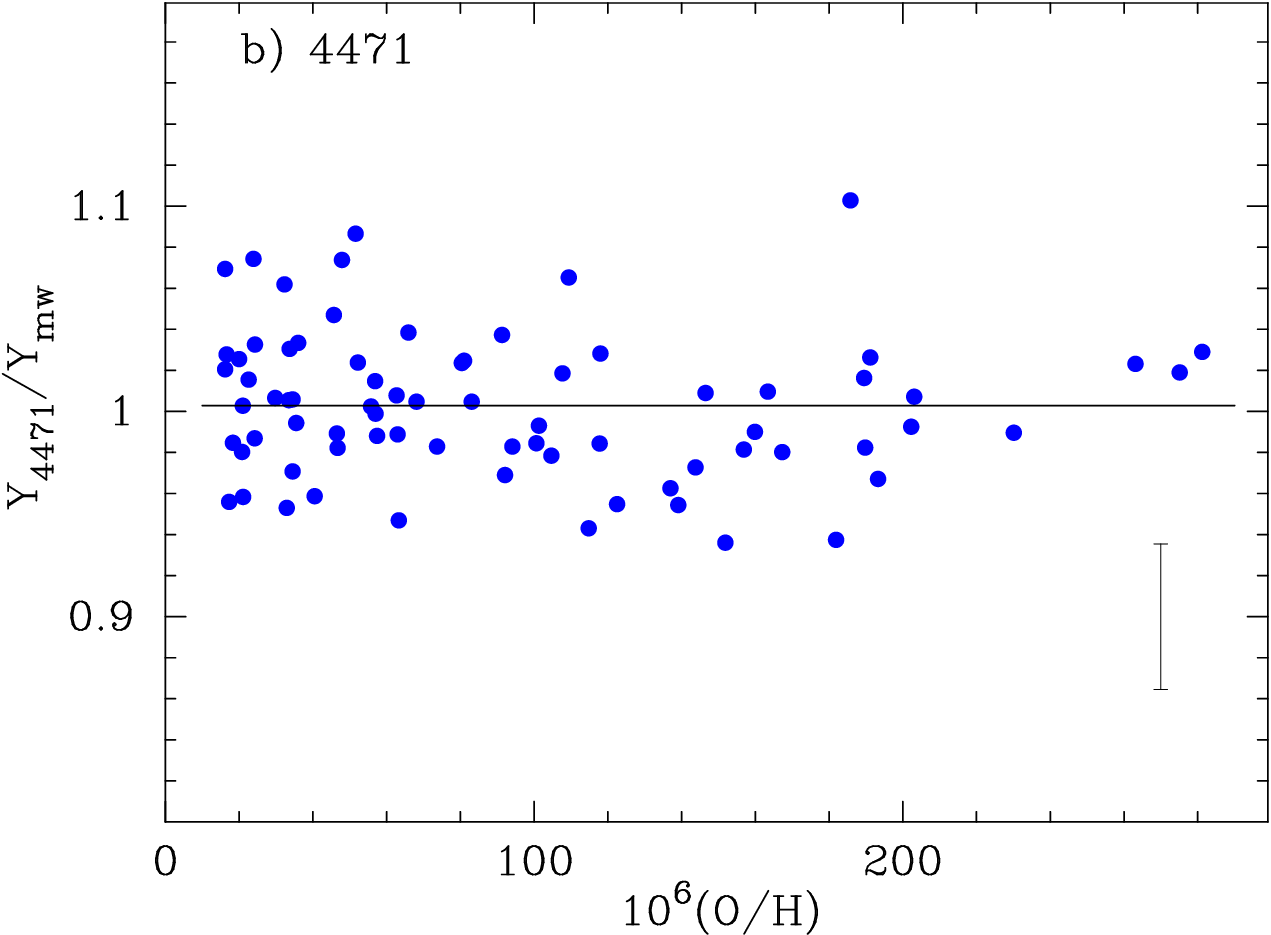}
\hspace{0.3cm}\includegraphics[width=4.0cm,angle=-90.]{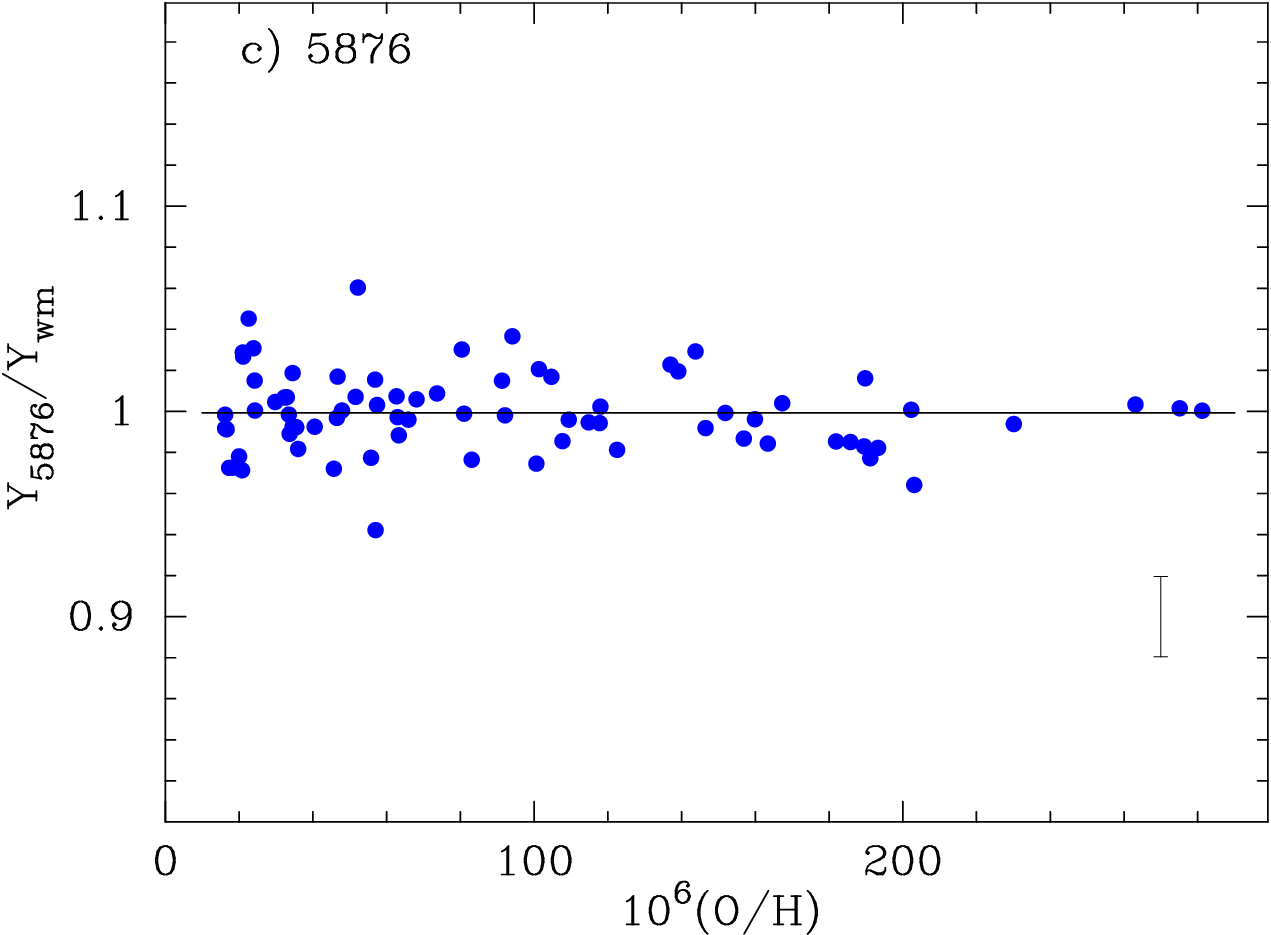}
}
\vspace{0.3cm}
\hbox{
\includegraphics[width=4.0cm,angle=-90.]{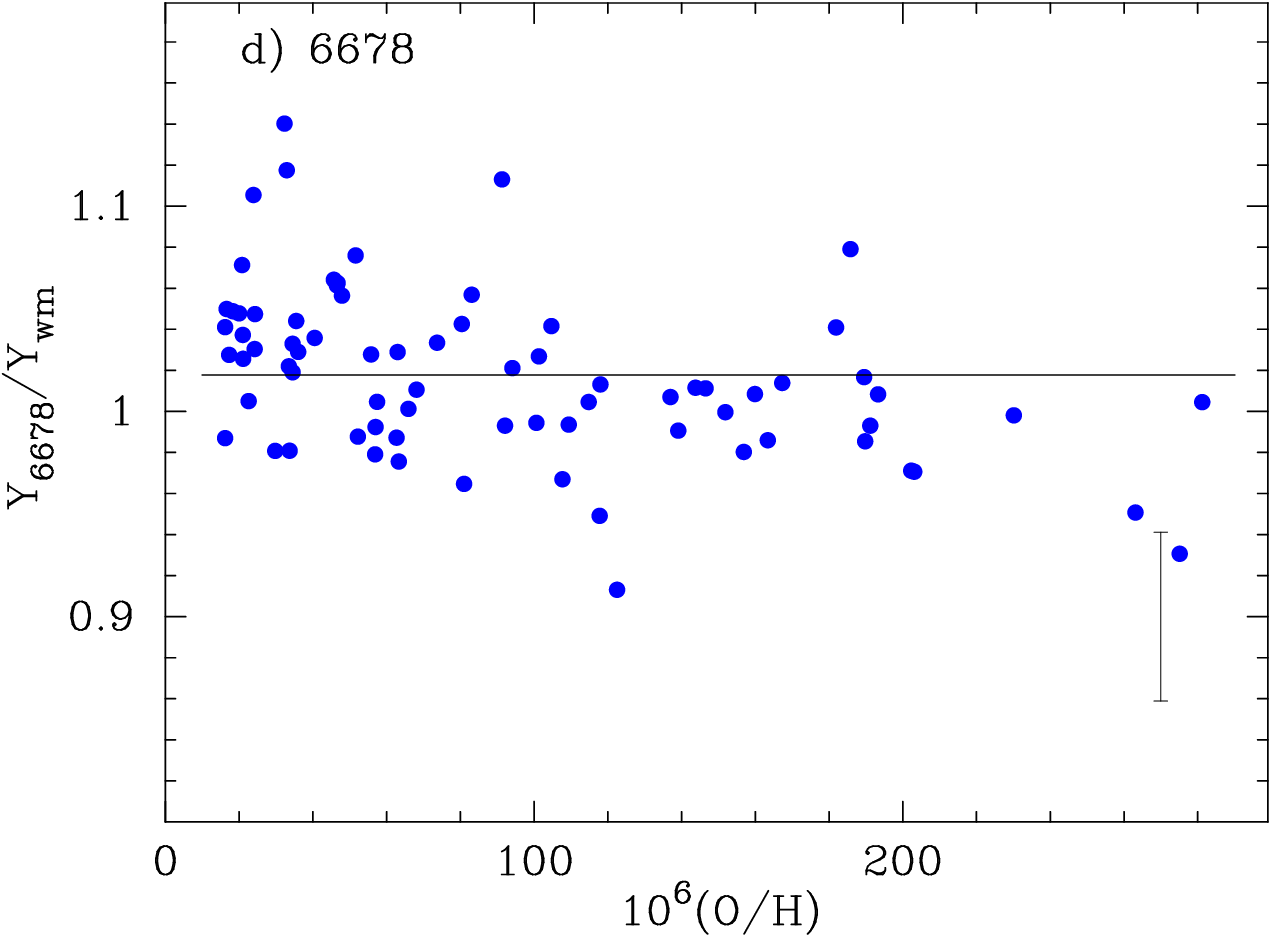}
\hspace{0.3cm}\includegraphics[width=4.0cm,angle=-90.]{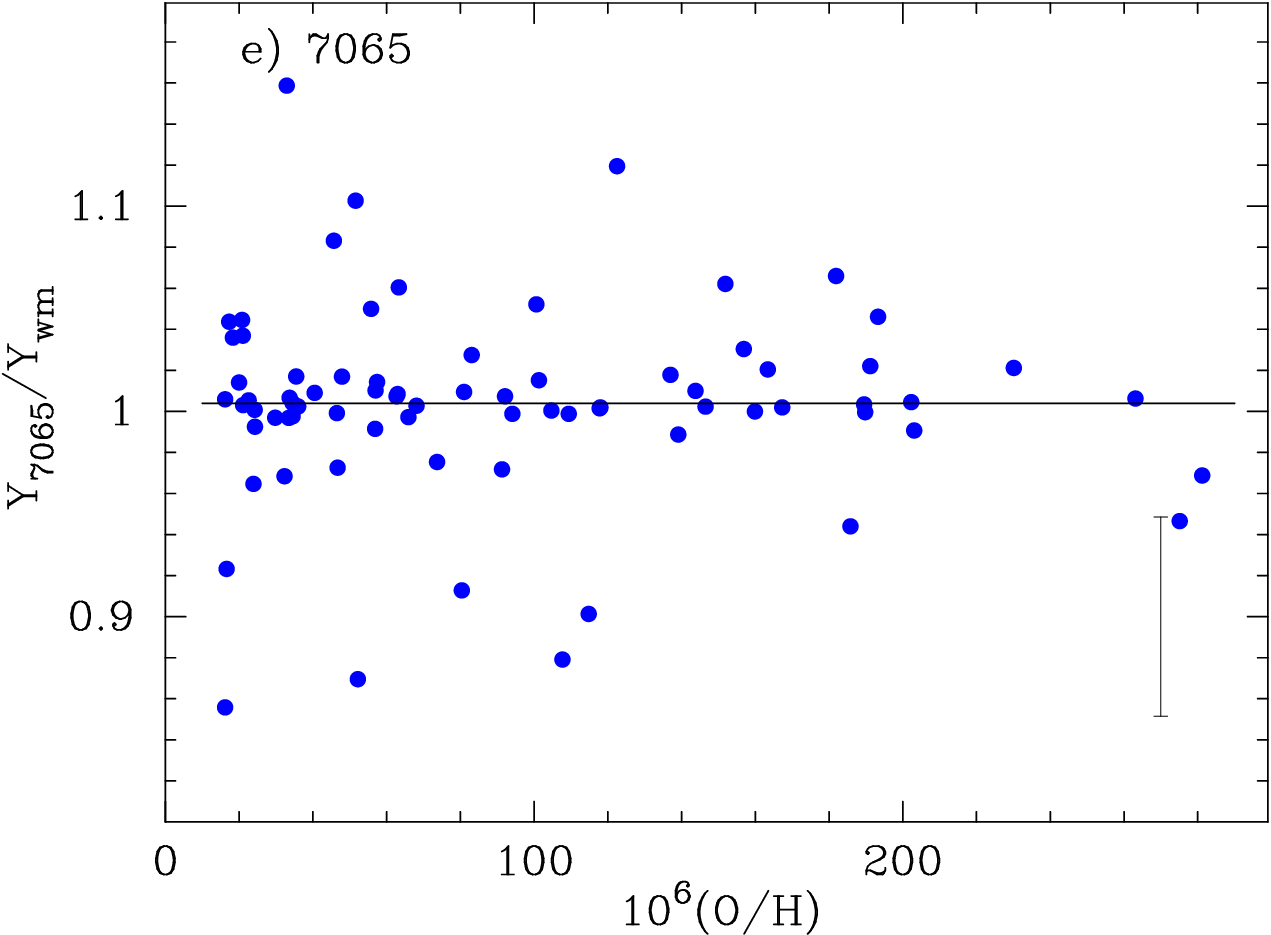}
\hspace{0.3cm}\includegraphics[width=4.0cm,angle=-90.]{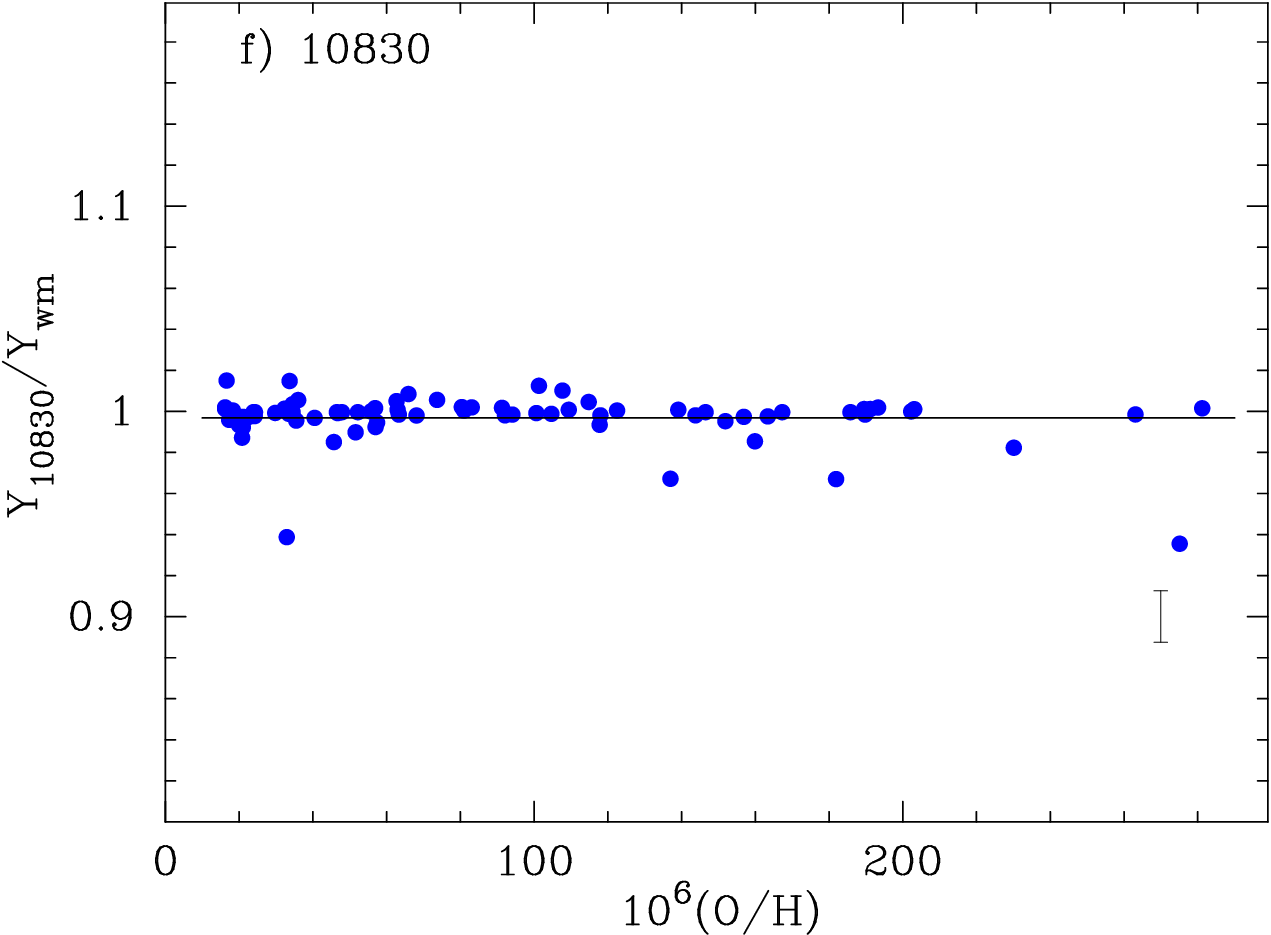}
}
\caption{Ratios of $Y$($\lambda$) derived from individual He~{\sc i} 
emission lines to the weighted mean helium abundance $Y_{\rm wm}$. 
All six He~{\sc i} emission lines were used for the $\chi^2$ 
minimisation and $Y_{\rm wm}$ determination. Horizontal
lines show a ratio of 1, and error bars are 1$\sigma$ dispersions.}
\label{fig3}
\end{figure*}

\begin{figure*}
\hbox{
\includegraphics[width=6.0cm,angle=-90.]{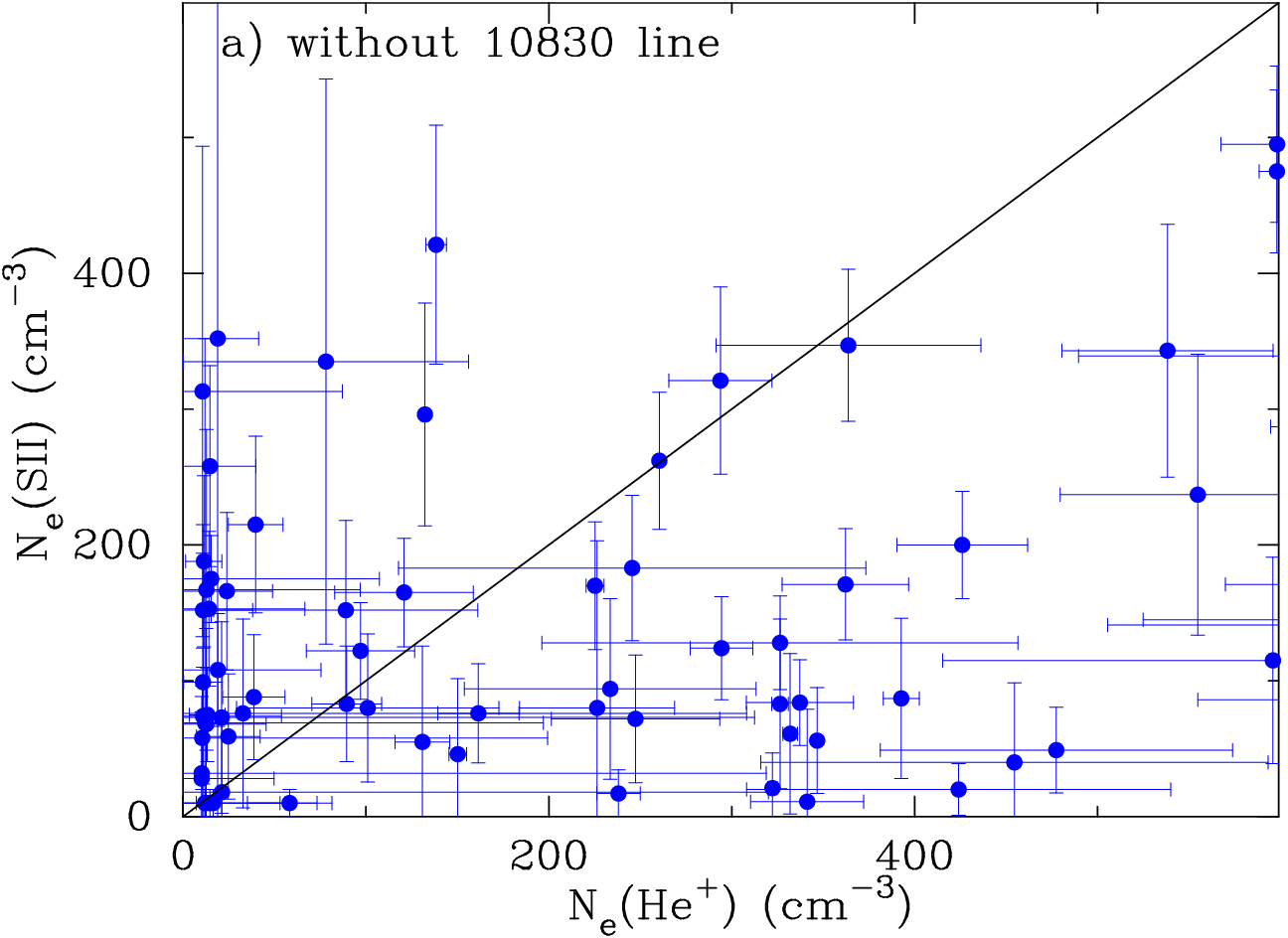}
\hspace{0.3cm}\includegraphics[width=6.0cm,angle=-90.]{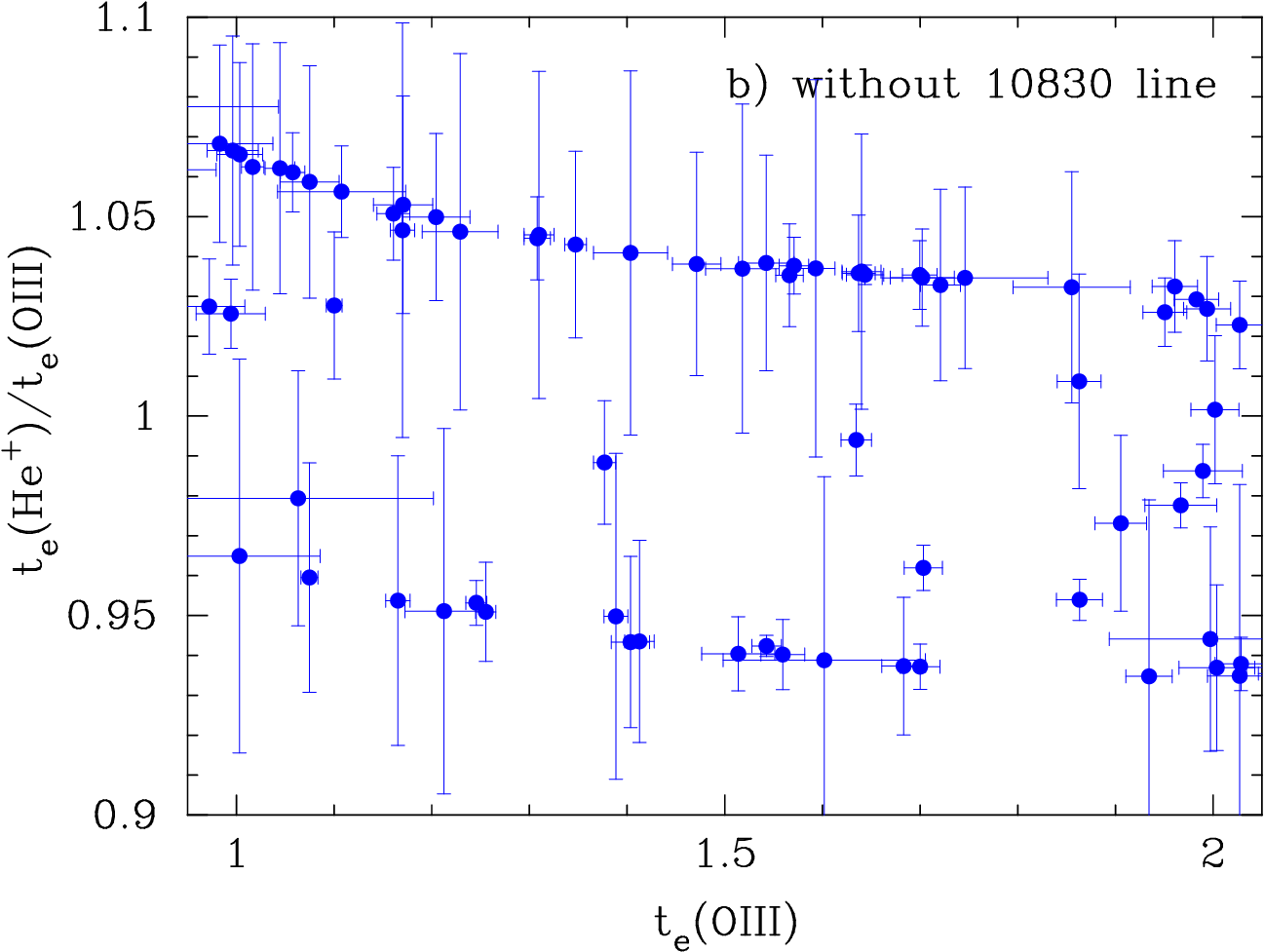}
}
\vspace{0.3cm}
\hbox{
\includegraphics[width=6.0cm,angle=-90.]{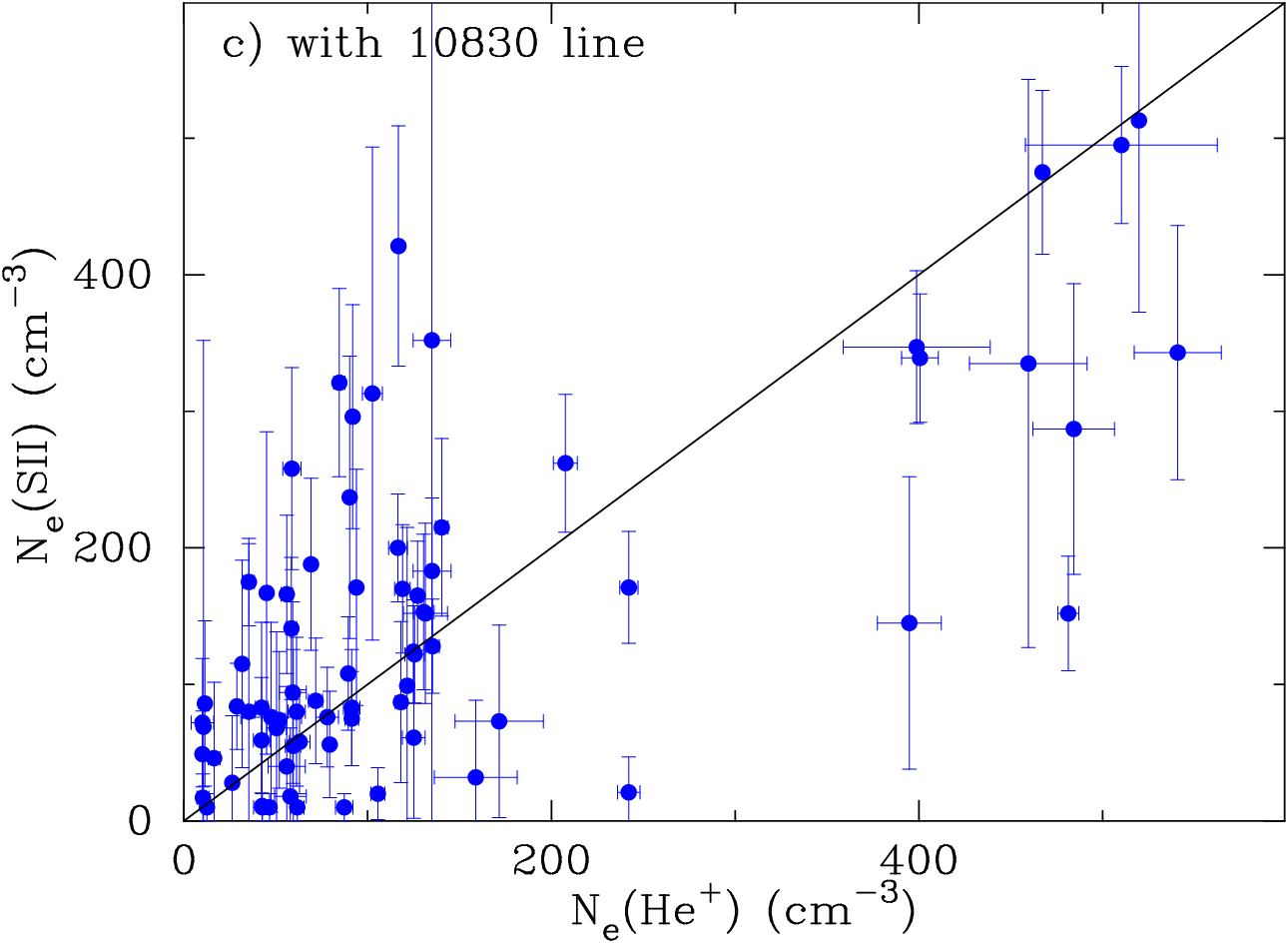}
\hspace{0.3cm}\includegraphics[width=6.0cm,angle=-90.]{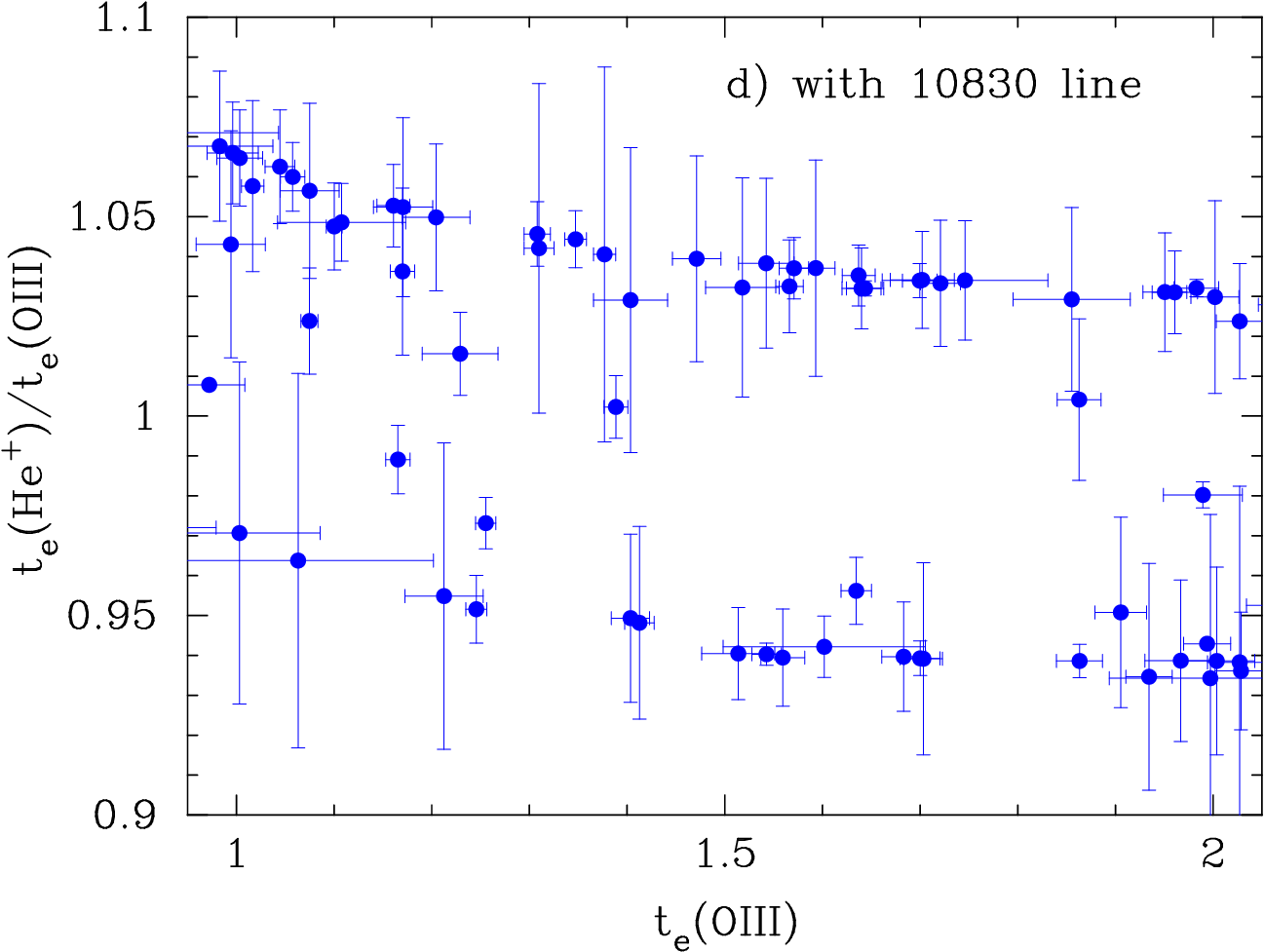}
}
\caption{(a) Electron densities $N_{\rm e}$(He$^+$) derived from the $\chi^2$ minimisation procedure  
versus $N_{\rm e}$(S~{\sc ii}) derived from the [S~{\sc ii}] line ratio. 
Only the five He~{\sc i}  $\lambda$3889, $\lambda$4471, $\lambda$5876, 
$\lambda$6678, and $\lambda$7065 optical emission lines 
have been used for the $\chi^2$ minimisation. The NIR He~{\sc i} $\lambda$10830 
emission line has been excluded.  $N_{\rm e}$(He$^+$) has been  
varied in the range 10 -- 600 cm$^{-3}$.
(b) Electron temperature ratios 
$T_{\rm e}$(He$^+$)/$T_{\rm e}$(O~{\sc iii}) derived from the $\chi^2$ minimisation procedure versus $T_{\rm e}$(O~{\sc iii}) derived from the 
O~{\sc iii} line ratio. Here, $t_{\rm e}$ = 10$^{-4}$$T_{\rm e}$. 
Only the five He~{\sc i}  $\lambda$3889, $\lambda$4471, $\lambda$5876, $\lambda$6678, and $\lambda$7065 optical emission lines 
have been used for the $\chi^2$ minimisation. The NIR He~{\sc i} $\lambda$10830 emission line has been excluded.  
$T_{\rm e}$(He$^+$) has been varied  
in the range 0.95 -- 1.05 of the $\widetilde{T}_{\rm e}$(He$^+$),
where $\widetilde{T}_{\rm e}$(He$^+$) is defined by Eq. \ref{tHeOIII}.
(c) and (d) Same as (a) and (b), respectively, except that the NIR He~{\sc i} $\lambda$10830 emission line is now included, so that all six He~{\sc i} $\lambda$3889, $\lambda$4471, $\lambda$5876, $\lambda$6678, 
$\lambda$7065, and $\lambda$10830 emission 
lines are used for the $\chi^2$ minimisation. 
1$\sigma$ errors bars are shown in all panels.}
\label{fig4}
\end{figure*}

\section{Determining the He abundance}\label{real}

\subsection{Corrections for dust extinction, fluorescent and collisional excitation, ionisation structure and underlying stellar absorption}

To carry out our empirical method to determine helium abundances in real H~{\sc ii} regions, several 
effects need to be considered and corrections applied.  

First, the Balmer decrement corrected for
the non-recombination contribution was used to simultaneously determine
the dust extinction and equivalent widths of underlying stellar 
hydrogen absorption lines, as described e.g. in \citet{ITL94}, and to 
correct line intensities for both effects. Second, since the
spectra of extragalactic H~{\sc ii} regions include both ionised gas 
and stellar emission, the underlying
stellar He~{\sc i} absorption lines should be taken into account
\citep[see e.g. ][]{I07}. Third, the He~{\sc i} emission lines
should be corrected for fluorescent excitation, parametrised 
by the optical depth $\tau$($\lambda$3889) of the He~{\sc i} $\lambda$3889
emission line. We used the correction factors for fluorescent excitation
derived by \citet{B99,B02}. 

To take into account the new set of He~{\sc i} emissivities of \citet{P13},
the collisional excitation of He~{\sc i} emission lines,
the non-recombination contribution to hydrogen emission-line intensities, and 
the correction for the ionisation structure of the H~{\sc ii} region, we have adopted the fits 
provided by \citet{I13}.

The equivalent width of the He~{\sc i} $\lambda$4471 absorption
line is chosen to be EW$_{\rm abs}$($\lambda$4471) = 0.4\AA, following
\citet{GD05}, \citet{IT10}, and \citet{I13}. 
The equivalent widths of the other absorption lines in the optical
range were fixed according to the ratios 
\begin{eqnarray}
{\rm EW}_{\rm abs}(\lambda 3889)/{\rm EW}_{\rm abs}(\lambda 4471)~\,& = &1.0,\nonumber \\
{\rm EW}_{\rm abs}(\lambda 5876)/{\rm EW}_{\rm abs}(\lambda 4471)~\,& = &0.8,\nonumber \\
{\rm EW}_{\rm abs}(\lambda 6678)/{\rm EW}_{\rm abs}(\lambda 4471)~\,& = &0.4,\nonumber \\
{\rm EW}_{\rm abs}(\lambda 7065)/{\rm EW}_{\rm abs}(\lambda 4471)~\,& = &0.4,\nonumber  \\
{\rm EW}_{\rm abs}(\lambda 10830)/{\rm EW}_{\rm abs}(\lambda 4471)& = &0.8. 
\label{ew}
\end{eqnarray}
The EW$_{\rm abs}$($\lambda$5876) / EW$_{\rm abs}$($\lambda$4471)
and EW$_{\rm abs}$($\lambda$6678) / EW$_{\rm abs}$($\lambda$4471) ratios
were set equal to the values 
predicted for these ratios by a Starburst99 \citep{L99} instantaneous
burst model with an age of 3 -- 4 Myr and a heavy-element
mass fraction $Z$ = 0.001 -- 0.008.
We note that the value chosen for the EW$_{\rm abs}$($\lambda$5876)/
EW$_{\rm abs}$($\lambda$4471) ratio is consistent with
the one given by \citet{GD05}. Since the
output high-resolution spectra in Starburst99 are calculated
only for wavelengths $<$~7000~\AA, we do not have a prediction
for the EW$_{\rm abs}$($\lambda$7065)/
EW$_{\rm abs}$($\lambda$4471) ratio. We set it to be
equal to 0.4, the value of the EW$_{\rm abs}$($\lambda$6678)/
EW$_{\rm abs}$($\lambda$4471) ratio. As for He~{\sc i} $\lambda$3889,
this line is blended with the hydrogen H8 $\lambda$3889 line. Therefore,
EW$_{\rm abs}$(He~{\sc i} $\lambda$3889) cannot be estimated from the 
Starburst99 models. We assumed the value shown in Eq. \ref{ew}.

No data are available for the 
EW$_{\rm abs}$ of the He~{\sc i} $\lambda$10830 NIR line. Therefore, we have set it to be
equal to the EW$_{\rm abs}$ of another strong line, the 
He~{\sc i} $\lambda$5876 optical emission line. This assumption does not introduce
an appreciable uncertainty in the He~{\sc i} $\lambda$10830 emission-line flux,
because the equivalent width of this line is generally high (Table \ref{tab2}).

\subsection{Starburst ages and ionisation correction factors}

The age $t_{\rm burst}$ of the starburst in the  H~{\sc ii} region needs to be derived. This is 
because the ionisation correction factors $ICF$(He) and the non-recombination 
contribution to hydrogen lines both depend on the starburst age 
\citep{I13}. As a first approximation, \citet{I13}
used the relation between $t_{\rm burst}$ and EW(H$\beta$) from the 
Starburst99 instantaneous burst models with a heavy-element mass fraction
$Z$ = 0.004 \citep{L99}. They fitted this relation by the expression
\begin{eqnarray}
 t_{\rm burst}&=&167.6w^3-2296w^2+12603w-35651 \nonumber \\
              &+&54976/w-43884/w^2+14208/w^3, \label{tb}
\end{eqnarray}
where $t_{\rm burst}$ is in Myr, $w$ = log EW(H$\beta$) 
and EW(H$\beta$) is in \AA. We adopt this relation in our analysis.
Eq. \ref{tb} does not take into account the contribution of old stellar 
populations in the underlying galaxy. The effect of an 
underlying galaxy on the $ICF$(He) was discussed by \citet{I13} who found 
it to be small for high-excitation H~{\sc ii} regions with 
EW(H$\beta$)$\geq$150\AA.
Since the relations for $Z$ = 0.001, $Z$ = 0.004, and 
$Z$ = 0.008 are similar for EW(H$\beta$) $\ga$ 100\AA, corresponding 
to $t_{\rm burst}$ $\la$ 4 Myr, 
we have adopted Eq. \ref{tb} for the entire range of oxygen
abundances in our sample galaxies. We have also adopted $t_{\rm burst}$ = 1 Myr
and 4 Myr, when the derived starburst age was $<$ 1 Myr or $>$ 4 Myr. For
$t_{\rm burst}$ in the range of 1 -- 4 Myr, we have linearly interpolated between 
 the fits given by \citet{I13}
to derive $ICF$(He) and to correct for the
non-recombination contribution to the intensities
of the hydrogen lines H$\alpha$, H$\beta$, H$\gamma$, and H$\delta$.

\subsection{Correction for the fraction of oxygen locked in dust grains}

Finally, the oxygen abundance should
be corrected for its fraction locked in dust grains. \citet{I06} found
the Ne/O abundance ratio in low-metallicity emission-line galaxies
to increase with increasing oxygen abundance. They interpreted this
trend by a larger fraction of oxygen locked in dust grains in galaxies
with higher O/H. Following \citet{I13}, we use the relation 

\begin{equation}
\Delta \left(\frac{\rm O}{\rm H}\right)_{\rm dust} =
10^{0.088(12+\log{\rm O/H})-0.616} \label{dustcorr}
\end{equation}
to derive the fraction of oxygen confined in dust.

This oxygen abundance locked in dust is added to the oxygen abundance 
in the gaseous phase, as derived from the emission-line fluxes.

\begin{figure*}
\hbox{
\includegraphics[width=6.0cm,angle=-90.]{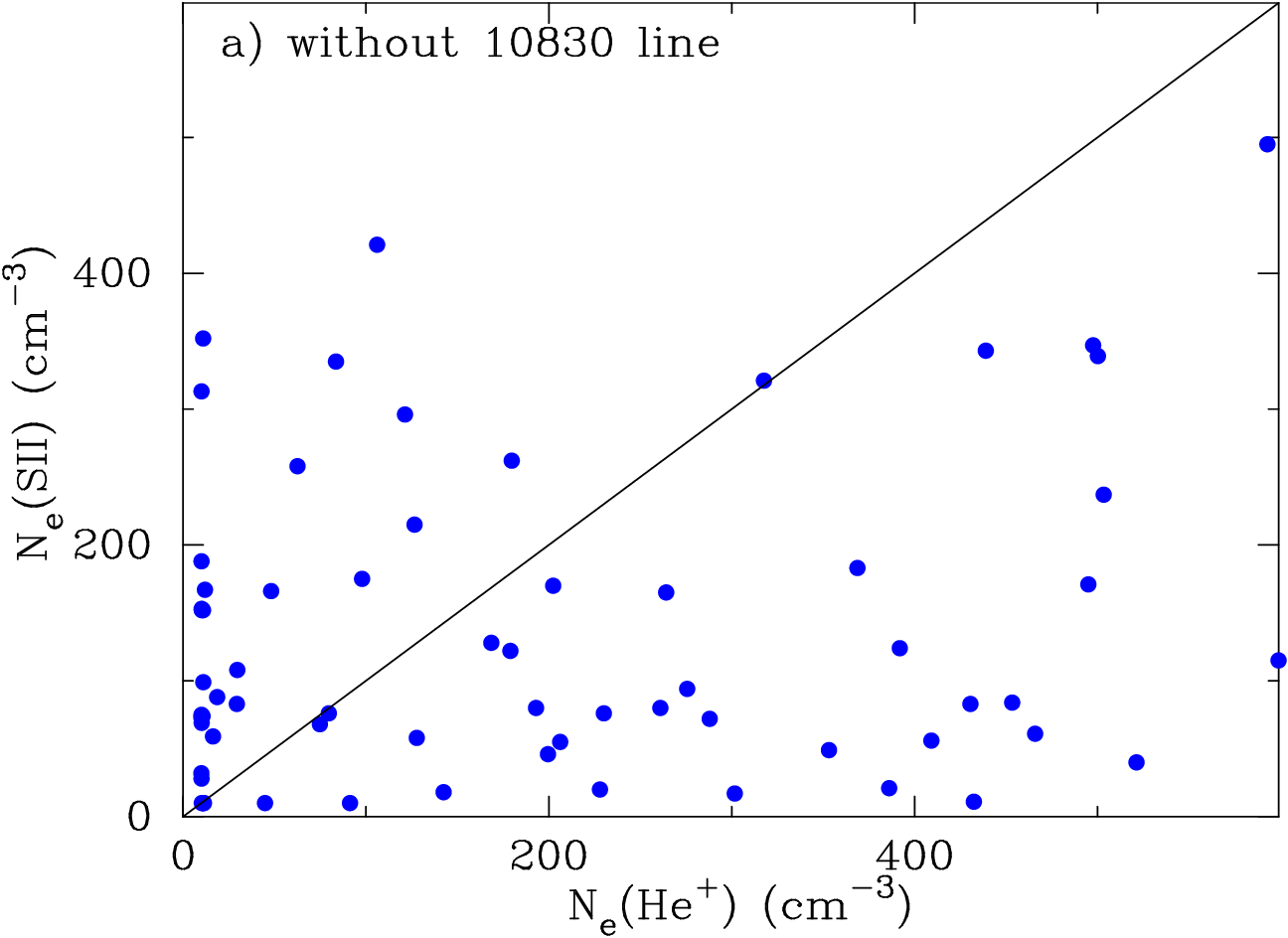}
\hspace{0.3cm}\includegraphics[width=6.0cm,angle=-90.]{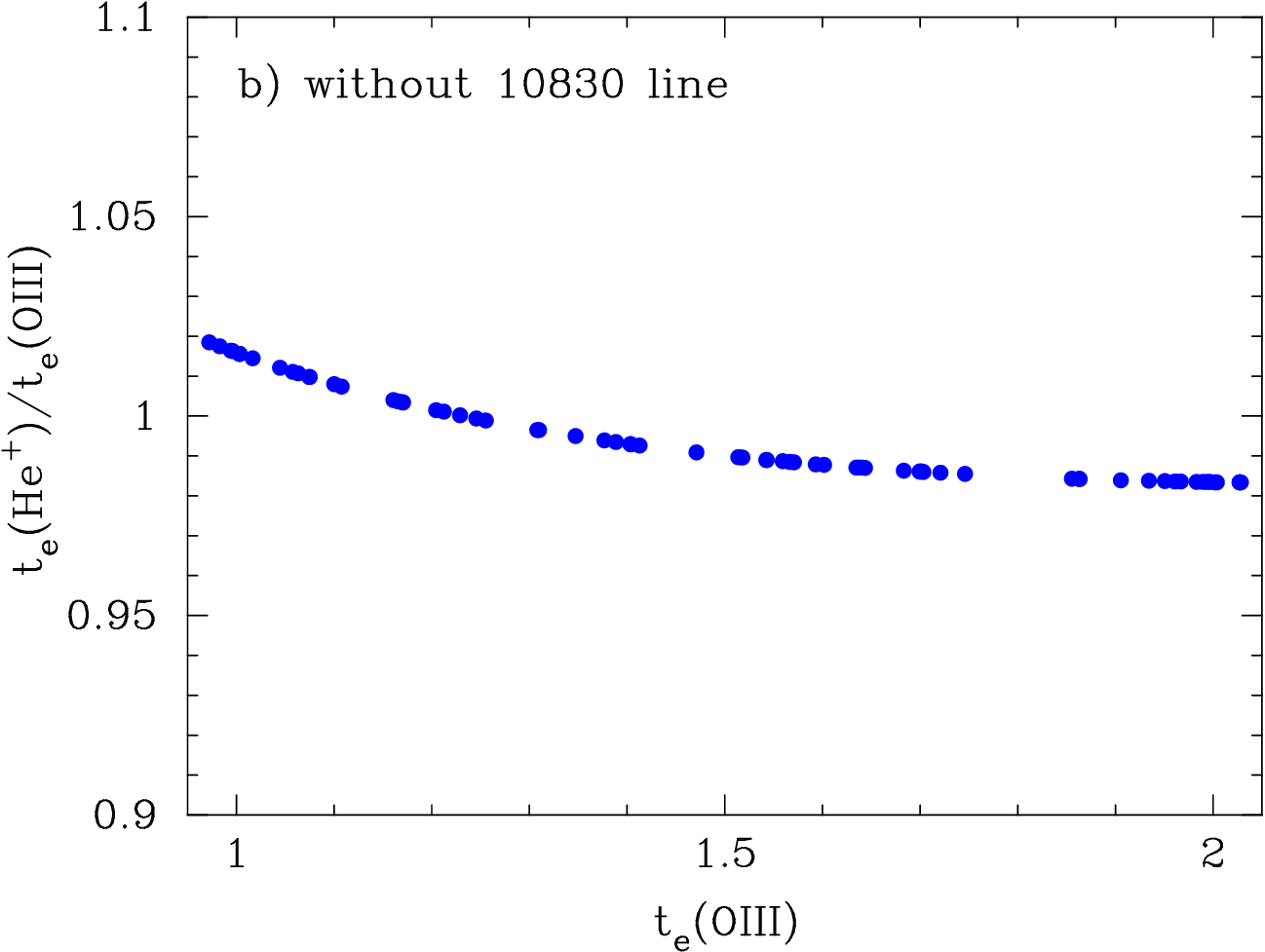}
}
\vspace{0.3cm}
\hbox{
\includegraphics[width=6.0cm,angle=-90.]{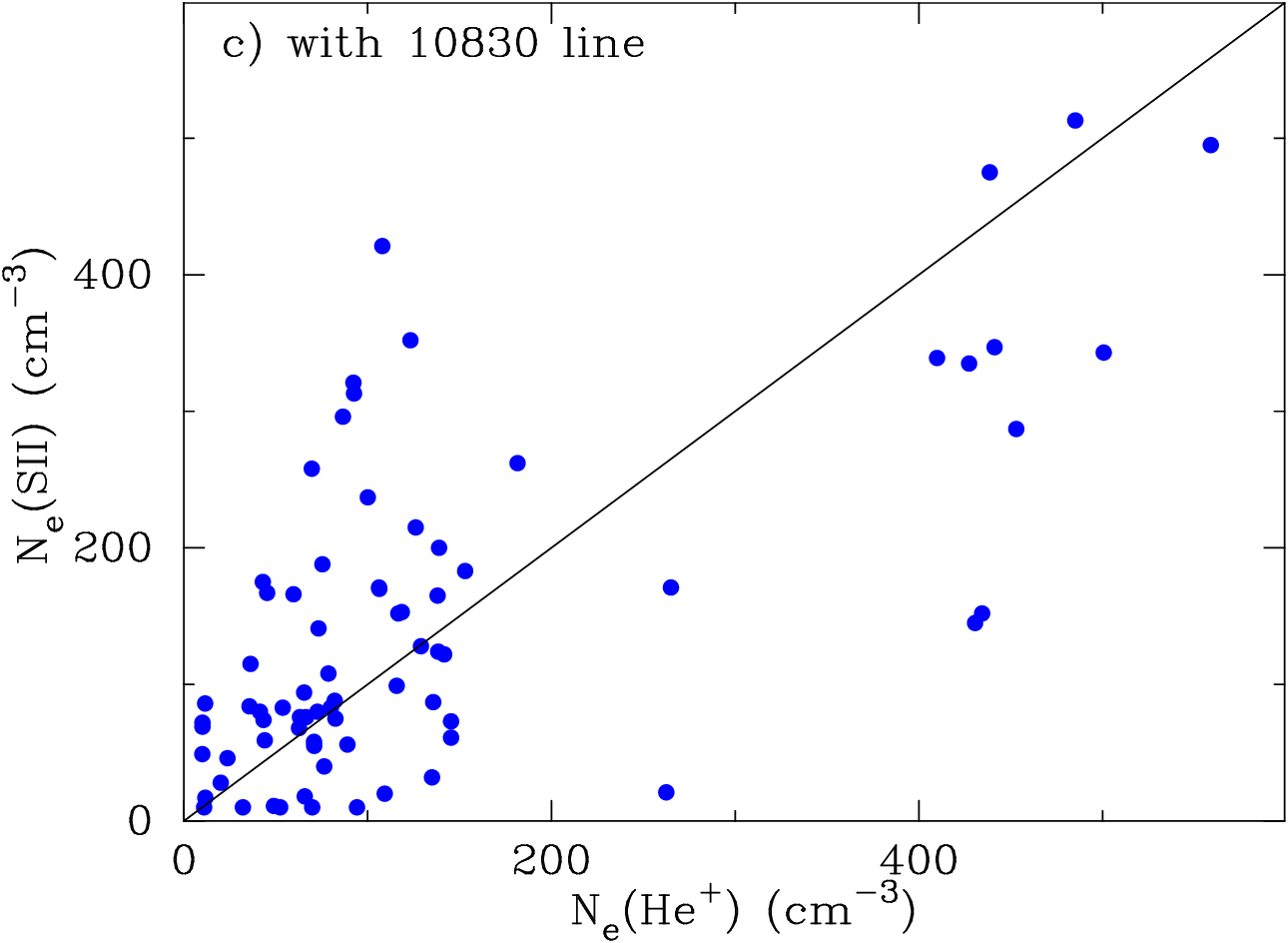}
\hspace{0.3cm}\includegraphics[width=6.0cm,angle=-90.]{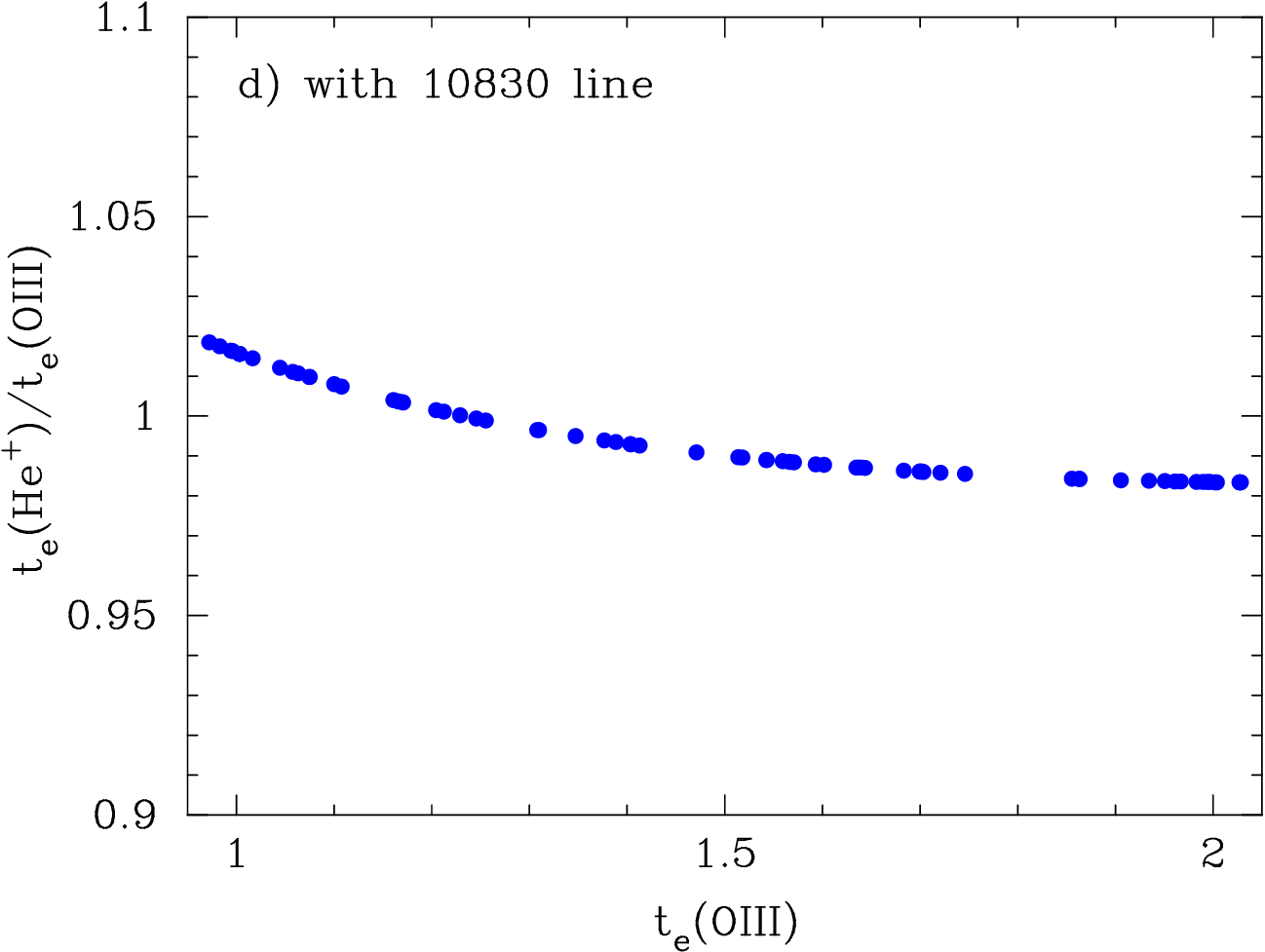}
}
\caption{Same as in Fig. \ref{fig4}, but the electron temperature 
$T_{\rm e}$(He$^+$) is set equal to $\widetilde{T}_{\rm e}$(He$^+$).}
\label{fig5}
\end{figure*}

\begin{figure*}
\hbox{
\includegraphics[width=6.0cm,angle=-90.]{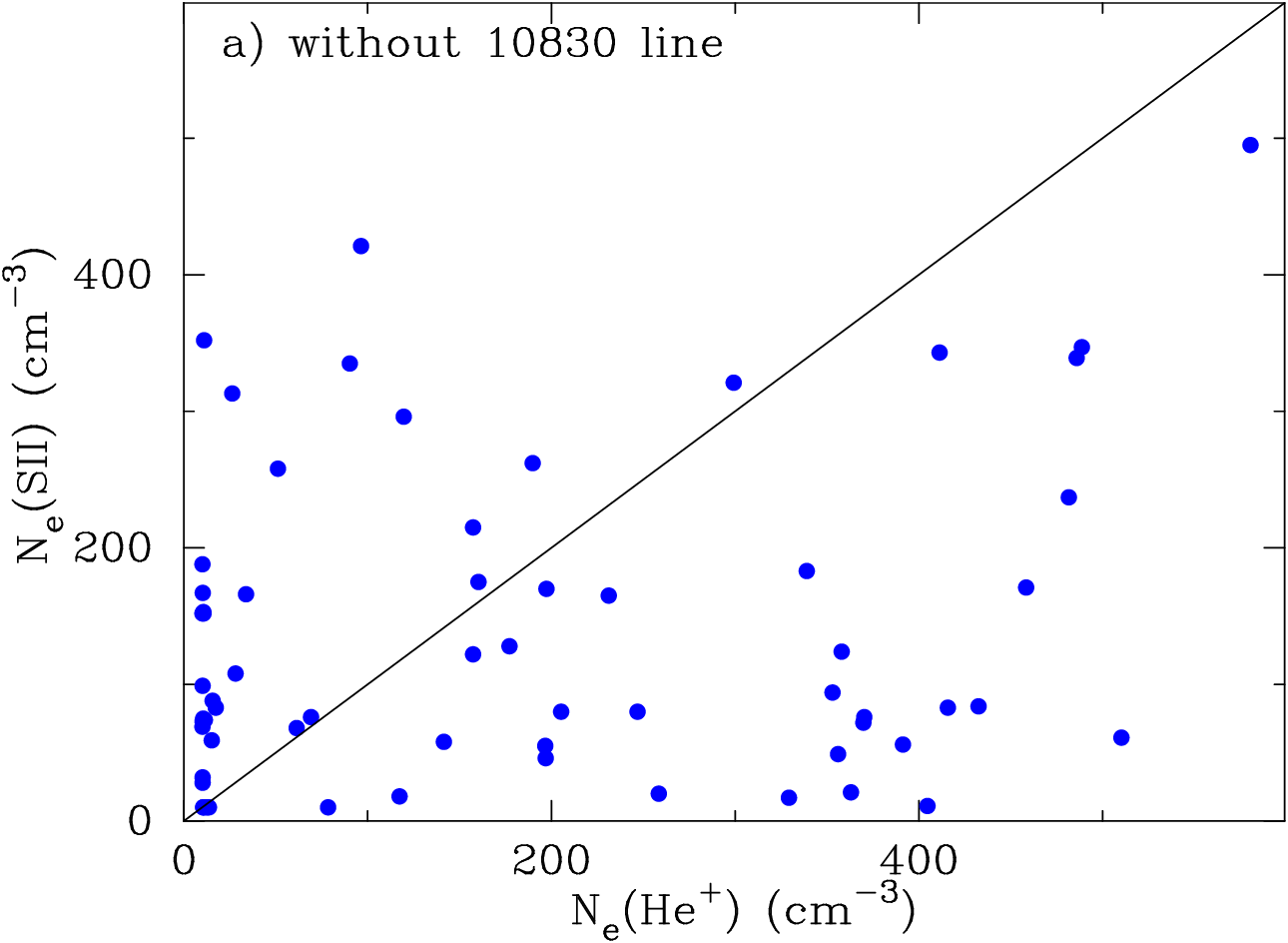}
\hspace{0.3cm}\includegraphics[width=6.0cm,angle=-90.]{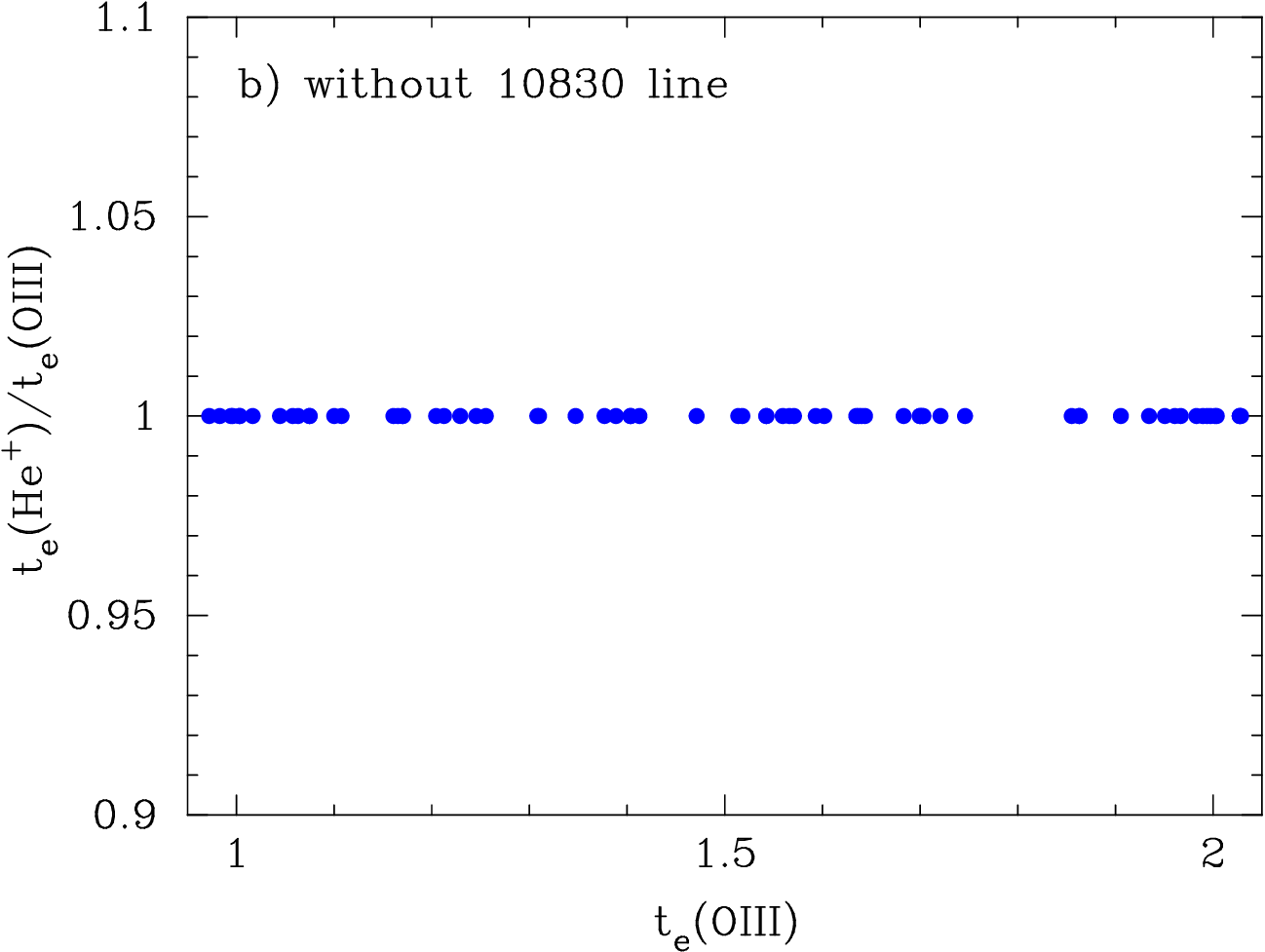}
}
\vspace{0.3cm}
\hbox{
\includegraphics[width=6.0cm,angle=-90.]{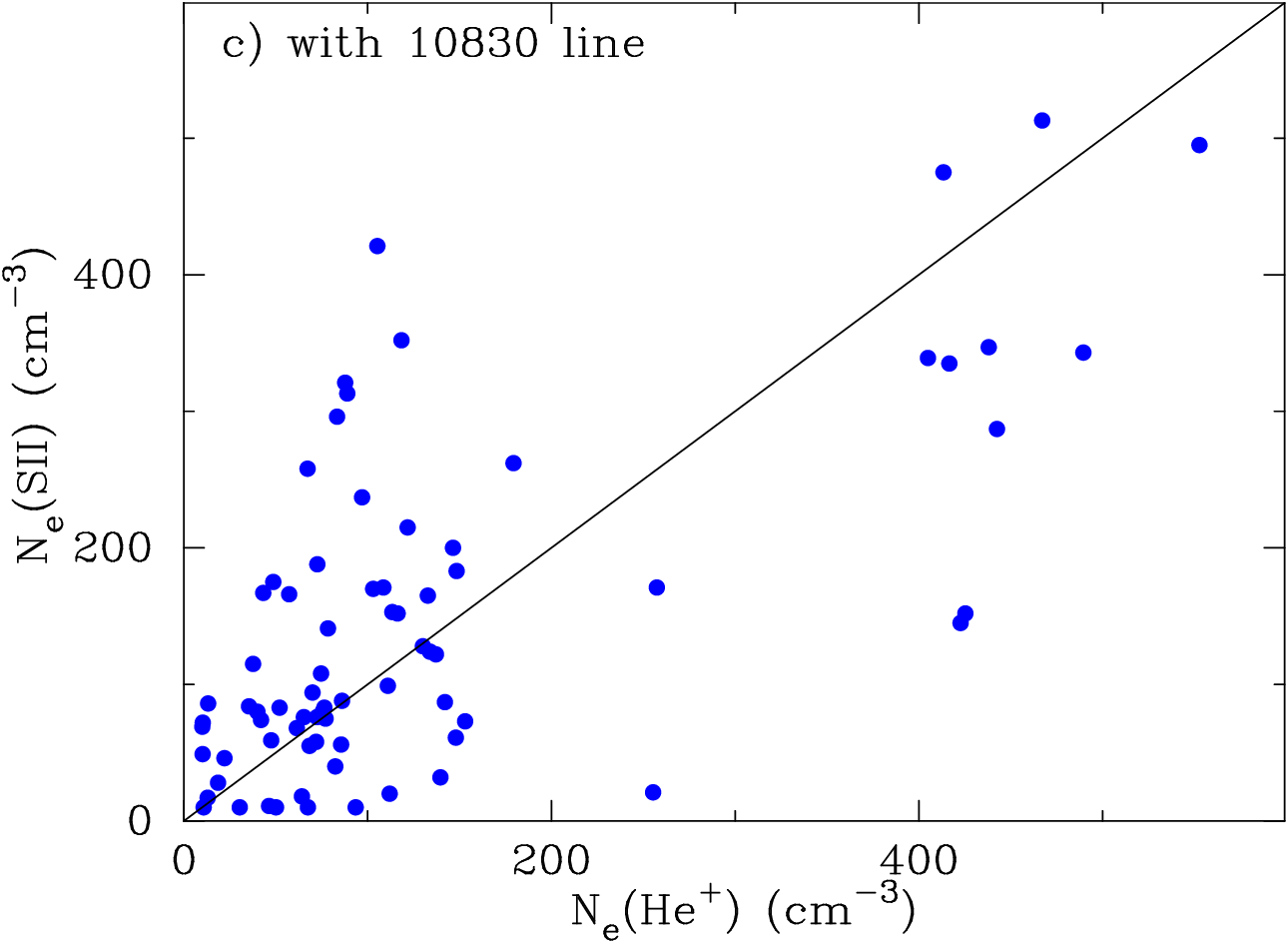}
\hspace{0.3cm}\includegraphics[width=6.0cm,angle=-90.]{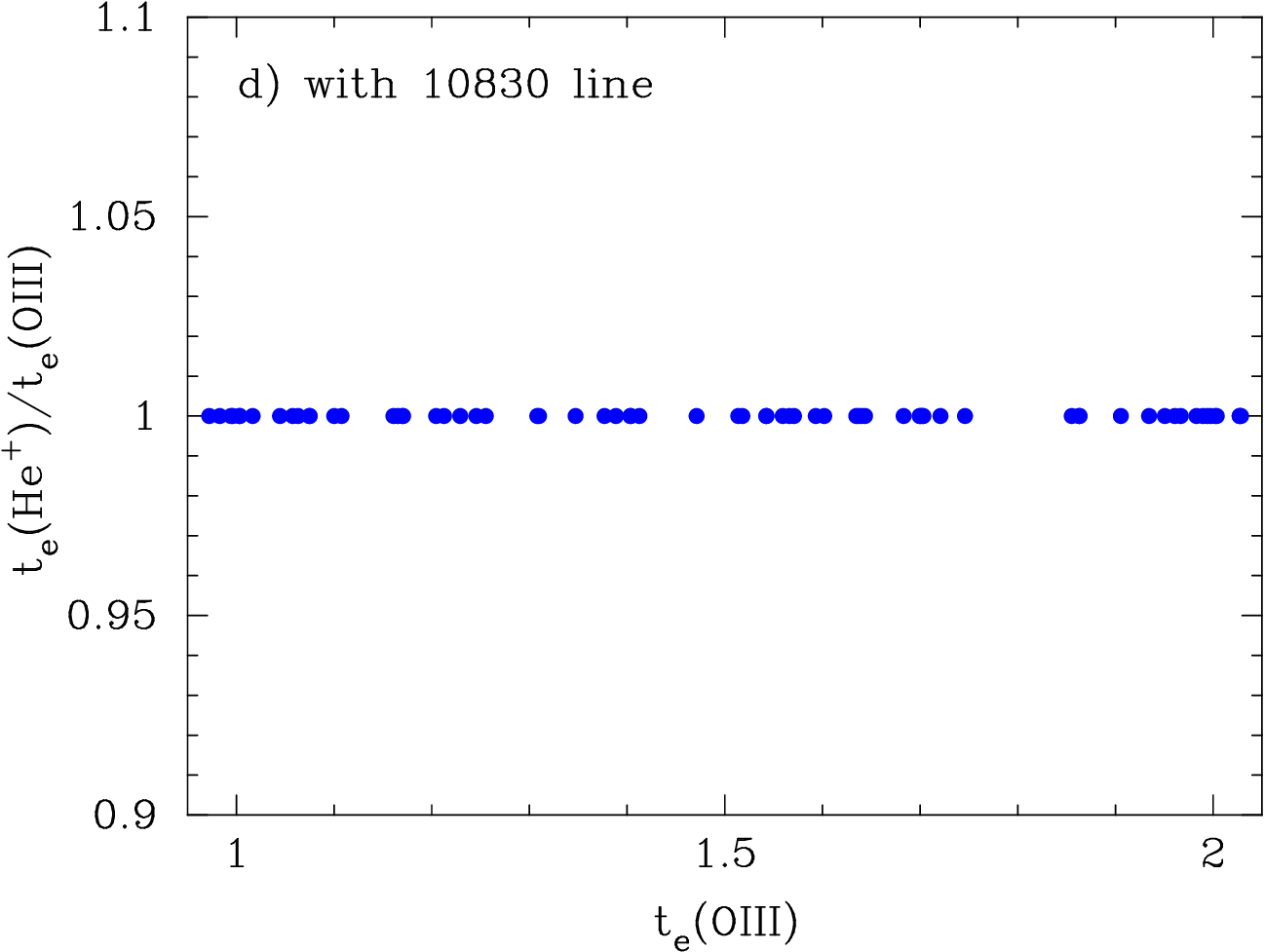}
}
\caption{Same as in Fig. \ref{fig4}, but the electron temperature 
$T_{\rm e}$(He$^+$) is set equal to $T_{\rm e}$(O~{\sc iii}).}
\label{fig6}
\end{figure*}


\subsection{Monte Carlo method}

To determine the He$^+$ abundance $y^+_i$ = He$^+_i$/H$^+$ , we have applied the 
Monte Carlo procedure described in \citet{IT04} and \citet{I07}. We 
randomly vary the electron temperature $T_{\rm e}$(He$^+$), the electron number
density $N_{\rm e}$(He$^+$) and the optical depth $\tau$($\lambda$3889) within a specified range, to minimise the quantity
\begin{equation}
\chi^2=\sum_i^n\frac{(y^+_i-y^+_{\rm wm})^2}{\sigma^2(y^+_i)}\label{eq1}.
\end{equation}

The weighted mean of the $y^+_i$, 
$y^+_{\rm wm}$, is defined by
\begin{equation}
y^+_{\rm wm}=\frac{\sum_i^n{y^+_i/\sigma^2(y^+_i)}}
{\sum_i^n{1/\sigma^2(y^+_i)}}\label{eq2},
\end{equation}
where $y^+_i$ is the He$^+$ abundance derived from the intensity of the 
He~{\sc i}
emission line labelled $i$, and $\sigma(y^+_i)$ is the statistical error
of $y^+_i$. 
Each 
$y^+_i$ has been assigned a weight inversely proportional
to that statistical error, since fluxes of weaker emission 
lines are more uncertain and therefore should be attributed lower weights.
The resulting $y^+_{\rm wm}$ is the empirically derived 
He$^+$ abundance in each model.

Additionally, in the case of the nebular He~{\sc ii} $\lambda$4686 
emission line, we have added the abundance of doubly ionised 
helium $y^{2+}$ $\equiv$ He$^{2+}$/H$^+$ to $y^+$ to calculate $y$.
Although the He$^{2+}$ zone is hotter than the He$^{+}$ zone, we 
have adopted $T_{\rm e}$(He$^{2+}$) = $T_{\rm e}$(He$^{+}$).
The last assumption has only a minor effect on the $y$ value, because 
except in a few high-excitation H~{\sc ii} regions, 
$y$$^{2+}$ is in general small ($\leq$ 1\% of $y^+$). 

The total He abundance $y$ is obtained from the expression 
$y$~=~$ICF$(He)$\times$($y^+$+$y$$^{2+}$), where
$ICF$(He)= H$^+$/(He$^+$+He$^{2+}$) is the ionisation correction factor for
He. It is converted to the
He mass fraction using equation
\begin{equation}
Y=\frac{4y(1-Z)}{1+4y}, \label{eq:Y}
\end{equation}
where $Z$ = $B$$\times$O/H is the heavy-element mass fraction.
The coefficient $B$ depends on O/$Z$, where O is the oxygen mass fraction.
\citet{M92} derived O/$Z$ = 0.66 and 0.41 for $Z$ = 0.001 and 0.02, 
respectively. The latter value is close to the most recent determination for the Sun 
\citep[0.43, using the abundances of ][]{A09}. Adopting 
the \citet{M92} values, \citet{I13} obtained $B$ = 18.2 and
27.7 for $Z$ = 0.001 and 0.02, respectively. 
$B$ values are then assumed to 
linearly scale with the oxygen abundance $A$(O) (=~12~+~log O/H) according to the relation 
\begin{equation}
B = 8.64A({\rm O})-47.44.\label{B}
\end{equation} 
We adopt this relation in our calculations.


We varied $N_{\rm e}$(He$^+$) 
in the range 10 -- 600 cm$^{-3}$.
The electron temperature $T_{\rm e}$(O~{\sc iii}) was
derived from the 
[O~{\sc iii}]$\lambda$4363/($\lambda$4959+$\lambda$5007) emission line 
intensity ratio, prior to determining $Y$. \citet{I13} showed that it is very 
close to the volume-emissivity-averaged value of the electron
temperature in the O$^{2+}$ zone calculated with CLOUDY grid models for the 
whole range of metallicities.
For variations of the electron temperature $T_{\rm e}$(He$^+$), we have followed
the prescriptions of  \citet{I13}. $T_{\rm e}$(He$^+$) was set in the three following ways: 

1) $T_{\rm e}$(He$^+$) was randomly varied
in the range (0.95 -- 1.05) of the temperature derived from the relation 
between volume-averaged temperatures $T_{\rm e}$(He$^+$) and $T_{\rm e}({\rm O}~\textsc{iii})$ in
CLOUDY models, which \citet{I13} fitted by the expression
\begin{equation}
\widetilde{T}_{\rm e}({\rm He}^+) = 2.51\times 10^{-6} T_{\rm e}({\rm O}~\textsc{iii})+0.8756+
1152/T_{\rm e}({\rm O}~\textsc{iii}).
\label{tHeOIII}
\end{equation}
This relation predicts 
$\widetilde{T}_{\rm e}$(He$^+$) $<$ $T_{\rm e}$(O~{\sc iii}) for hotter H~{\sc ii} regions
and $\widetilde{T}_{\rm e}$(He$^+$) $>$ $T_{\rm e}$(O~{\sc iii}) for cooler H~{\sc ii} 
regions;  

2) $T_{\rm e}$(He$^+$) = $\widetilde{T}_{\rm e}$(He$^+$);

\noindent and 

3) $T_{\rm e}$(He$^+$) = $T_{\rm e}$(O~{\sc iii}).

As for the optical depth $\tau$($\lambda$3889), we varied it randomly in the range 
0 -- 5.

\subsection{The role of the He~{\sc i} $\lambda$10830\AA\ NIR line in the primordial helium abundance determination}

We derive the He$^+$ abundance $y^+_i$ = He$^+_i$/H$^+$ using the strongest He~{\sc i}  emission lines that are available. We consider two cases : 
1) the case where all six strongest helium emission lines are used: the five $\lambda$3889, $\lambda$4471,
$\lambda$5876, $\lambda$6678, $\lambda$7065 optical lines and the 
NIR He~{\sc i} $\lambda$10830\AA\ line; 2) the case which makes use only 
of the five strongest optical helium emission lines, excluding the NIR 
He~{\sc i} $\lambda$10830\AA\ line.  The latter case is the one we have 
considered in the past in all our previous work on the determination of 
the primordial helium abundance \citep[][ and references therein]{IT10,I13}. 
Comparing the two cases will allow us 
to assess the impact of the use of the NIR line on helium abundance determinations.

\subsubsection{A small dispersion about the mean helium abundance} 

In Fig. \ref{fig3}, we show for the whole sample the ratios of $Y$s derived 
from individual
lines to the weighted mean $Y_{\rm wm}$ as a function of oxygen abundance O/H. 
The sample consists of 75 optical spectra and 45 NIR spectra in 45 H~{\sc ii} 
regions of 43 galaxies. The number of optical spectra is greater than that
of NIR spectra, because some H~{\sc ii} regions were observed 
several times in the optical range.
All six lines were used in the $\chi^2$ minimisation
and determination of $Y_{\rm wm}$. It is seen that, for all lines, the 
$Y_i$/$Y_{\rm wm}$ values scatter around 1 (indicated by a horizontal line in each
panel), as should be the case. However, the dispersions of the points about the value 1 are quite different,
depending on the emission line. The He~{\sc i} $\lambda$10830\AA\ emission line, the one with the highest 
intensity, shows the smallest dispersion 
(panel f). On the other hand, the two 
He~{\sc i} $\lambda$3889\AA\ and He~{\sc i} $\lambda$7065\AA\ lines (panels a and e) show the highest dispersions. There are reasons for these high dispersions: the former
line is blended with the H8 hydrogen line and the determination of its
flux is subject to large uncertainties, while the latter line is generally
the weakest of all six lines. This is the first virtue of  
the He~{\sc i} $\lambda$10830\AA\ NIR line: it plays a very important role in the determination of the 
helium abundance. Because of its very small dispersion about the weighted mean $Y_{\rm wm}$, the latter quantity is determined to a large extent by this line.

\subsubsection{Sensitivity to the electron density}

We compare here the derived electron densities and temperatures of the He$^+$ 
zone compared to those derived for the S$^+$ and O$^{2+}$ regions, respectively.  
  
In Fig. \ref{fig4}a and \ref{fig4}b we show respectively the relations
$N_{\rm e}$(S~{\sc ii}) -- $N_{\rm e}$(He$^+$) and 
$T_{\rm e}$(He$^+$)/$T_{\rm e}$(O~{\sc iii}) -- $T_{\rm e}$(O~{\sc iii}), in the
case where He~{\sc i} $\lambda$10830\AA\ is not used in  the minimisation of 
$\chi^2$ (Eq. \ref{eq1}) and the determination of $Y_{\rm wm}$.  
$T_{\rm e}$(He$^+$) has been randomly varied in the range 
(0.95 -- 1.05)$\widetilde{T}_{\rm e}$(He$^+$). In agreement with  
\citet{I13}, we found no correlation between the number densities 
$N_{\rm e}$(S~{\sc ii}) and $N_{\rm e}$(He$^+$) (Fig. \ref{fig4}a). This lack of 
correlation also holds for the derived temperatures. The 
$T_{\rm e}$(He$^+$)/$T_{\rm e}$(O~{\sc iii}) ratio does not scatter around the 
value $\widetilde{T}_{\rm e}$/$T_{\rm e}$(O~{\sc iii}), but tends to be at 
either the lower (0.95$\times$$\widetilde{T}_{\rm e}$/$T_{\rm e}$(O~{\sc iii}))
or upper boundary (1.05$\times$$\widetilde{T}_{\rm e}$/$T_{\rm e}$(O~{\sc iii})) 
of the range adopted for temperature variations, with a few points in between (Fig. \ref{fig4}b).

We next examine the situation when the He~{\sc i} $\lambda$10830\AA\ 
emission line is included in the minimisation of $\chi^2$ and the 
determination of $Y_{\rm wm}$ (Fig. \ref{fig4}c,d). 
The temperature's behavior remains about the same (Fig. \ref{fig4}d). 
Apparently, our 
data is not accurate enough to allow a more precise derivation of the 
temperature $T_{\rm e}$(He$^+$) within the 
adopted small range of $\pm$5\% around $\widetilde{T}_{\rm e}$.
However, we note that the $T_{\rm e}$(He$^+$)/$T_{\rm e}$(O~{\sc iii}) ratios
do appear to be slightly more evenly distributed between the lower and upper boundaries, 
especially for H~{\sc ii} regions with $T_{\rm e}$(O~{\sc iii}) $<$ 15000K. For H~{\sc ii}
regions with $T_{\rm e}$(O~{\sc iii}) $>$ 18000K, there are also points scattered between 
the two boundaries, although they are less in number as compared to the case when the 
He~{\sc i} $\lambda$10830\AA\  line is not included. These hot H~{\sc ii} regions play an 
important role in the determination of $Y_{\rm p}$ as they are the ones with the lowest metallicities.
Therefore, on statistical ground, the average $T_{\rm e}$(He$^+$) for our  
H~{\sc ii} region sample is similar to $\widetilde{T}_{\rm e}$, which is what is desired. 
Of course, the range of temperature variations can be enlarged, but
this is difficult to justify physically because photoionised H~{\sc ii} region 
models (e.g., those produced with the CLOUDY code) show that the temperatures in 
H$^+$ and He$^+$ zones cannot differ by more than a 
few percent from the temperature derived from the [O~{\sc iii}] emission lines.

On the other hand,
there is now a clear correlation between $N_{\rm e}$(S~{\sc ii})
and $N_{\rm e}$(He$^+$) (Fig. \ref{fig4}c),
despite the relatively high dispersion. This is a consequence of the high 
sensitivity of the flux of the He~{\sc i} $\lambda$10830\AA\ line 
to the density of the H~{\sc ii} region.
The correlation between the number 
densities does not change appreciably when the
electron temperature $T_{\rm e}$(He$^+$) is set to either 
$\widetilde{T}_{\rm e}$(He$^+$) (Fig. \ref{fig5}) or  
$T_{\rm e}$(O~{\sc iii}) (Fig. \ref{fig6}). 

Hence the second virtue of the He~{\sc i} $\lambda$10830\AA\ emission line is: 
the number density $N_{\rm e}$(He$^+$) is much better constrained
when it is taken into account in the determination of helium
abundances. We note, that temperature variations such as those shown  
in Figs. \ref{fig4} -- \ref{fig6} do change the primordial helium mass
fraction $Y_{\rm p}$ (considered in the next section), but only by a little 
amount, not exceeding 0.3\%.

\begin{figure}
\hbox{
\includegraphics[width=6.0cm,angle=-90.]{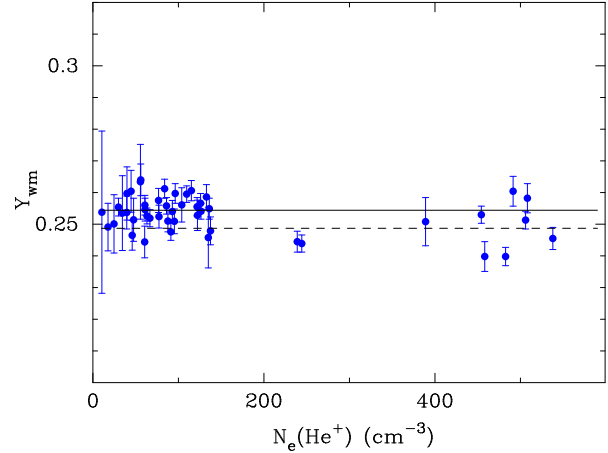}
}
\caption{Relation between the derived number density $N_{\rm e}$(He$^+$)
and helium mass fraction for low-metallicity H~{\sc ii} regions, with
O/H $\leq$ 10$^{-4}$. All six He~{\sc i} emission 
lines $\lambda$3889, $\lambda$4471, $\lambda$5876, $\lambda$6678, 
$\lambda$7065, and $\lambda$10830 have been used for the $\chi^2$ minimisation.
The solid and dashed lines indicate the mean $Y$ value for H~{\sc ii} regions with
$N_{\rm e}$(He~{\sc i}) $\leq$ 200 cm$^{-3}$, and that for  H~{\sc ii} regions with
$N_{\rm e}$(He~{\sc i}) $>$ 200 cm$^{-3}$, respectively.}
\label{fig7}
\end{figure}

\subsubsection{Density effects on the helium abundance}

It is seen in Figs. \ref{fig4}c, \ref{fig5}c, and \ref{fig6}c that the
number density $N_{\rm e}$(He$^+$) in several H~{\sc ii} regions can be high,
greater than 200 cm$^{-3}$. The near-infrared spectrum of one of those 
high-density H~{\sc ii} regions is shown in Fig. \ref{fig1}a. It is
characterised by a very high He~{\sc i} 
$\lambda$10830/P$\gamma$ $\lambda$10940 flux ratio (see also Table \ref{tab2}).
Most of these high-density H~{\sc ii} regions have low oxygen
abundance O/H $<$ 10$^{-4}$, corresponding to 12~+~logO/H~$<$~8.0. Some
of them are among the lowest-metallicity galaxies known, with 12~+~logO/H~$\leq$~7.6.
We check whether the helium mass fraction $Y$ derived in the 
high-density H~{\sc ii} regions differs from that in the low-density H~{\sc ii}
regions. In Fig. \ref{fig7} we show the dependence of the weighted mean helium
mass fraction $Y_{\rm wm}$ on the electron number density
$N_{\rm e}$(He$^+$). Only low-metallicity galaxies, with O/H $<$ 10$^{-4}$, 
are shown. We find that the average $Y_{\rm wm}$ in the high-density H~{\sc ii}
regions with $N_{\rm e}$(He$^+$) $>$ 200 cm$^{-3}$ (dashed horizontal line) is 
$\ga$2\% lower than that in the low-density H~{\sc ii} regions with 
$N_{\rm e}$(He$^+$) $<$ 200 cm$^{-3}$ (solid horizontal line).

$Y_{\rm wm}$s are not expected to vary in H~{\sc ii} regions 
with similar oxygen abundances, like those shown in Fig.~\ref{fig7}. 
Yet, Fig.~\ref{fig7} shows that, in the mean, high-density H~{\sc ii} regions 
have lower $Y_{\rm wm}$ than low-density H~{\sc ii} regions. This is because 
the sensitivity of He~{\sc i} line fluxes to density is not the same for different
lines. In fact, at a given density, collisional excitation effects for 
the He~{\sc i} $\lambda$10830\AA\ emisson line are $\sim$ 10 times stronger 
than for the He~{\sc i} $\lambda$5876\AA\ emisson line.
We have assumed in all our calculations that the density of the H~{\sc ii}
region is constant. However, this is not the case in reality. If some
inhomogeneities and density gradients are present in the H~{\sc ii} regions, 
then the observed volume-averaged fluxes of those He~{\sc i} lines that are more
density-dependent would be more characteristic of the H~{\sc ii} region denser 
parts than of its less dense parts.
In particular, He~{\sc i} $\lambda$10830\AA\ is such a density-dependent line. 
It would tend to be more characteristic of the densest parts of an H~{\sc ii} region. This would lead 
to a overcorrection for collisional excitation as compared to the case of an uniform H~{\sc ii} region, 
and result in an underestimation of $Y_{\rm wm}$. To avoid this effect, we will consider in our final sample 
only H~{\sc ii} regions with an electron density $<$ 200 cm$^{-3}$. 
The choice of this cut-off is motivated by examination of Fig. \ref{fig7} which tells us that
the mean $Y_{\rm wm}$ of H~{\sc ii} regions with densities greater than $\sim$200 cm$^{-3}$ is 
systematically lower than that for H~{\sc ii} regions with $N_{\rm e}$(He$^+$) $\la$ 200 cm$^{-3}$. 
Furthermore, the flux of the He~{\sc i} $\lambda$10830\AA\ emission line 
due to collisional excitation in
H~{\sc ii} regions with $N_{\rm e}$(He$^+$) $>$ 200 cm$^{-3}$ 
exceeds the recombination flux, indicating that this line is
very sensitive to density inhomogeneities. On the other hand, the flux of the 
He~{\sc i} $\lambda$5876\AA\ emission line due to collisional excitation is
at most only 10-20\% of the total flux.


It is also possible
that the difference in $Y_{\rm wm}$ is due to the uncertainties in the 
emissivities of the He~{\sc i} lines and their analytical fits. 
In particular, the analytical fits \citep[see Fig. 1 in ][]{I13} at higher 
number densities deviate more from the exact 
values of emissivities than those at lower number densities (see also
the discussion in Sect. \ref{syst}). 

\section{Primordial He mass fraction $Y_{\rm p}$}\label{primo}

As in our previous work \citep[see ][ and references therein]{I07,I13,IT10},
we determine the primordial He mass fraction
$Y_{\rm p}$ by fitting the data points in the $Y$ -- O/H
plane with a linear regression line of the
form \citep{PTP74,PTP76,P92}
\begin{equation}
Y = Y_p + \frac{{\rm d}Y}{{\rm d}({\rm O/H})} ({\rm O/H}).               \label{eq:YvsO}
\end{equation}

To derive the parameters of the linear regressions,
 we used the maximum-likelihood method \citep{Pr92},
 which takes into account the errors in 
$Y$ and O/H for each object.

\subsection{Inclusion of the He~{\sc i} $\lambda$10830\AA\ line}

\begin{figure*}
\hbox{
\includegraphics[width=6.0cm,angle=-90.]{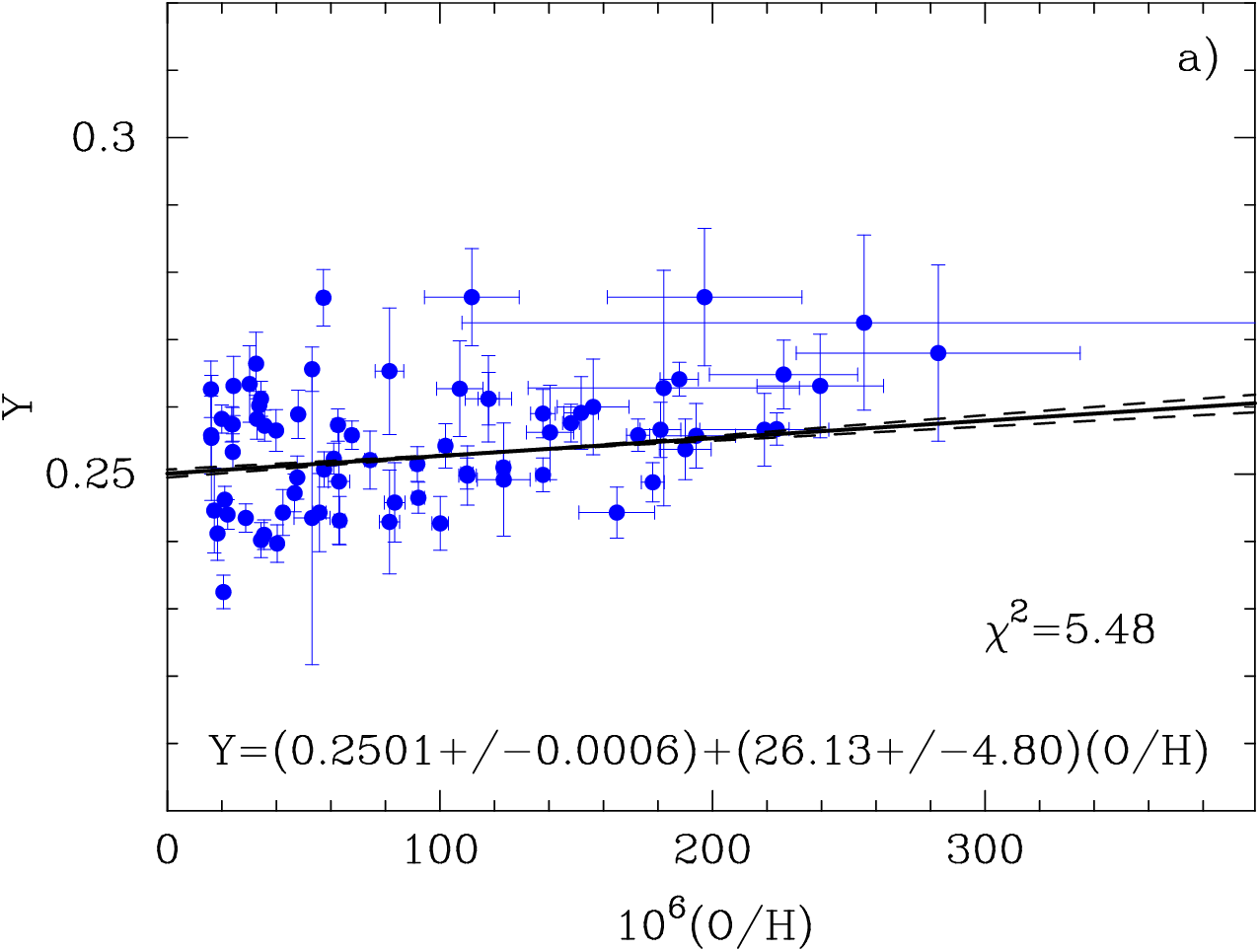}
\hspace{0.3cm}\includegraphics[width=6.0cm,angle=-90.]{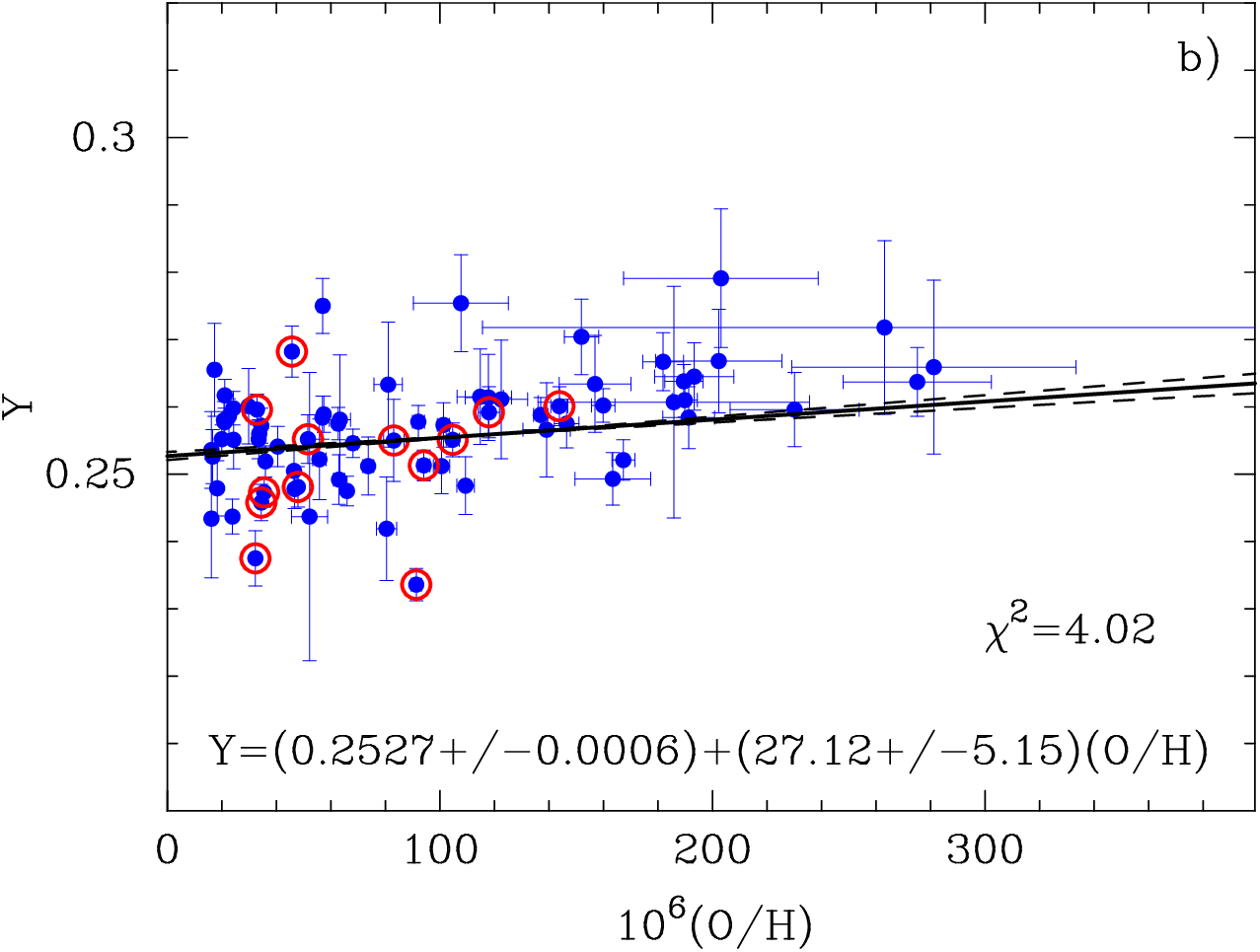}
}
\caption{(a) $Y$ -- O/H linear regression for the 75 spectra
of 45 H~{\sc ii} regions. Only the five 
He~{\sc i} optical emission lines $\lambda$3889, $\lambda$4471, $\lambda$5876, 
$\lambda$6678, and $\lambda$7065 have been used for the $\chi^2$
minimisation and determination of $Y$. 
$T_{\rm e}$(He$^+$) was varied in the range 0.95 -- 1.05 of the 
$\widetilde{T}_{\rm e}$(He$^+$) value. The equation of the linear regression and the value of the $\chi^2$ are given at the bottom of each panel.
(b) Same as (a), but all six 
He~{\sc i} emission lines $\lambda$3889, $\lambda$4471, $\lambda$5876, 
$\lambda$6678, $\lambda$7065, and $\lambda$10830 have been used for the $\chi^2$
minimisation and determination of $Y$. The points representing high-density H~{\sc ii} regions, with
$N_{\rm e}$(He$^+$) $>$ 200 cm$^{-3}$, are encircled in (b).}
\label{fig8}
\end{figure*}

\begin{figure*}
\hbox{
\includegraphics[width=6.0cm,angle=-90.]{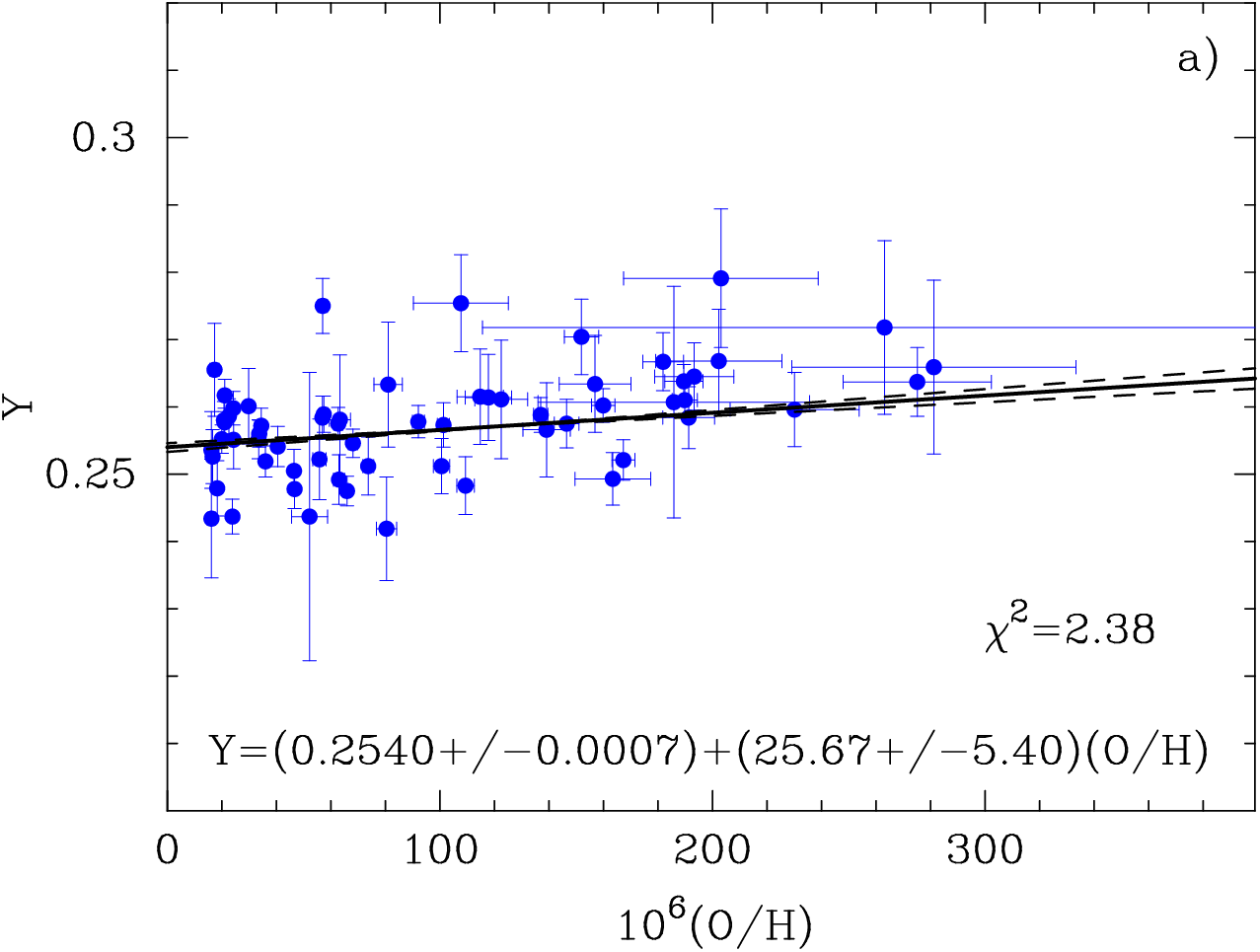}
\hspace{0.3cm}\includegraphics[width=6.0cm,angle=-90.]{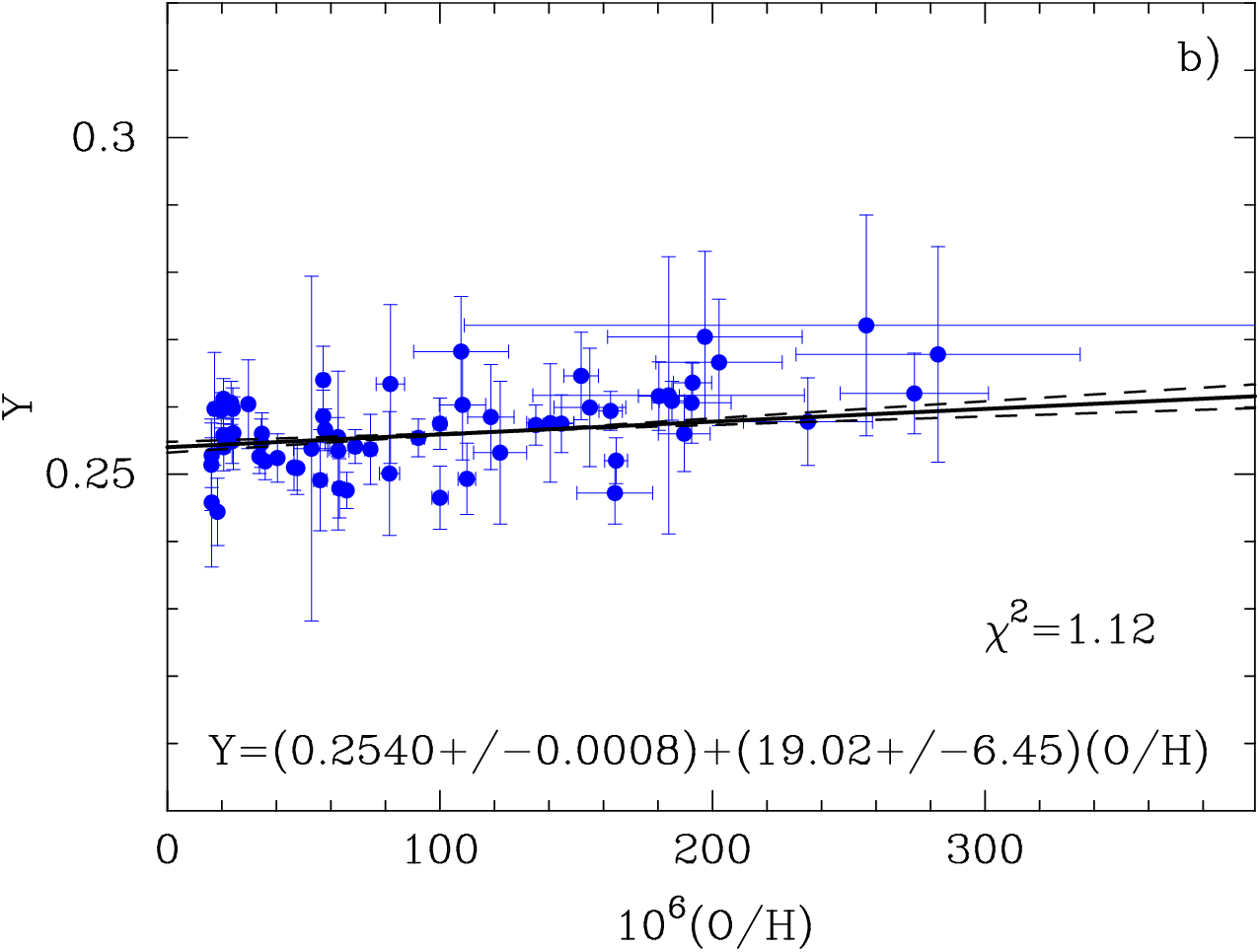}
}
\caption{(a) Same as Fig. \ref{fig8}b, but high-density H~{\sc ii} regions with
$N_{\rm e}$(He$^+$) $>$ 200 cm$^{-3}$ have been excluded. 
(b) Same as (a), but while all six 
He~{\sc i} emission lines $\lambda$3889, $\lambda$4471, $\lambda$5876, 
$\lambda$6678, $\lambda$7065, and $\lambda$10830 have been used for the $\chi^2$
minimisation, only the four He~{\sc i} emission lines $\lambda$4471, $\lambda$5876, 
$\lambda$6678, and $\lambda$10830 have been used for the determination of $Y$.}
\label{fig9}
\end{figure*}

We consider linear regressions $Y$ -- O/H for the entire sample, 
adopting an electron 
temperature $T_{\rm e}$(He$^+$) which is randomly varied in the range 
(0.95 -- 1.05)$\widetilde{T}_{\rm e}$(He$^+$) so as to minimise $\chi^2$, defined by Eq.5. 
$\widetilde{T}_{\rm e}$(He$^+$) is given by Eq.~\ref{tHeOIII}.
We have performed regressions for both the case where 
He~{\sc i} $\lambda$10830\AA\ is not used in the determination of $Y$ (Fig. \ref{fig8}a), 
and the case where that line is used  (Fig. \ref{fig8}b).
Comparison of Figs. \ref{fig8}a and \ref{fig8}b shows that 
the inclusion of the He~{\sc i} $\lambda$10830\AA\ emission line in the 
determination of $Y_{\rm wm}$ reduces the 
scatter of the points. The $\chi^2$ per degree of freedom for the regression line 
in Fig. \ref{fig8}b is 4.02, while it is 5.48 in Fig. \ref{fig8}a.
The $Y$ -- O/H relation  becomes tighter when high-density
H~{\sc ii} regions with $N_{\rm e}$(H$^+$) $>$ 200 cm$^{-3}$ are excluded
from the determination of $Y_{\rm wm}$. With the exclusion of these
H~{\sc ii} regions, the $\chi^2$ per degree of 
freedom becomes 2.38 (Fig. \ref{fig9}a). 

Thus, the third virtue of the He~{\sc i} $\lambda$10830
emission line is to make the $Y$ -- O/H relation much tighter
as compared to all previous studies.

\begin{figure*}
\hbox{
\includegraphics[width=12.0cm,angle=-90.]{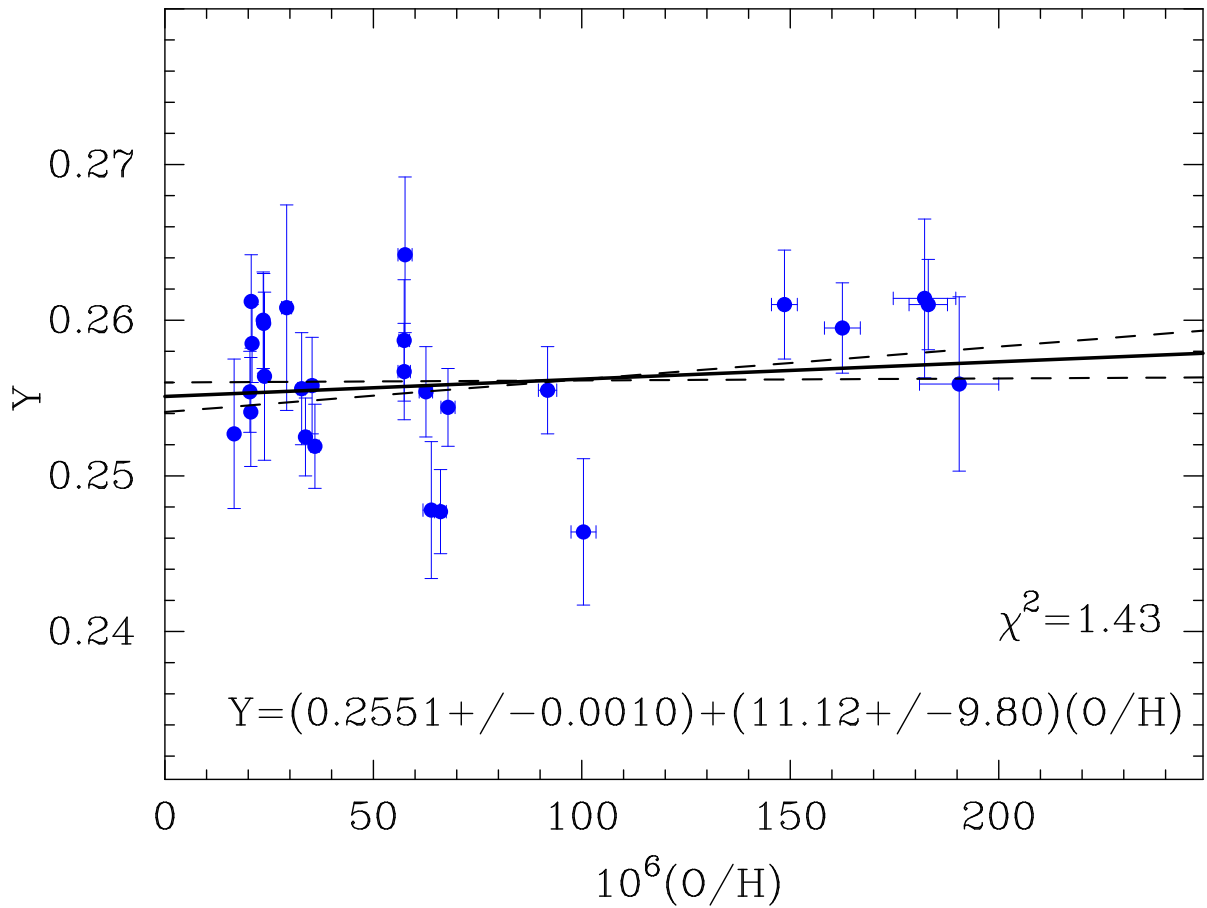}
}
\caption{Same as in Fig. \ref{fig8}b, but only 28 H~{\sc ii} regions with 
EW(H$\beta$) $\geq$ 150\AA, an excitation parameter 
O$^{2+}$/O $\geq$ 0.8 and a 1$\sigma$ error in $Y$ $\leq$3\% are
included.}
\label{fig10}
\end{figure*}

\subsection{Exclusion of the He~{\sc i} $\lambda$3889 and $\lambda$7065 lines 
and of high-density H~{\sc ii} regions}

The $Y$ -- O/H relation becomes 
even tighter, with a $\chi^2$ of only 1.12 (Fig. \ref{fig9}b),
when the two most uncertain He~{\sc i} $\lambda$3889 and $\lambda$7065 
emission lines (Fig. \ref{fig2}) are excluded
from the determination of $Y_{\rm wm}$.
The He~{\sc i} $\lambda$3889 line is uncertain because of contamination by the 
hydrogen H8 $\lambda$3889 line, and the He~{\sc i} $\lambda$7065 emission 
line is uncertain because of its weakness.  
The parameters derived for the sample shown in Fig. \ref{fig9}b, 
with the additional inclusion of the high-density H~{\sc ii} regions, are presented
in Table \ref{tab3}. The uncertainties of these parameters are propagated 
in the derivation of errors of the He mass fraction $Y$.

\subsection{Final NIR sample}

We have further restricted the final NIR sample to those H~{\sc ii} regions with 
high EW(H$\beta$) and high excitation parameters $x$=O$^{2+}$/O. Both these parameters 
are higher in younger starbursts.
In their analysis of a large sample of several hundred 
galaxies, \citet{I13} have found that the weighted mean $Y_{\rm wm}$ increases with
decreasing EW(H$\beta$) and decreasing $x$
(their Figs. 10a and 11). These trends are unphysical and    
suggest that objects with low
EW(H$\beta$) and low $x$ should not be used for the determination of 
$Y_{\rm p}$. The main reason for not including objects with low EW(H$\beta$) is because 
of the larger contribution of the light of the underlying galaxy to their optical continuum, so that 
EW(H$\beta$) is no longer an accurate starburst age indicator. The starburst age would then be overestimated,
resulting in overestimated values of $ICF$(He) and $Y_{\rm p}$. 
On the other hand, \citet{I13} have shown that no trend is apparent for galaxies with
EW(H$\beta$) $\geq$ 150\AA\ and $x$ $\geq$ 0.8, limits which we adopt for our sample. \citet{I13} found 
that the $ICF$(He) derived for these galaxies using starburst ages based on EW(H$\beta$) are consistent 
with those derived using    
SED fitting (their Fig. 12). There is a further advantage in using only high-excitation H~{\sc ii} regions: 
He abundances for most objects with EW(H$\beta$) $\geq$ 150\AA\ and 
$x$ $\geq$ 0.8 are derived with an accuracy better than 3\% because
of a stronger [O~{\sc iii}]~$\lambda$4363 emission, resulting in    
more accurate electron temperatures and derived abundances.

Therefore, our final sample (hereafter the NIR sample) consists only   
of those H~{\sc ii} regions that have EW(H$\beta$)~$\geq$150\AA, 
excitation ratios O$^{2+}$/O~$\geq$~0.8
and $\sigma$($Y_{\rm wm}$)/$Y_{\rm wm}$~$\leq$~3\%. These selection criteria give 
a total sample of 28 H~{\sc ii} regions. 
The linear regression for the NIR sample, excluding He~{\sc i} $\lambda$3889 and
$\lambda$7065 from the determination of $Y_{\rm wm}$, is shown in 
Fig. \ref{fig10}. 
We derive a primordial helium abundance mass fraction 
$Y_{\rm p}$ = 0.2551$\pm$0.0010 
(statistical), with a $\chi^2$ per degree of freedom of 1.43. Although the 
dispersion about the regression line is considerably smaller than that, for 
example, of the HeBCD sample analyzed by \citet{IT10}, the statistical error 
on the $Y_{\rm p}$ (0.0010) derived from the NIR sample is similar to the one 
derived for the HeBCD sample. This is because the size of the NIR sample 
(28 galaxies) is also considerably smaller than that of the HeBCD sample 
(86 galaxies).

For comparison, we have also derived $Y_{\rm p}$ for a sample where only the density 
limit has been set (i.e. the 13 H~{\sc ii}
regions with $N_{\rm e}$ $>$ 200 cm$^{-3}$ have been excluded), and where 
the conditions $x$$\geq$0.8 and $\sigma$($Y_{\rm wm}$)/$Y_{\rm wm}$~$\leq$~3\% are not imposed 
(Fig. \ref{fig9}b). This results in a sample that is more than twice as large as the previous sample 
(62 instead of 28 H {\sc ii}
regions), giving $Y_{\rm p}$ = 0.2540$\pm$0.0008, with a somewhat lower $\chi^2$ of 1.12. 
This value is consistent, within a 1$\sigma$ error, with the value $Y_{\rm p}$ = 0.2551$\pm$0.0010, 
obtained for the final sample. We prefer however the final sample value as it is not subject to 
the unphysical trends discussed above.





\subsection{Systematic effects}\label{syst}
 
We now estimate the systematic errors in our helium abundance determination. We have considered 
three sources of systematic errors:
1) uncertainties of the He~{\sc i} emissivities and their fits; 2) uncertainties
in the fits of the ionisation correction factors $ICF$(He); and 3) 
uncertainties due to the correction for the
non-recombination contribution to hydrogen-line intensities.

Concerning the uncertainties in the He~{\sc i} emissivities, we used the results of   
\citet{P09} who performed Monte Carlo calculations to estimate uncertainties
of their old He~{\sc i} emissivities, not used here. They found that ``optimistic'' estimates of the 
uncertainties in the case of low-density extragalactic H {\sc ii} regions with
$N_{\rm e}$ = 100 cm$^{-3}$ are far below 1\% for most of He {\sc i} emission
lines \citep[Table 2 in ][]{P09}, except for the 
He {\sc i} $\lambda$10830\AA, for which the 
uncertainty can be as high as 2\%. This gives a weighted emissivity uncertainty of less 
than 2\% for all lines. \citet{P09} also found the emissivity 
uncertainties to be higher in denser H {\sc ii} regions because of the uncertainties introduced by 
the collisional excitation parameters. They
are thus $\sim$ 5-10 times higher at $N_{\rm e}$ = 10$^{4}$ cm$^{-3}$ than at $N_{\rm e}$ = 10$^{2}$ cm$^{-3}$, 
the latter value approximating better the density of our H {\sc ii} regions.
 \citet{P12} later analysed the uncertainties of the new emissivities that are used here,  
and found that the \citet{P09} ``optimistic'' uncertainty estimates for the old emissitivities become  
"realistic" estimates for the new ones. 

Regarding the analytical fits of emissivities,
they are accurate to better than 1\% in the ranges $T_{\rm e}$ = (1-2)$\times$10$^4$K
and $N_{\rm e}$ $\leq$ 100 cm$^{-3}$. Uncertainties increase however for denser
H {\sc ii} regions \citep[Fig. 1 in ][]{I13}. Since the majority of our 
H {\sc ii} regions have low densities, we have adopted uncertainties of 1\% 
for the He~{\sc i} emissivities and their fits.

As for the ionisation correction factors $ICF$(He), they do not deviate from unity 
by more than $\sim$ 2\% for all the H {\sc ii} regions in our
sample (Table \ref{tab3}). That deviation 
depends on both the starburst age and the excitation parameter $x$=O$^{+2}$/O.
From modeling photoionised H {\sc ii} regions with varying starburst 
ages and excitation parameters, \citet{I13}
estimated the $ICF$ uncertainties to be $\sim$ 0.25\%, which we adopt. 
Using the same photoionised H {\sc ii} region models
\citet{I13} found a dispersion of 0.5\% for the
non-recombination intensities of hydrogen lines, varying starburst ages and metallicities  
(their Fig. 4). We thus adopt uncertainties of 0.5\% for these quantities.

To estimate the systematic uncertainty introduced in $Y_{\rm p}$ by these 
effects, we have run Monte Carlo simulations varying the
above three sources of systematic errors. We find that these uncertainty 
sources are responsible for a $\leq$0.75\% systematic error in $Y_{\rm p}$,
or 0.0019 in $Y_{\rm p}$. Combining quadratically this systematic error with 
the statistical error of 0.0010 
derived for the regression
fit (Fig. \ref{fig10}), we obtain a total uncertainty
of $\sqrt{0.0019^2+0.0010^2}$ = 0.0022. 

Thus,  our final determination is: 
\begin{equation}
Y_{\rm p}=0.2551\pm0.0022\ {\rm (statistical + systematic)}. \label{Yp}
\end{equation}

The newly derived $Y_{\rm p}$ using the NIR He~{\sc i} line is in good agreement
with the values derived by \citet{IT10} and \citet{A12}. However,
this agreement is somewhat fortuitous because different sets of He~{\sc i}
emissivities and corrections for non-recombination excitation of hydrogen
lines have been used in these various studies.

   \begin{figure}
   \centering
   \includegraphics[width=5.5cm,angle=-90]{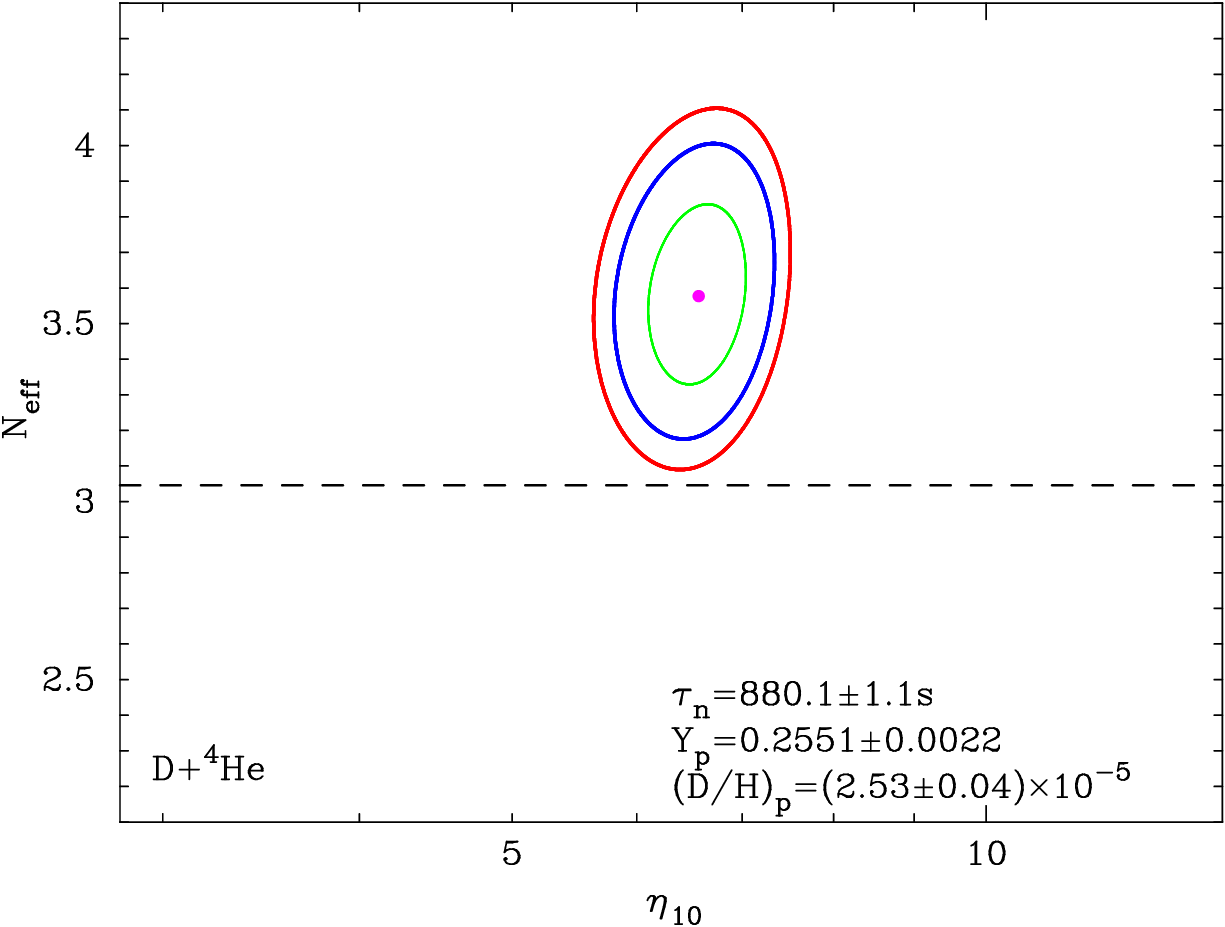}
      \caption{Joint fits to the baryon-to-photon number ratio, 
$\eta_{10}$=10$^{10}$$\eta$, and the effective number of light neutrino 
species $N_{\rm eff}$, using a $\chi^2$ analysis with the code developed by 
\citet{Fi98} and \citet{Li99}. The primordial value of the He abundance has
been set to $Y_{\rm p}$ = 0.2551 (Fig. \ref{fig9}) and that of (D/H)$_{\rm p}$ 
is taken from \citet{C14}. The neutron lifetime is taken to be 
$\tau_{\rm n}$ = 880.1 $\pm$ 1.1s \citep{B12}. The filled circle corresponds to $\chi^2$ = 
$\chi^2_{\rm min}$ = 0.
Ellipses from the inside out correspond respectively to confidence levels of 68.3\% 
($\chi^2$ -- $\chi^2_{\rm min}$ = 2.30), 95.4\% 
($\chi^2$ -- $\chi^2_{\rm min}$ = 6.17) and 99.0\% 
($\chi^2$ -- $\chi^2_{\rm min}$ = 9.21). The SBBN value $N_{\rm eff}$ = 3.046
is shown with a dashed line.}
         \label{fig11}
   \end{figure}

\section{Cosmological implications}\label{cosmo}

The primordial He mass fraction we have derived 
is higher (at the 3.4 $\sigma$ level) than
the SBBN value of 0.2477$\pm$0.0001 inferred from analysis of the 
temperature fluctuations of the cosmic microwave background (CMB) radiation 
observed by the {\sl Planck} satellite \citep{A13}, in the context of the 
standard spatially flat six-parameter $\Lambda$CDM model. This may indicate 
small deviations from the standard rate of Hubble expansion in the early Universe.
These deviations may be caused by an extra contribution to the total energy 
density of the Universe from a "dark radiation" component, for example, 
additional species of neutrinos such as "sterile" neutrinos \citep{D14}. 
The different species of weakly interacting light relativistic particles can 
be conveniently characterised by the parameter $N_{\rm eff}$, the effective number 
of neutrino species. 

We have used the statistical $\chi^2$ technique, with the code
described by \citet{Fi98} and \citet{Li99}, to analyse the constraints 
that the measured He and D
abundances put on the baryon-to-photon number ratio $\eta$ and $N_{\rm eff}$. 
We have not included the abundance of $^7$Li as a constraint because the discrepancy between 
the low lithium abundances measured in metal-poor halo stars in the Milky Way and the predictions of SBBN has not yet been resolved satisfactorily \citep{St12}. 
We have adopted a deuterium abundance (D/H)$_{\rm p}$ = (2.53$\pm$0.04)$\times$10$^{-5}$ \citep{C14}, and the most recently published value for the neutron lifetime 
$\tau_{\rm n}$ = 880.1 $\pm$ 1.1 s \citep{B12}.

The joint fits of $\eta$ and $N_{\rm eff}$ are shown 
in Figure \ref{fig11}. With two degrees of freedom ($\eta_{10}$ and $N_{\rm eff}$), 
deviations at the 68.3\% confidence level (CL) correspond to 
$\chi^2$ -- $\chi^2_{\rm min}$ = 2.30,
those at the 95.4\% CL to $\chi^2$ -- $\chi^2_{\rm min}$ = 6.17, and those  
at the 99.0\% CL to $\chi^2$ -- $\chi^2_{\rm min}$ = 9.21. These confidence levels  
are shown in Figure \ref{fig11} by ellipses from the inside out.

With $Y_{\rm p}$ = 0.2551$\pm$0.0022,
the minimum $\chi^2_{\rm min}$ = 0 is obtained for $\eta_{10}$ = 6.57, corresponding to 
$\Omega_{\rm b}h^2$ = 0.0240$\pm$0.0017 (68\% CL), $\pm$0.0028  (95.4\% CL) and
$\pm$0.0034  (99.0\% CL), 
and to $N_{\rm eff}$ = 3.58$\pm$0.25 (68\% CL), $\pm$0.40  (95.4\% CL) and $\pm$0.50  
(99.0\% CL).  Our derived value of $\Omega_{\rm b}h^2$ is in agreement, within the errors, with the ones 
derived from the WMAP-7 and Planck CMB data of respectively 0.0222$\pm$0.0004 \citep{K11} and 0.0221$\pm$0.00033 (68\% CL) \citep{A13}.

However, our derived value of $N_{\rm eff}$ 
is higher than the SBBN value of 3.046 at the 99\% CL, implying 
devations from SBBN. 
We note that, while the primordial helium abundance is not as precise a 
baryometer as deuterium, it sets  tight constraints on the effective number 
of neutrino species. These constraints
are similar to or are tighter than those derived using the CMB and 
galaxy clustering
power spectra. For example, using these two sets of data, \citet{Ko11}
derived $N_{\rm eff}$ = 4.34$^{+0.86}_{-0.88}$ at the 68\% confidence level.
\citet{K11} analysed joint {\sl WMAP}-7 and 
South Pole Telescope (SPT) data on CMB temperature fluctuations 
derived $N_{\rm eff}$ = 3.85 $\pm$ 0.62 (68\% CL).
Adding low-redshift measurements of the Hubble constant $H_0$ using the Hubble
Space Telescope, and the baryon acoustic oscillations (BAO) using
SDSS and 2dFGRS, \citet{K11} obtained
$N_{\rm eff}$ = 3.86 $\pm$ 0.42 (68\% CL).
On the other hand, \citet{A13}, using the data of the {\sl Planck} mission, 
derived $N_{\rm eff}$ = 
3.30 $\pm$ 0.27 (68\% CL), while \citet{Dv14} obtained
$N_{\rm eff}$=3.86$\pm$0.25 by combining data from the {\sl Planck} and ACT/SPT 
temperature spectra, {\sl WMAP}-9 polarization, $H_0$, baryon acoustic oscillation 
and local cluster abundance measurements with BICEP2 data.
Thus, there appears to be general agreement between the $N_{\rm eff}$
obtained in this paper and the values derived by other researchers with 
different methods: a non-standard value is preferred at the 99\% CL.


\section{Conclusions}\label{summary}

We present for the first time spectroscopic observations 
in the near-infrared (NIR) range of the
high-intensity density-sensitive He~{\sc i} $\lambda$10830\AA\ emission line
for a large sample of 45 low-metallicity high-excitation H~{\sc ii} regions
in star-forming dwarf galaxies. Using this NIR line flux in combination 
with existing spectroscopic data in the optical range of the same 
H~{\sc ii} regions, we have obtained a new determination of the primordial 
He abundance. At the same time, we have also shown the importance of the 
He~{\sc i} NIR line for improving the accuracy of He abundance determinations.

Our main results are summarised below.

1. We demonstrate that the use of the He~{\sc i} $\lambda$10830\AA\
emission line greatly improves the determination of the physical conditions in 
the H~{\sc ii} regions because of the strong dependence of its flux on the electron 
number density.

2. We find that the linear regressions $Y$ -- O/H used for determination
of the primordial He mass fraction $Y_{\rm p}$ are much tighter than those
studied previously if the NIR He~{\sc i} $\lambda$10830\AA\ emission line is included,  
rather than relying only on the optical He~{\sc i} lines. 

3. Using the linear regression $Y$ -- O/H for a sample of 28 highest-excitation H~{\sc ii}
regions, we have derived a primordial He mass fraction
$Y_{\rm p}$ = 0.2551 $\pm$ 0.0022. This is higher than the
standard big bang nucleosynthesis (SBBN) value of 0.2477 $\pm$ 0.0001
inferred from the temperature fluctuations of the microwave background 
radiation. This difference possibly indicates deviations from the standard rate of Hubble expansion in 
the early Universe, and hence the possible presence of dark radiation.  

4. Using our derived He primordial abundance together with the most recently derived primordial abundance of D, and the $\chi^2$ technique, we found that the best 
agreement between the abundances of these light elements is achieved in the BBN 
model with a baryon mass fraction $\Omega_{\rm b} h^2$ = 0.0240$\pm$0.0034
(99\% CL) and an effective number of neutrino species $N_{\rm eff}$ = 
3.58$\pm$0.50 (99\% CL). Both the $\Omega_{\rm b} h^2$ and 
$N_{\rm eff}$ values agree with those inferred from the 
temperature fluctuations of the microwave background radiation. A non-standard value of $N_{\rm eff}$ is preferred at the 99\% CL, implying the possible existence of additional types of neutrino species.

\section*{Acknowledgments}

Y.I.I. thanks the University of Virginia for warm hospitality. T.X.T. 
acknowledges the financial support of NASA/JPL  grant RSA1463350. 
This study is based on observations made with ESO Telescopes at the La Silla 
Paranal Observatory under programmes 59.A-9004(D),
64.H-0467(A), 65.I-0097(A), and 67.B-0287(B).
It is also based on observations with the Large Binocular Telescope (LBT)
and the Apache Point Observatory (APO) 3.5m telescope. 
The LBT is an international collaboration among institutions in the United 
States, Italy and Germany. LBT Corporation partners are: The University of 
Arizona on behalf of the Arizona university system; Istituto Nazionale di 
Astrofisica, Italy; LBT Beteiligungsgesellschaft, Germany, representing the 
Max-Planck Society, the Astrophysical Institute Potsdam, and Heidelberg 
University; The Ohio State University, and The Research Corporation, on 
behalf of The University of Notre Dame, University of Minnesota and 
University of Virginia. The APO 3.5m telescope 
is owned and operated by the Astrophysical Research Consortium.
Funding for the Sloan Digital Sky Survey (SDSS) and SDSS-II has
been provided by the Alfred P. Sloan Foundation, the Participating
Institutions, the National Science Foundation, the U.S. Department
of Energy, the National Aeronautics and Space Administration, the
Japanese Monbukagakusho, the Max Planck Society, and the
Higher Education Funding Council for England.

\bsp

\label{lastpage}


\begin{thebibliography}{}

\bibitem[Abazajian et al.(2009)]{Ab09} Abazajian, K., et al. 2009, ApJS,
182, 543
\bibitem[Ade et al.(2013)]{A13} Planck Collaboration: Ade, P. A. R., et al. 
2013, A\&A, in press; preprint arXiv:1303.5076
\bibitem[Asplund et al.(2009)]{A09} Asplund, M., Grevesse, 
N., \& Sauval, A.~J. \& Scott, P. 2009, ARAA, 47, 481
\bibitem[Aver et al.(2010)]{A10} Aver, E., Olive, K. A., \&
Skillman, E. D. 2010, J. Cosmology Astropart. Phys., 05, 003A
\bibitem[Aver et al.(2011)]{A11} Aver, E., Olive, K. A., \&
Skillman, E. D. 2011, J. Cosmology Astropart. Phys., 03, 043A
\bibitem[Aver et al.(2012)]{A12} Aver, E., Olive, K. A., \&
Skillman, E. D. 2012, J. Cosmology Astropart. Phys., 04, 004A
\bibitem[Aver et al.(2013)]{Av13} Aver, E., Olive, K. A., Porter, R. L., 
\& Skillman, E. D. 2013, J. Cosmology Astropart. Phys., 11, 017A
\bibitem[Benjamin et al.(1999)]{B99} Benjamin, R. A.,
Skillman, E. D., \& Smits, D. P. 1999, ApJ, 514, 307
\bibitem[Benjamin et al.(2002)]{B02} Benjamin, R. A.,
Skillman, E. D., \& Smits, D. P. 2002, ApJ, 569, 288
\bibitem[Beringer et al. (2012)]{B12} Beringer, J., et al. (Particle Data
Group) 2012, Phys. Rev. D, 86, 010001
\bibitem[Cardelli et al.(1989)]{C89} Cardelli, J. A., Clayton, G. C., \&
Mathis, J. S. 1989, ApJ, 345, 245
\bibitem[Cooke et al.(2014)]{C14} Cooke, R. J., Pettini, M., Jorgenson, R. A.,
Murphy, M. T., \& Steidel, C. C. 2014, ApJ, 781, 31
\bibitem[Di Valentino et al.(2013)]{D14} Di Valentino, E., Melchiorri, A., 
\& Mena, O. 2013, J. Cosmology Astropart. Phys., 11, 018D
\bibitem[Dvorkin et al.(2014)]{Dv14} Dvorkin, C., Wyman, M., Rudd, D. H.,
\& Hu, W. 2014, preprint arXiv:1403.8049

\bibitem[Ferland et al.(1998)]{F98} Ferland, G. J., Korista, K. T.,
Verner, D. A., Ferguson, J. W., Kingdon, J. B., \& Verner, E. M. 1998,
PASP, 110, 761
\bibitem[Ferland et al.(2013)]{F13} Ferland, G. J., et al. 2013, Rev. Mexicana
 Astron. Astrofis., 49, 137
\bibitem[Fiorentini et al.(1998)]{Fi98} Fiorentini, G., Lisi, E., Sarkar, S., 
\& Villante, F. L. 1998, Phys. Rev. D, 58, 063506
\bibitem[Gonz\'alez Delgado et al.(2005)]{GD05} Gonz\'alez Delgado, R. M., 
Cervi\~no, M., Martins, L. P., Leitherer, C., \& Hauschildt, P. H. 
2005, MNRAS, 357, 945
\bibitem[Guseva et al.(2003)]{G03} Guseva, N. G., Papaderos, P.,
Izotov, Y. I., Green, R. F., Fricke, K. J., Thuan, T. X., \& Noeske, K. G.
  2003, A\&A, 2003, 407, 105
\bibitem[Guseva et al.(2011)]{G11} Guseva, N. G., Izotov, Y. I., 
Stasi\'nska, G., Fricke, K. J., Henkel, C., \& Papaderos, P. 2011, A\&A,
529, A149
\bibitem[Hummer \& Storey(1992)]{HS92} Hummer, D. G., \& Storey, P. J. 1992,
MNRAS, 254, 277
\bibitem[Iocco et al.(2009)]{Io09} Iocco, F., Mangano, G., Miele, G., 
Pisanti, O., \& Serpico, P. D. 2009, Phys.Rept., 472, 1
\bibitem[Izotov \& Thuan(1998)]{IT98} Izotov, Y. I., \& Thuan, T. X. 1998, 
ApJ, 500, 188
\bibitem[Izotov \& Thuan(2004)]{IT04} Izotov, Y. I., \& Thuan, T. X. 2004, ApJ, 602, 200
\bibitem[Izotov \& Thuan(2010)]{IT10} Izotov, Y. I., \& Thuan, T. X. 2010,
ApJ, 710, L67
\bibitem[Izotov et al.(1994)]{ITL94} Izotov, Y. I.,
Thuan, T. X., \& Lipovetsky, V. A. 1994, ApJ, 435, 647
\bibitem[Izotov et al.(1997)]{ITL97} Izotov, Y. I.,
Thuan, T. X., \& Lipovetsky, V. A. 1997, ApJS, 108, 1
\bibitem[Izotov et al.(1999)]{I99} Izotov, Y. I., Chaffee, F. H., Foltz,
C. B., Green, R. F., Guseva, N. G., \& Thuan, T. X. 1999, ApJ, 527, 757
\bibitem[Izotov et al.(2001)]{ICG01} Izotov, Y. I.,
Chaffee, F. H., \& Green, R. F. 2001, ApJ, 562, 727
\bibitem[Izotov et al.(2004)]{I04} Izotov, Y. I., Papaderos, P., 
Guseva, N. G., Fricke, K. J., \& Thuan, T. X. 2004, A\&A, 421, 539
\bibitem[Izotov et al.(2006)]{I06} Izotov, Y. I., Stasi\'nska, G., Meynet, G.,
Guseva, N. G., \& Thuan, T. X. 2006, A\&A, 448, 955
\bibitem[Izotov et al.(2007)]{I07} Izotov, Y. I., Thuan, T. X., \& 
Stasi\'nska, G. 2007, ApJ, 662, 15
\bibitem[Izotov et al.(2009)]{I09} Izotov, Y. I., Guseva, N. G., Fricke, K. J.,
\& Papaderos, P. 2009, A\&A, 503, 61
\bibitem[Izotov et al.(2011)]{I11b} Izotov, Y. I., Guseva, N. G.,
Fricke, K. J., \& Henkel, C. 2011, A\&A, 533, A25
\bibitem[Izotov et al.(2013)]{I13} Izotov, Y. I., Stasińska, G., 
\& Guseva, N. G. 2013, A\&A, 558, A57
\bibitem[Keisler et al.(2011)]{K11} Keisler, R., Reichardt C. L., Aird, K. A.,
et al. 2011, ApJ, 743, 28
\bibitem[Komatsu et al.(2011)]{Ko11} Komatsu, E., et al. 2011, ApJS, 192, 18
\bibitem[Leitherer et al.(1999)]{L99} Leitherer, C., Schaerer, D.,
Goldader, J. D., Gonzalez Delgado, R. M., Robert, C., Kune D. F.,
de Mello, D. F., Devost, D., \& Heckman, T. M. 1999, ApJS, 123, 3
\bibitem[Lisi et al.(1999)]{Li99} Lisi, E., Sarkar, S., \& Villante, F. L. 
1999, Phys. Rev. D, 59, 123520
\bibitem[Maeder(1992)]{M92} Maeder, A. 1992, A\&A, 264, 105
\bibitem[Noterdaeme et al.(2012)]{N12} Noterdaeme, P., L\'opez, S., Dumont, V.,
Ledoux, C., Molaro, P., \& Petitjean, P. 2012, A\&A, 542, L33
\bibitem[Pagel et al.(1992)]{P92} Pagel, B. E. J., Simonson, E. A.,
Terlevich, R. J., \& Edmunds, M. G. 1992,
MNRAS, 255, 325
\bibitem[Peimbert \& Torres-Peimbert(1974)]{PTP74} Peimbert, M., \&
Torres-Peimbert, S. 1974, ApJ, 193, 327
\bibitem[Peimbert \& Torres-Peimbert(1976)]{PTP76} Peimbert, M., \&
Torres-Peimbert, S. 1976, ApJ, 203, 581
\bibitem[Peimbert et al.(2007)]{Pe07} Peimbert, M., Luridiana, V.,
\& Peimbert, A. 2007, ApJ, 666, 636
\bibitem[Pettini \& Cooke(2012)]{PC12} Pettini, M., \& Cooke, R. 2012,
MNRAS, 425, 2477 
\bibitem[Porter et al.(2009)]{P09} Porter, R. L., Ferland, G. J., 
MacAdam, K. B., \& Storey, P. J. 2009, MNRAS, 393, L36
\bibitem[Porter et al.(2012)]{P12} Porter, R. L., Ferland, G. J.,
Storey, P. J., \& Detisch, M. J. 2012, MNRAS, 425, L28 
\bibitem[Porter et al.(2013)]{P13} Porter, R. L., Ferland, G. J.,
Storey, P. J., \& Detisch, M. J. 2013, MNRAS, 433, L89 
\bibitem[Press et al.(1992)]{Pr92} Press, W. H., Teukolsky, S. A., 
Vetterling, W. T.,\& Flannery, B. P., 1992, Numerical Recipes in C, The Art of 
Scientific Computing /Second Edition/, Cambrige University Press
\bibitem[Steigman(2005)]{St05} Steigman, G. 2005, Phys. Scr., T121, 142
\bibitem[Steigman(2006)]{St06} Steigman, G. 2006, Int. J. Mod. Phys. E, 
15, 1
\bibitem[Steigman(2012)]{St12} Steigman, G. 2012, preprint arXiv:1208.0032
\bibitem[Wilson et al.(2004)]{W04} Wilson, J. C., et al. 2004, Proc. SPIE,
5492, 1295

\end{thebibliography}
\end{document}